\documentclass{ws-acs}
\usepackage{url}
\usepackage{multirow}
\usepackage{tikz}
\usepackage{caption}
\usepackage{subcaption}
\usepackage{placeins}

\usepackage
	{todonotes}

\usepackage[
    colorlinks=true,
    bookmarksnumbered=true,bookmarksopen=true,
	unicode=true,colorlinks=true,linktoc=all,
	linkcolor=blue,citecolor=blue,filecolor=blue,urlcolor=blue,
	pdfstartview=FitH,
	hypertexnames=false
]{hyperref}
\usepackage{doi}
\usepackage{bookmark}

\begin{document}
\makeatletter
\def\@biblabel#1{[#1]}
\makeatother

\markboth{Vincent Labatut}{Complex Network of a Graphic Novel}

%
\catchline{}{}{}{}{}
%

\title{Complex Network Analysis of a Graphic Novel: \\ The Case of the Bande Dessinée \textit{Thorgal}}

\author{VINCENT LABATUT}
\address{Laboratoire Informatique d'Avignon (LIA) \\ Avignon Université \\ 339 chemin des Meinajariès -- BP 1228 \\ F-84911 Avignon Cedex 9 -- France \\ vincent.labatut@univ-avignon.fr}


\maketitle


\begin{abstract}
The task of extracting and analyzing character networks from works of fiction, such as novels and movies, has been the object of a number of recent publications. However, only a very few of them focus on graphic novels, and even fewer on European graphic novels. In this article, we focus on \textit{Thorgal}, a bande dessinée, i.e. a comic of the French-Belgian tradition. We manually annotate all the volumes of this series, in order to constitute a corpus allowing us to extract its character network. We perform a descriptive analysis of the network structure and compare it to real-world and fictional social networks. We also study the effect of character filtering over the network structure. Finally, we leverage complex network analysis tools to answer two research questions from the literature, related to the similarity between \textit{Thorgal} and the \textit{Saga of Icelanders}; and to the position of women in the series. Our data and source code are both publicly available online.
\end{abstract}


\section{Introduction}
\label{sec:Introduction}
Character networks are a specific type of complex network built by identifying the characters involved in a narrative as well as their interactions, and modeling them through vertices and edges, respectively. As such, they constitute a model of the narrative, that can be leveraged to perform a number of tasks, such as summarizing a narrative~\cite{Bost2018a}, identifying its main characters~\cite{Rochat2014a}, and separating its storylines~\cite{Weng2009}. They can also be used to answer various questions related to corpora of narratives, e.g. automatically classifying narratives by genre~\cite{Hettinger2015}, or to specific narratives, e.g. assessing literary theories~\cite{Elson2010} and estimating the level of historicity of a narrative~\cite{MacCarron2012}.

This modeling tool has been applied to a number of works of fiction in recent years, spanning all the main forms of these cultural objects: novels~\cite{Elson2010}, tales~\cite{Schmidt2021}, myths~\cite{MacCarron2012}, fables~\cite{Markovic2018}, plays~\cite{Stiller2003}, movies~\cite{Lee2016e}, TV series~\cite{Bost2016b}, video games~\cite{Rochat2017}, and graphic novels~\cite{Alberich2002} (see~\cite{Labatut2019} for a complete review). However, there are only very few publications regarding the later form, and even fewer deal with \textit{European} graphic novels. This may be because extracting character networks from comics is a difficult task to automate~\cite{Labatut2019}. Yet, considering that the comic form has its own specificities but also mixes different elements from other forms~\cite{Chute2008}, one may wonder how it compares to them, narratively speaking. From the modeling perspective, this amounts to finding which features are common to comic-based character networks, and what makes them different from those extracted from other forms. Of course, answering this question would require a larger corpus of comic-based character networks.

In this article, we add our contribution to this corpus by focusing on \textit{Thorgal}, a bande dessinée, i.e. a comic of French-Belgian tradition. We manually annotate all the volumes of this series, in order to constitute a dataset allowing us to extract its character network. We perform a descriptive analysis of the network structure and compare it to real-world and fictional social networks. We also study how the process of filtering out minor characters, which is largely spread in the literature, affects the network structure. After this, we leverage complex network analysis tools to answer two research questions identified in the literature, and specifically related to \textit{Thorgal}. The first concerns its hypothesized similarity with the \textit{Íslendingasögur}, a Scandinavian corpus of medieval literature~\cite{Thiry2019}. The second is to know whether \textit{Thorgal} actually implements a form of soft masculinity~\cite{Desfontaine2018}, by studying the position of women in the series.

The rest of the article is organized as follows. In Section~\ref{sec:RelatedWork}, we review the literature dealing with the character networks of graphic novels. In Section~\ref{sec:ExtractionMethods}, we describe the methods that we use to annotate our dataset and to extract our character networks. In Section~\ref{sec:Dataset}, we introduce the \textit{Thorgal} series and our related research questions. We also describe the raw dataset obtained after our annotation process. Section~\ref{sec:DescriptiveAnalysis} is dedicated to the descriptive analysis of the character networks, whereas Section~\ref{sec:ResearchQuestions} focuses on the remaining research questions. We conclude in Section~\ref{sec:Conclusion} by summarizing our findings and contributions, and discussing the limitations and possible extensions of our work.

\section{Related Work}
\label{sec:RelatedWork}
There are only a very few published articles dealing with the extraction and analysis of character networks from graphic novels. As far as we know, the articles discussed in this section and listed in Table~\ref{tab:SurveyList} constitute a comprehensive survey. Some of them deal with American super-hero comic books, Japanese mangas, Korean webtoons, or European bande dessinée. 

\begin{table}[htb!]
    \caption{List of articles dealing with the extraction of character networks based on graphic novels.}
    \label{tab:SurveyList}
    \scriptsize
    \begin{tabular}{p{2.1cm} l l p{3.5cm}}
        \toprule\noalign{\smallskip}
        \textbf{Bibliographic} & \multicolumn{2}{l}{\textbf{Graphic Novel}} & \textbf{Application}  \\
        \noalign{\smallskip}\cline{2-3}\noalign{\smallskip}
        \textbf{References} & \textbf{Type} & \textbf{Title} &  \\
        \noalign{\smallskip}\colrule\noalign{\smallskip}
        Alberich \textit{et al}.~\cite{Alberich2002} \newline Gleiser~\cite{Gleiser2007} & Comics & \textit{Marvel Comics} & Study the level of realism of the social network \\
        \noalign{\smallskip}\colrule\noalign{\smallskip}
        Murakami \textit{et al}. \cite{Murakami2011} & Mangas & \textit{Dragon Ball} vol.32 & Detect the most important characters \\
        \noalign{\smallskip}\colrule\noalign{\smallskip}
        Murakami \textit{et al}. & Mangas & \textit{Naruto} vol.46--50 & Build character family trees \\
         \cite{Murakami2018,Murakami2020} &  & \textit{Dragon Ball} vol.29--34 & / kinship graphs \\
         &  & \textit{Boys Over Flowers} vol.16--20 &  \\
         &  & \textit{Space Brothers} vol.1--5 &  \\
        \noalign{\smallskip}\colrule\noalign{\smallskip}
        Rochat \& Triclot & Bandes & \textit{Saga of the Metabarons} vol.1--8 & Narrative classification / \\
         \cite{Rochat2017} & Dessinées & \textit{Worlds of Aldebaran} vol.6--10 & Study of character roles \\
        \noalign{\smallskip}\cline{2-3}\noalign{\smallskip}
         & Mangas & \textit{Akira} vol.1--3 (French Edition) & \\
         &  & \textit{Gunnm} vol.1--7 &  \\
        \noalign{\smallskip}\colrule\noalign{\smallskip}
        Lee \& Kim \cite{Lee2020f} & Webtoons & Collection of 35 webtoons & Predict user preference \\
        \noalign{\smallskip}\botrule
    \end{tabular}
\end{table}

Alberich \textit{et al}.~\cite{Alberich2002} study the \textit{Marvel Comics} social network. They first extract a bipartite graph using the \textit{Marvel Chronology Project}\footnote{\url{http://www.chronologyproject.com/}}, a fan-made database covering the 1961--2002 period, by linking each significant character to the comic books in which they appear. Then, they perform a one-mode projection to get a character-to-character network modeling same-book co-appearances. They compare the resulting unweighted co-occurrence network to a null random model and to real-world collaboration networks, such as the ones resulting from scientific co-authorship or from movie castings. Some of the studied topological properties are realistic, in the sense that they are similar to those of real-world networks (existence of a giant component, small average distance). Others are not, though: compared to the null model, the average degree is even smaller than in realistic collaboration networks, underlining the intensive re-use of the same characters in the books. Similarly, the exponent of the power law-distributed node degree is smaller than in other collaboration networks, confirming the artificiality of the network: unlike in real life networks, where the number of collaborations remains limited because of time constraints, the \textit{Marvel} network is dominated by a few widely connected characters.

Gleiser~\cite{Gleiser2007} uses the same data and extraction method, and extends the work of Alberich \textit{et al}.~\cite{Alberich2002} by considering other network properties. First, the \textit{Marvel} network turns out to be disassortatively mixed by degree, suggesting, like the previously discussed small power law exponent, the presence of large hubs within the network as an evidence of its artificial nature. Second, the author introduces edge weights corresponding to the number of books in which the characters co-appear. The resulting weight distribution follows a power law with a relatively small exponent, suggesting that only a few characters interact frequently. Filtering weak edges reveals a few communities centered around popular characters. Third, the clustering coefficient turns out to be much lower for widely connected characters than for more isolated ones. Besides, the central characters, mostly super-\textit{heroes}, as opposed to super-\textit{villains}, tend to be occasionally connected to each other, forming a \textit{rich-club} of heroes able to triumph over evil, in accordance with the \textit{Comics Authority Code}.

Murakami \textit{et al}.~\cite{Murakami2011} extract character networks from mangas. First, they manually annotate character occurrences in each panel. Second, they use character co-occurrence as a proxy for interaction. The intensity of this interaction is measured by a score that depends on the number of involved characters (the fewer, the better). They also assume an interaction, albeit weaker, between characters appearing in two consecutive panels. Third, they built a static network by summing these scores over the narrative to get edge weights. They apply their extraction method to Toriyama's \textit{Dragon Ball} vol.32, and leverage vertex strength and edge weight to identify automatically the most important characters. A survey conducted over readers confirms the relevance of this approach. In~\cite{Murakami2018,Murakami2020}, the same team takes advantage of the text present in the speech balloons, in order to identify the nature of the character interactions. Both characters and speech text are detected manually. They assume speech is addressed to characters in the same frame, or the next one. They look for predefined kinship words in the text, such as \textit{son} or \textit{father}, and infer the relation between the speaker and the addressee. They mainly look for family relationships (father/son, spouse, etc.) over the whole narrative. They filter out infrequent relationships, considered as noise. They eventually produce a sort of family tree. They apply their method to 4 different mangas this time: \textit{Naruto} (vol.46--50), \textit{Dragon Ball} (vol.29--34), \textit{Boys Over Flowers} (vol.16--20), and \textit{Space Brothers} (vol.1--5).

Rochat \& Triclot~\cite{Rochat2017} do not focus specially on comics, but more generally on a \textit{genre} of narrative: they study a corpus of science-fiction novels, video games, movies, TV series and graphic novels. The latter include mangas \textit{Akira} (vol.1--3) and \textit{Gunnm} (vol.1--7), and bandes dessinées \textit{The Saga of The Meta-Barons} (vol.1--8) and \textit{The worlds of Aldebaran} (vol.6--10). They identify character occurrences manually, and insert an edge between two characters when they appear within two consecutive pages. Its weight is $2$ if both characters occur within the same page, and $1$ for two distinct pages. They sum over the whole book to get a static network, and filter out edges whose weight is smaller than $3$, as these are not considered as significant. Although not explicitly stated in the paper, their data reveal that the authors focus only on the main characters. The authors identify $4$ types of character networks, based on their general structure: unicentric, acentric, polycentric, and kernel networks (we detail these in the Supplementary Material, cf. Section~\ref{sec:AddAnalyNetTypes}). They additionally categorize the characters using $3$ classes (scientist, politician, technician) and study how these are distributed over the networks.

Lee \& Kim~\cite{Lee2020f} use an embedding-based method to predict user preference regarding webtoons. Webtoons are a narrative medium popular in Korea, corresponding to online comics with music and possibly animation. A webtoon is not divided into distinct pages, but is rather a single long strip that can be conveniently read on a smartphone using infinite scrolling. It is generally serially published, and therefore split in distinct episodes. The authors propose a method to learn a multimodal representation of the narrative based not only on the character network, but also on its textual, visual and audio content. They extract a dynamic character network by considering each episode as a time slice. In each episode, they manually identify the characters and their verbal interactions. They assume that all occurring characters address each other, and consequently connect them all in the considered time slice. The numbers of utterances are used as edge weights. They do not directly use the network, but rather project it into an embedding space using their own method~\cite{Lee2020}. They also embed the textual content using a Long Short-Term Memory neural network (LSTM), and visual and audio features using a Convolutional Neural Network (CNN). They combine these representations and use the result to assess user preference over a corpus of $35$ webtoons.

We conclude this review with two general observations. First, only one work deals with European bandes dessinées, and it does not provide a thorough analysis of the extracted networks, as those are used as part of a larger collection of narratives. Second, all authors extract their character networks through co-occurrence based approaches, differing only in the nature of the selected window: whole comic books for Alberich \textit{et al}.~\cite{Alberich2002} and Gleiser~\cite{Gleiser2007}, pairs of panels for Murakami \textit{et al}.~\cite{Murakami2011}, pairs of pages for Rochat \& Triclot~\cite{Rochat2017}, and episodes for Lee \& Kim~\cite{Lee2020f}. By comparison, we adopt a scene-based approach, which we deem more appropriate to the series studied in this article. 

\section{Network Extraction Method}
\label{sec:ExtractionMethods}
A comic \textit{page} is composed of a set of chronologically ordered \textit{panels}, generally separated by an empty space called \textit{gutter}. A panel itself can contain characters, background, but also text that can be contained in speech balloons or directly integrated in the drawings, and even iconography. As such, the automatic processing of comics is related to document image understanding, but with additional constraints regarding organization and structure~\cite{Hirata2016}. The problem of extracting information from comics is very difficult, as information can appear under many forms, and is subject to extreme variations from one artist to the other (or even within the same artist's work). It requires solving a number of lower level problems~\cite{Augereau2018}, which are themselves hard, and for the most part not satisfactorily treated by state-of-the-art methods. These include identifying panel boundaries, ordering panels, identifying speech bubbles and captions, extracting their textual content, detecting out-of-bubble text, identifying characters and recognizing their poses, attributing speech to characters, detecting and interpreting iconography and graphical conventions. For this reason, a part of the character network extraction process is performed manually in all the works described in Section~\ref{sec:RelatedWork}, and this is also the case for us.

In this article, we adopt the general method described in~\cite{Labatut2019} to extract character networks from any type of narrative. It is three-stepped: 1) identify character occurrences throughout the narrative; 2) detect interactions between these characters; and 3) perform some form of temporal integration to obtain a graph. We perform the first and second steps manually (Section~\ref{sec:ExtractionMethodsAnnotation}) whereas the third one is automated (Section~\ref{sec:ExtractionMethodsGraph}).

\subsection{Manual Annotation}
\label{sec:ExtractionMethodsAnnotation}
The annotation step is generally not described in a very detailed way in the literature, if mentioned at all (cf. Section~\ref{sec:RelatedWork}). Yet, the method and nature of the annotation have a direct effect on the constituted dataset, and therefore the results obtained later when processing it. Therefore, this step should not be overlooked. Moreover, the task is not as straightforward as it may seem at first glance, as it requires dealing with various specific situations. These are the reasons why we summarize our annotation approach in this section, and review the most typical of these situations. We illustrate them using excerpts from the \textit{Thorgal} series, which is studied later in this article. 

\begin{figure*}[t!]
    \centering
    \begin{tikzpicture}
        \node[anchor=south west,inner sep=0,align=left] (image) at (0,0) {
	        \begin{subfigure}[t]{0.49\textwidth}
                 \includegraphics[height=5.6cm]{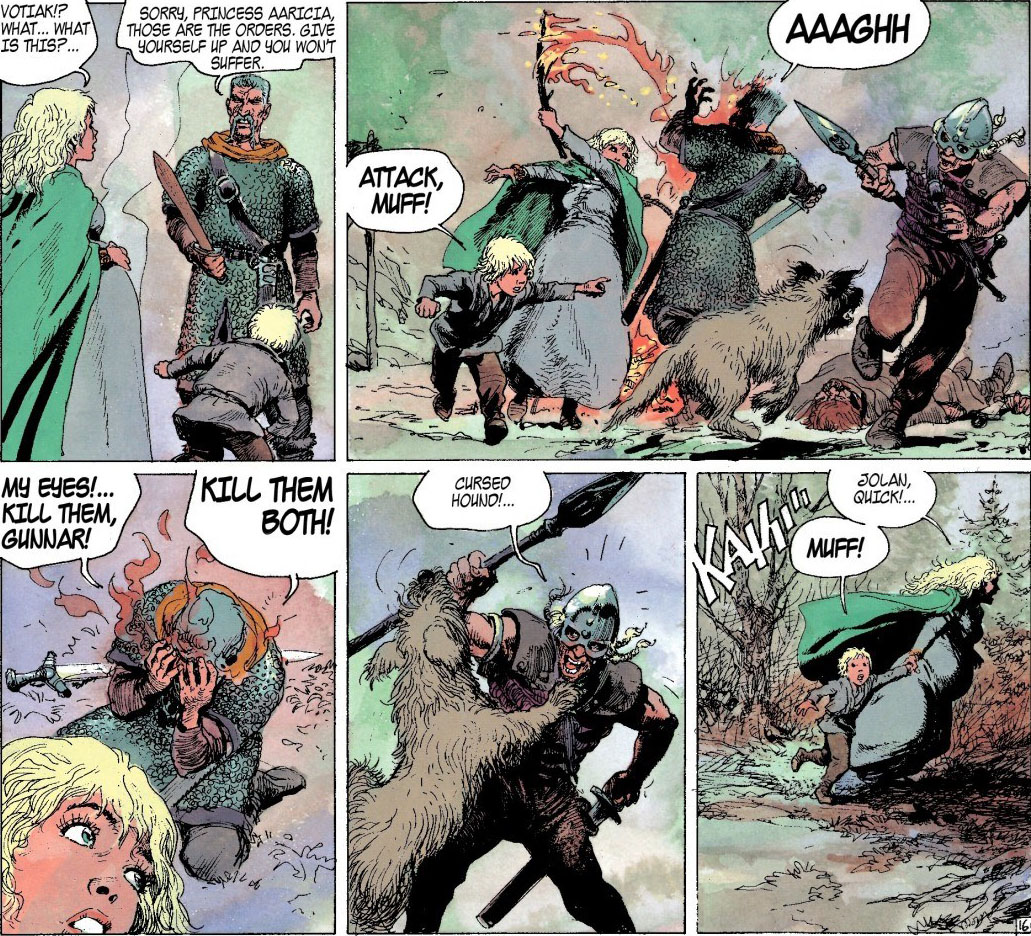}
                \phantomsubcaption\label{fig:BdExAnimal}
            \end{subfigure}\hspace{1mm}
	        \begin{subfigure}[t]{0.49\textwidth}
                \includegraphics[height=5.6cm]{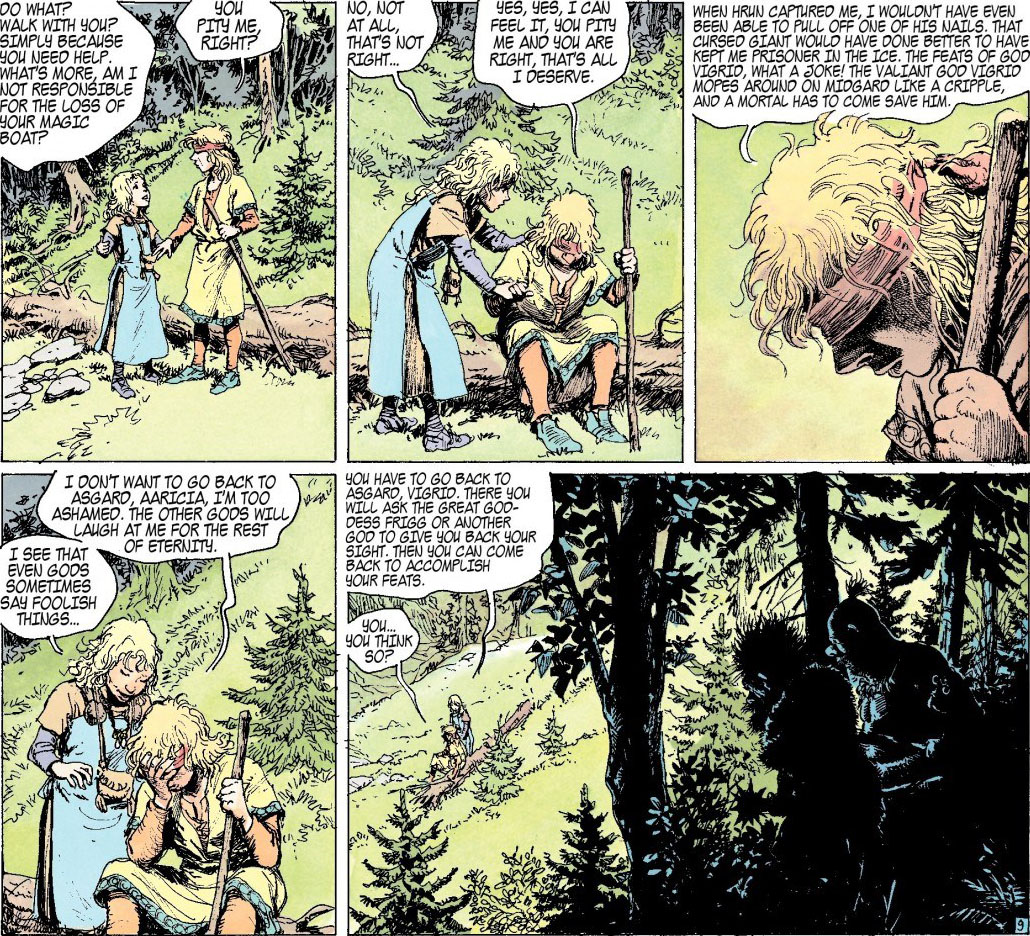}
                \phantomsubcaption\label{fig:BdExUnilateral}
            \end{subfigure}\\[-3mm]
	        \begin{subfigure}[t]{0.49\textwidth}
                 \includegraphics[height=5.6cm]{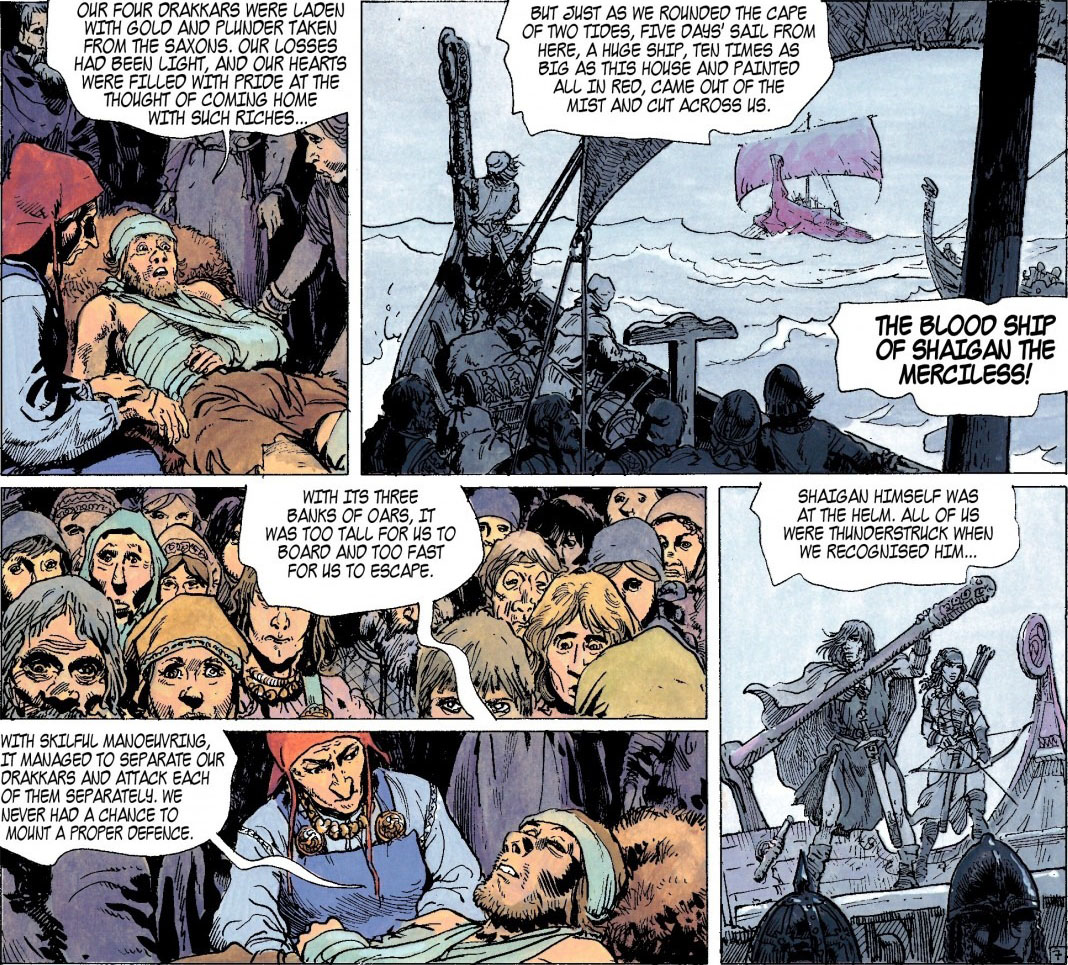}
                \phantomsubcaption\label{fig:BdExNarration}
            \end{subfigure}\hspace{1mm}
	        \begin{subfigure}[t]{0.49\textwidth}
                \includegraphics[height=5.6cm]{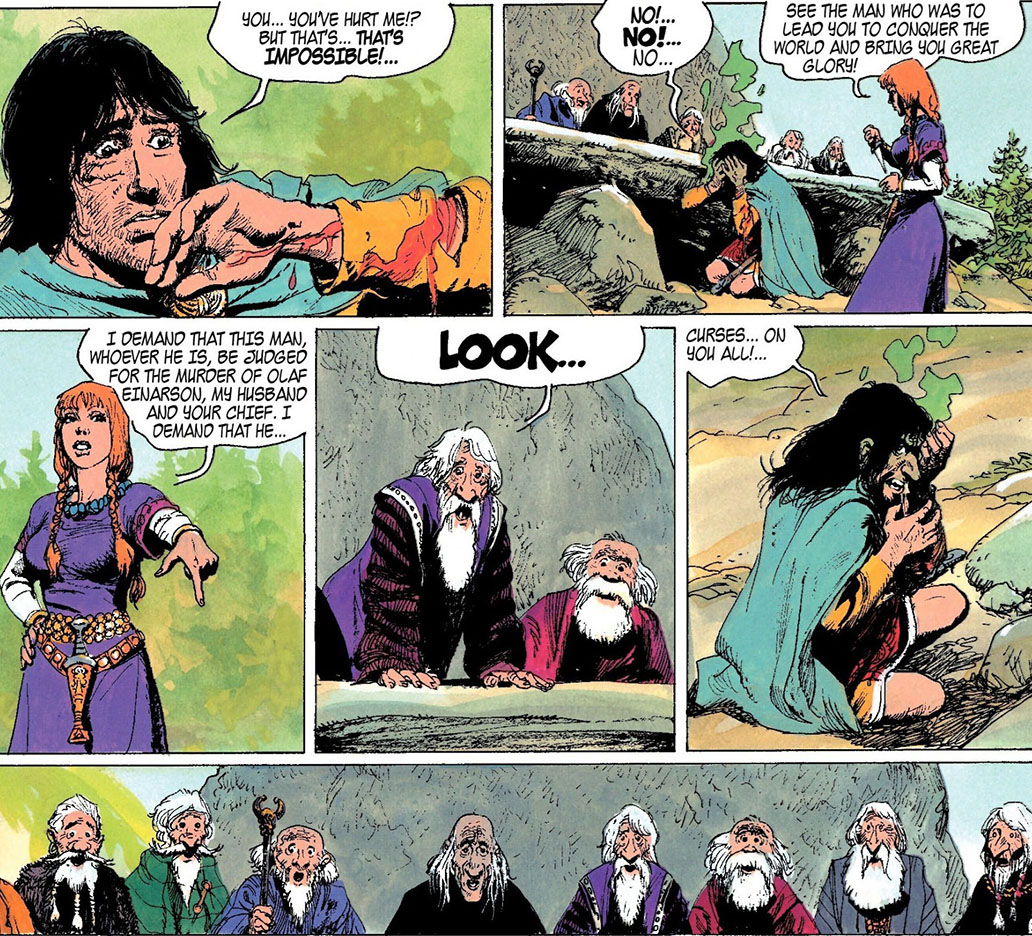}
                \phantomsubcaption\label{fig:BdExImpers}
            \end{subfigure}
        };
        \node[anchor=west, circle, fill=white, inner sep=0.5pt] at (0.08,6.55) {\fontsize{8}{8}\selectfont{}\textbf{a)}};
        \node[anchor=west, circle, fill=white, inner sep=0.5pt] at (6.49,6.55) {\fontsize{8}{8}\selectfont{}\textbf{b)}};
        \node[anchor=west, circle, fill=white, inner sep=0.5pt] at (0.08,0.75) {\fontsize{8}{8}\selectfont{}\textbf{c)}};
        \node[anchor=west, circle, fill=white, inner sep=0.5pt] at (6.49,0.75) {\fontsize{8}{8}\selectfont{}\textbf{d)}};
    \end{tikzpicture}
    \vspace{-0.9cm}
    \caption{(a)~Muff, the dog of Jolan (Thorgal's son), has a significant impact on the story, which illustrates how animals can be considered as characters. (b)~Two thieves watch Aaricia and Vigrid from the distance, thereby performing a remote and unilateral action. (c)~Raider Erik tells his misadventure (blueish panels) to the inhabitants of his village, which we consider here as a collective character. (d)~Volsung of Nichor is exposed as Thorgal's impersonator before the village's Thing; we represent him as Volsung even when he is mistaken as Thorgal by the other characters.}
\end{figure*}

We define the notion of \textit{character} as any sentient entity able to have a meaningful interaction. This includes humans of course, but also supernatural creatures such as ghosts, as well as certain animals (e.g. Figure~\ref{fig:BdExAnimal}), magical artifacts (e.g. talking swords), or machines (e.g. robots). We ignore dead entities, unless they are supernatural and able to interact with some character (e.g. zombies, vampires, ghosts). We mean \textit{interaction} in a very broad sense here, including active perception (watching, listening), even if unilateral (e.g. Figure~\ref{fig:BdExUnilateral}). Sometimes, an event involves a group of undetermined or undifferentiated people, akin to extras in a movie: we consider them as a \textit{collective character} and represent them using a single vertex in the graph (e.g. the audience in Figure~\ref{fig:BdExNarration}). When a character impersonates another person, we use his real identity (e.g. Figure~\ref{fig:BdExImpers}). We refer to the characters using their full names when they are mentioned in the graphic novel, otherwise we use a brief description and a unique id to avoid any ambiguity.

\begin{figure*}[t!]
    \centering
    \begin{tikzpicture}
        \node[anchor=south west,inner sep=0,align=left] (image) at (0,0) {
	        \begin{subfigure}[t]{0.49\textwidth}
                 \includegraphics[height=5.6cm]{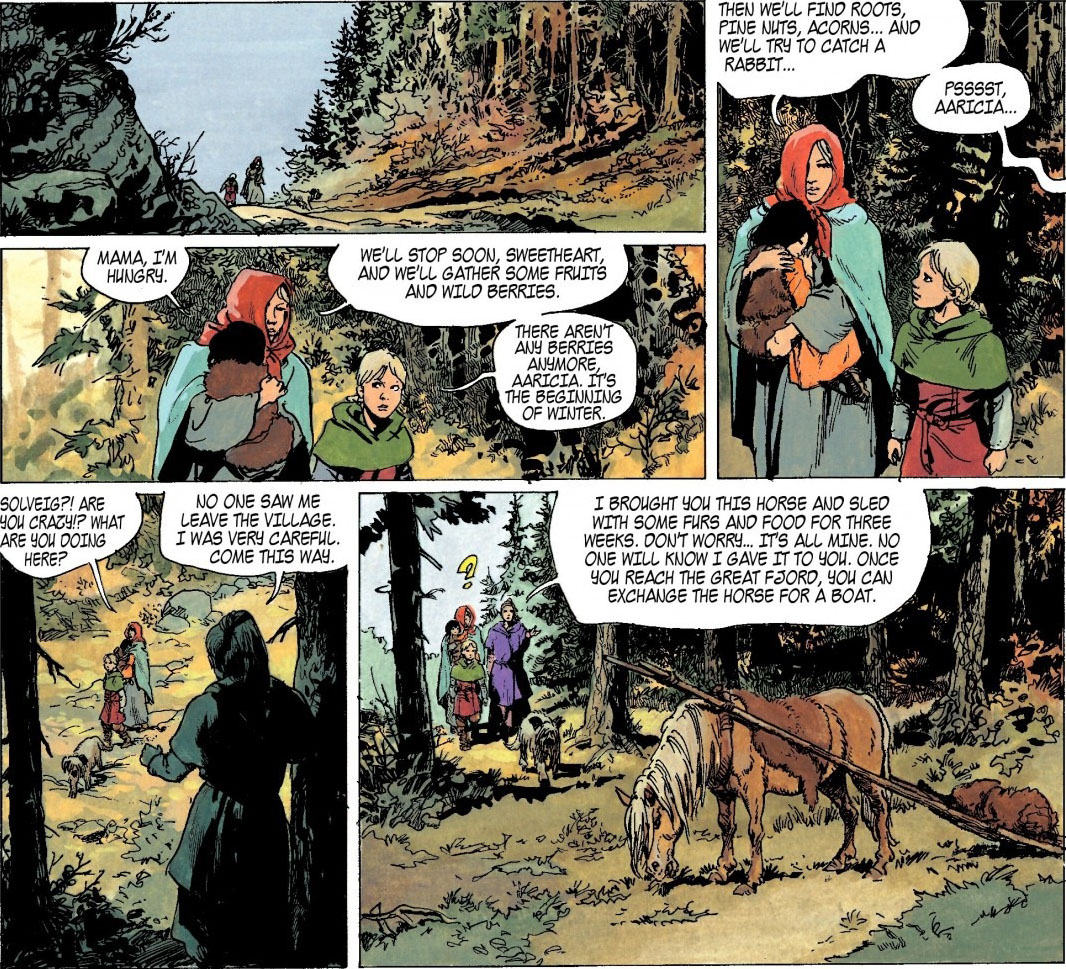}
                \phantomsubcaption\label{fig:BdExEnters}
            \end{subfigure}\hspace{1mm}
	        \begin{subfigure}[t]{0.49\textwidth}
                \includegraphics[height=5.6cm]{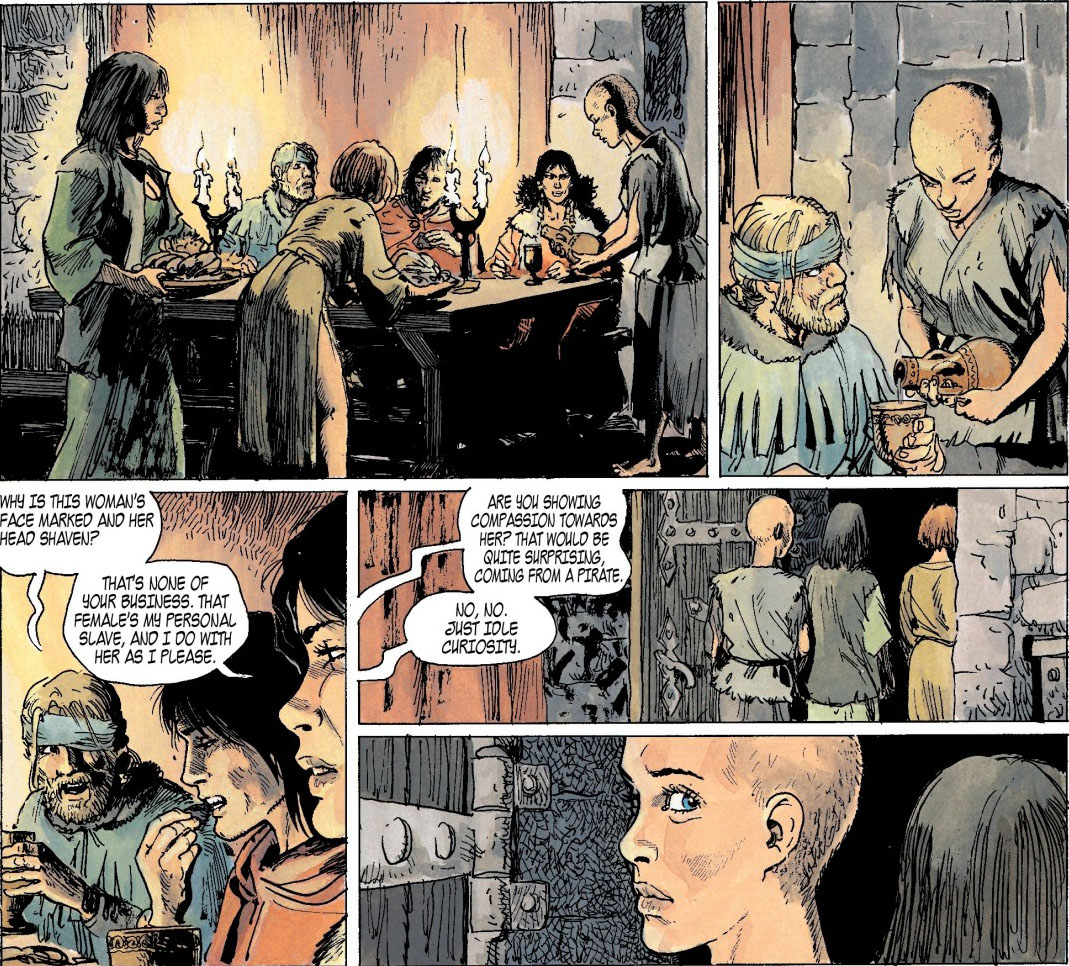}
                \phantomsubcaption\label{fig:BdExLeaves}
            \end{subfigure}\\[-3mm]
	        \begin{subfigure}[t]{0.49\textwidth}
                 \includegraphics[height=5.6cm]{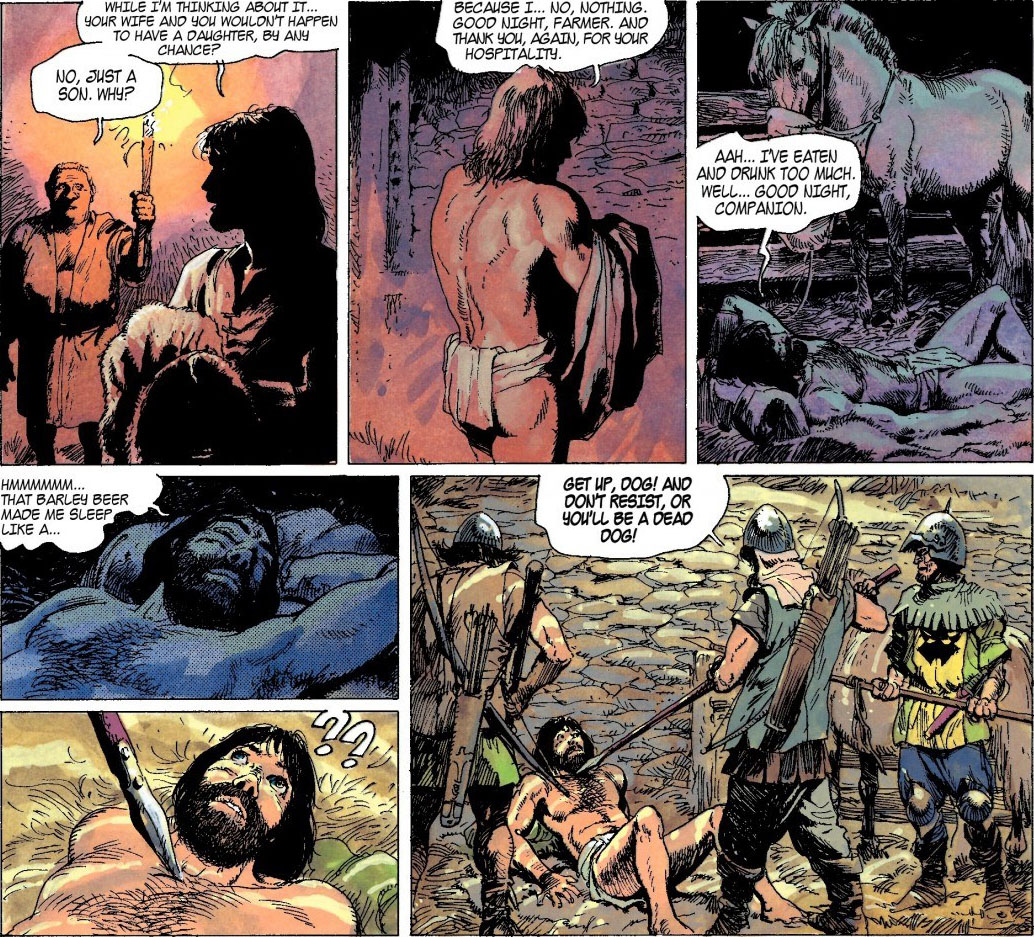}
                \phantomsubcaption\label{fig:BdExEllipsis}
            \end{subfigure}\hspace{1mm}
	        \begin{subfigure}[t]{0.49\textwidth}
                \includegraphics[height=5.6cm]{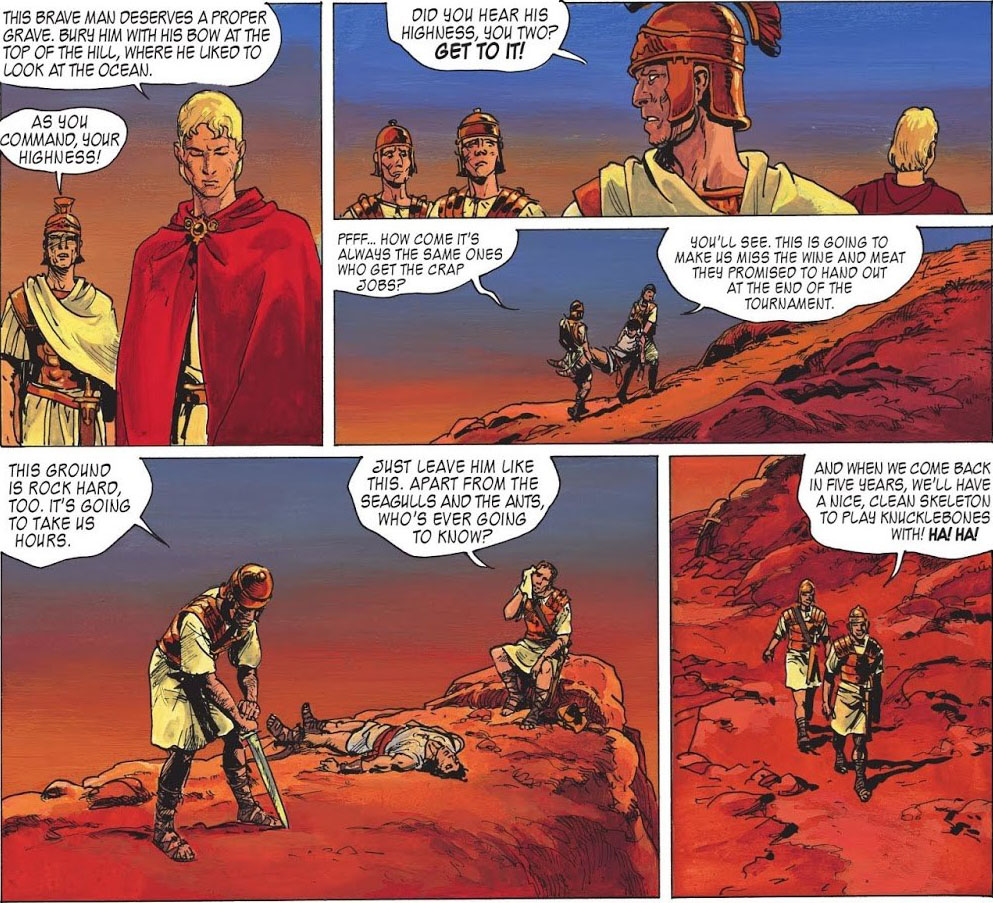}
                \phantomsubcaption\label{fig:BdExUnconsc}
            \end{subfigure}
        };
        \node[anchor=west, circle, fill=white, inner sep=0.5pt] at (0.08,6.55) {\fontsize{8}{8}\selectfont{}\textbf{a)}};
        \node[anchor=west, circle, fill=white, inner sep=0.5pt] at (6.49,6.55) {\fontsize{8}{8}\selectfont{}\textbf{b)}};
        \node[anchor=west, circle, fill=white, inner sep=0.5pt] at (0.08,0.75) {\fontsize{8}{8}\selectfont{}\textbf{c)}};
        \node[anchor=west, circle, fill=white, inner sep=0.5pt] at (6.49,0.75) {\fontsize{8}{8}\selectfont{}\textbf{d)}};
    \end{tikzpicture}
    \vspace{-0.9cm}
    \caption{(a)~A group composed of Aaricia (Thorgal's wife) and her children is joined by Solveig, which starts a new scene according to our definition. (b)~Aaricia serves dinner to a group, and her departure from the room also starts a new scene. (c)~Thorgal goes to sleep in one panel and it is already morning in the next one, which illustrates the concept of ellipsis. (d)~Two soldiers carry unconscious Thorgal, believing him dead: we still consider that there is an interaction between these three characters. }
\end{figure*}

\begin{figure*}[t!]
    \centering
    \begin{tikzpicture}
        \node[anchor=south west,inner sep=0,align=left] (image) at (0,0) {
	        \begin{subfigure}[t]{0.49\textwidth}
                 \includegraphics[height=5.6cm]{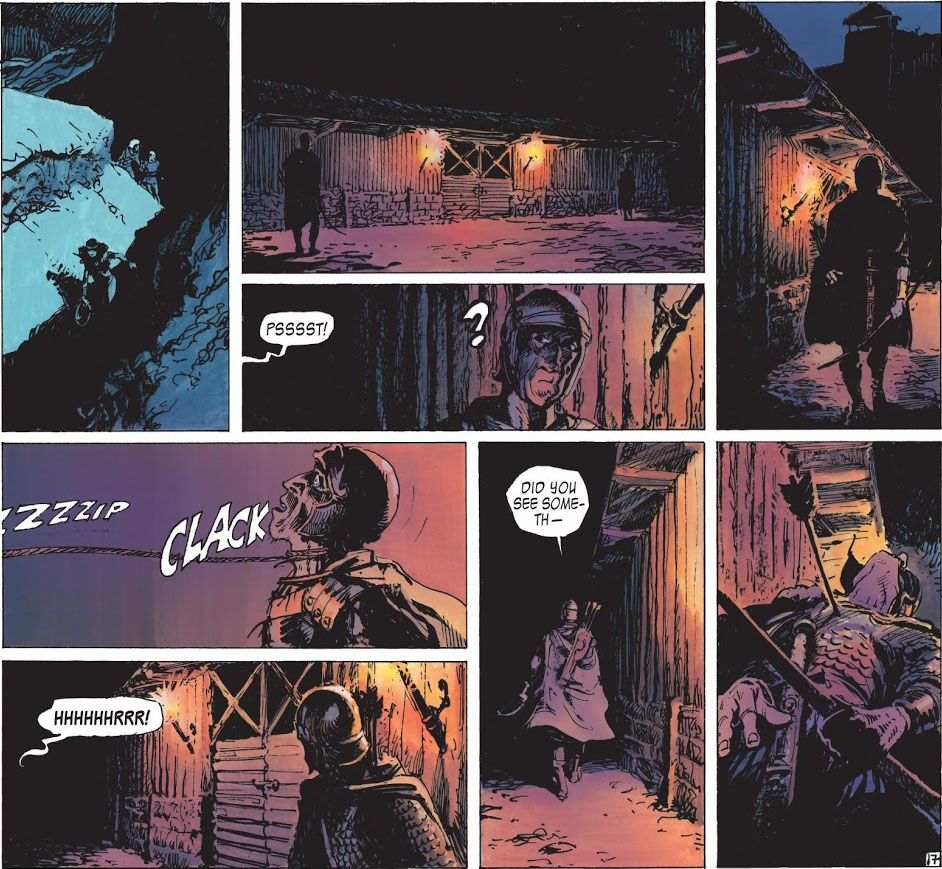}
                \phantomsubcaption\label{fig:BdExInvisible}
            \end{subfigure}\hspace{1mm}
	        \begin{subfigure}[t]{0.49\textwidth}
                \includegraphics[height=5.6cm]{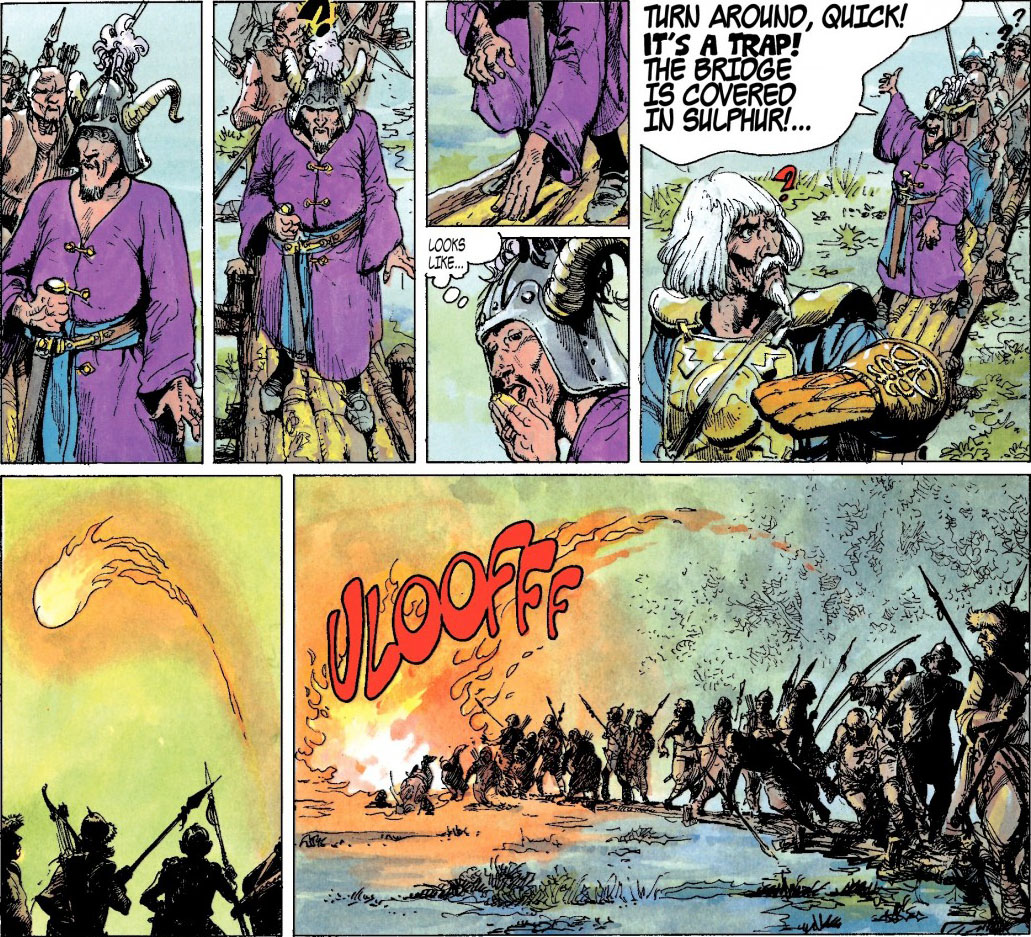}
                \phantomsubcaption\label{fig:BdExDistant}
            \end{subfigure}\\[-3mm]
	        \begin{subfigure}[t]{0.49\textwidth}
                 \includegraphics[height=5.6cm]{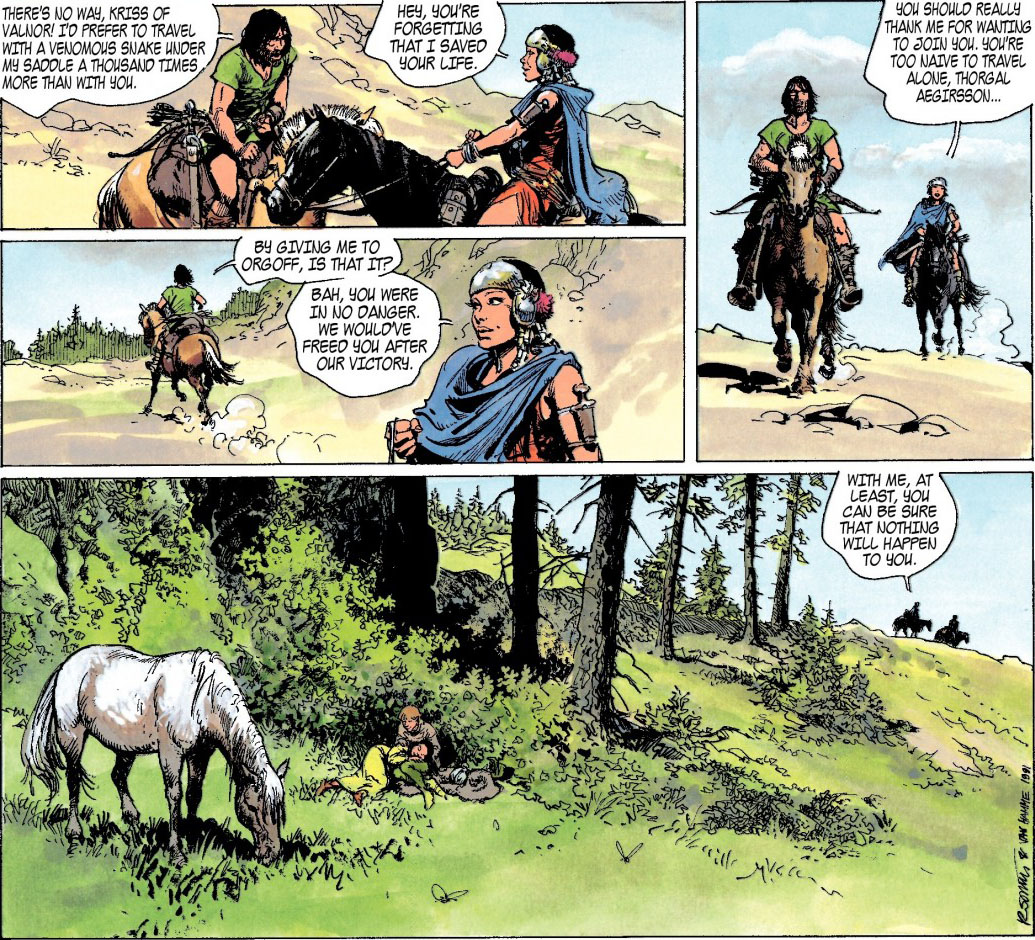}
                \phantomsubcaption\label{fig:BdExGroups}
            \end{subfigure}\hspace{1mm}
	        \begin{subfigure}[t]{0.49\textwidth}
                \includegraphics[height=5.6cm]{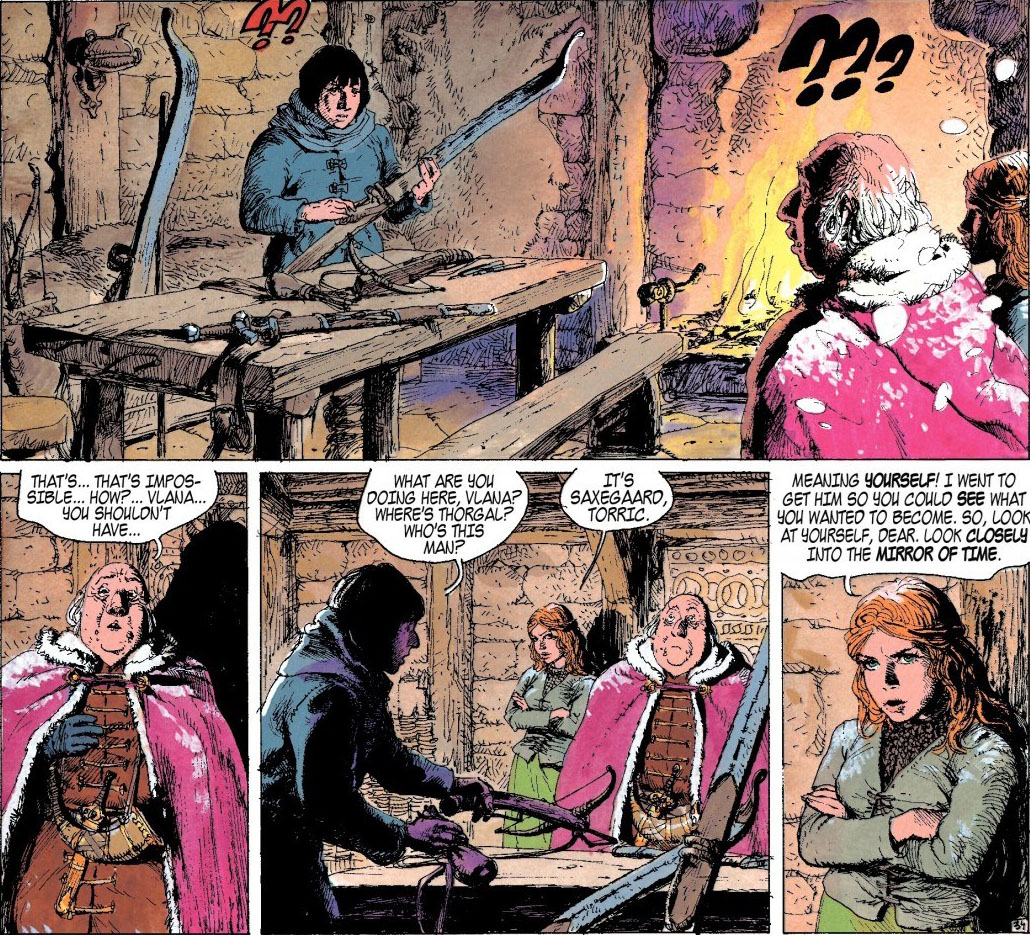}
                \phantomsubcaption\label{fig:BdExMultiple}
            \end{subfigure}
        };
        \node[anchor=west, circle, fill=white, inner sep=0.5pt] at (0.08,6.55) {\fontsize{8}{8}\selectfont{}\textbf{a)}};
        \node[anchor=west, circle, fill=white, inner sep=0.5pt] at (6.49,6.55) {\fontsize{8}{8}\selectfont{}\textbf{b)}};
        \node[anchor=west, circle, fill=white, inner sep=0.5pt] at (0.08,0.75) {\fontsize{8}{8}\selectfont{}\textbf{c)}};
        \node[anchor=west, circle, fill=white, inner sep=0.5pt] at (6.49,0.75) {\fontsize{8}{8}\selectfont{}\textbf{d)}};
    \end{tikzpicture}
    \vspace{-0.9cm}
    \caption{(a)~Kriss of Valnor explicitly appears only in the first panel, but we can infer that she participates in the rest of the scene. (b)~Example of remote interaction between an off-panel assailant and a group of characters. (c)~Scene with two independent groups of characters: Thorgal \& Kriss of Valnor (shown in the first panel), and Shenko \& Floriana (last panel). (d)~Scene involving twice the same character, Torric, at different ages.}
\end{figure*}

We manually determine which characters are interacting based on our understanding of the situation, and use this information to identify scene boundaries. We define a \textit{scene} as a sequence of consecutive panels describing some action involving the same group of characters for an uninterrupted period of time. Consequently, if a character involved in a scene leaves the group (e.g. Figure~\ref{fig:BdExEnters}), or if a new character joins the action (e.g. Figure~\ref{fig:BdExLeaves}), then a new scene begins. Similarly, an ellipsis (i.e. omitting a part of the story to suggest the passage of time) also starts a new scene (e.g. Figure~\ref{fig:BdExEllipsis}). We consider unconscious or asleep characters to be involved in the scene, as long as they are the object of some other character's action (e.g. Figure~\ref{fig:BdExUnconsc}). Note that a character can still be a part of the interacting group even if not explicitly shown in some panel, provided her involvement can be inferred from the surrounding panels (e.g. Figure~\ref{fig:BdExInvisible}).

One could find it surprising that we use the drama concept of \textit{unity of time} to define the notion of scene, but not that of \textit{unity of space}. This is because we want to include, in our broad definition of an action, remote acts such as watching someone from the distance unbeknownst to them (e.g. Figure~\ref{fig:BdExEnters}), or throwing something at someone while remaining hidden (e.g. Figure~\ref{fig:BdExDistant}). Note that some authors take even more remote actions into account, e.g. Alexander considers postal correspondence, in novels~\cite{Alexander2021}. It is worth stressing that the unilateral nature of certain actions (e.g. \textit{speaking to}, by opposition to \textit{chatting with}) could be used to extract \textit{directed} links. However, for the sake of simplicity, we discard this information in this study, and consequently extract undirected links.

Our approach makes it possible to have several groups of characters involved in separate actions at the same time, resulting in parallel and possibly overlapping scenes, as illustrated by Figure~\ref{fig:BdExGroups}. We treat flashbacks, hallucinations, and dreams, which are usually represented using different graphical codes than the rest of the story (e.g. blue tone in Figure~\ref{fig:BdExNarration}), just as regular scenes. There is an exception though, when the action is told by a visible narrator. We then consider that there are two separate interacting groups (and therefore two overlapping scenes): the narrator and his audience vs. the characters involved in the story told. Note that some characters can belong to both groups (e.g. a narrator telling a story involving himself). More generally, it is possible that several instances of the same character belong to the same interaction group (e.g. through time travel as in Figure~\ref{fig:BdExMultiple}).

\subsection{Graph Building}
\label{sec:ExtractionMethodsGraph}
We perform the third step of the extraction process, temporal integration, in an automatic way, resulting in a static graph representing the whole narrative. We define edges between the characters that interact during the same scene, and integrate over the narrative. 

Our method compares to those described in Section~\ref{sec:RelatedWork} in the following ways. First, like them, we detect characters manually. However, they all infer character interactions based on co-occurrences, whereas we do not make this assumption and identify proper interactions manually. Second, by leveraging the scene to break down the narrative, we use a different temporal granularity. Alberich \textit{et al}.~\cite{Alberich2002} and Gleiser~\cite{Gleiser2007} consider character co-occurrences at the level of the whole comic book, and Lee \& Kim~\cite{Lee2020f} at the level of the whole webtoon episode. This would not be relevant considering our dataset, because the number of characters appearing in a single volume ($34$ in average) is large enough to make the extracted graph too dense to be useful. The approach adopted by Murakami \textit{et al}.~\cite{Murakami2011} amounts to using a two-panel sliding window with a single-panel overlap, which generally constitutes a finer granularity than ours, as scenes are often longer than two panels. However, such a window is not narratively sound, as two consecutive panels can belong to completely different subplots. Rochat \& Triclot~\cite{Rochat2017} prefer to use a two-page sliding window with a single-page overlap, which generally corresponds to a rougher granularity than ours. Using such a page-based window is questionable when applied to novels, since writers do not control the pagination of the end product, as noted in~\cite{Labatut2019}. But it could be relevant for graphic novels, as the elaboration of the page layout is an important part of the story telling. For instance, scenes often start and end at the first and last panels of a page (not necessarily the same page). Authors also frequently use the last panel of a page to build suspense, as the reader has to turn the page to known what happens next (see for instance Figure~\ref{fig:BdEx4} in the Supplementary Material).

\section{Dataset Description}
\label{sec:Dataset}
As mentioned before, in this article we study the character network extracted from the \textit{Thorgal} series of graphic novels. In this section, we first introduce the series itself (Section~\ref{sec:DataSeries}), and list the research questions that arose from its study in the literature. We then describe the main characteristics of the dataset resulting from our annotation process, and discuss the resulting statistics (Section~\ref{sec:DataProp}). Next, we explain how we filter the character network to retain only meaningful characters (Section~\ref{sec:DataFiltering}). In addition, we also compare our dataset to other graphic novel datasets in the Supplementary Material (Section~\ref{sec:AddDataCompar}).

\subsection{The \textit{Thorgal} Series}
\label{sec:DataSeries}
Our dataset is based on the Franco-Belgian series \textit{Thorgal}, which was originally created by writer Jean Van Hamme and artist Grzegorz Rosiński in 1977. It takes place during a period reminiscent of the European medieval age, and mixes elements of fantasy, mythology, and science-fiction. It mainly follows the adventures of Thorgal (shown in Figure~\ref{fig:BdExEllipsis}), originally a baby adopted by a Viking chief, and his family. The main secondary characters are his wife Aaricia (Figure~\ref{fig:BdExUnilateral}), his elder son Jolan (Figure~\ref{fig:BdExAnimal}), his daughter Louve (Figure~\ref{fig:BdExEnters}), and his antagonist Kriss of Valnor (Figure~\ref{fig:BdExGroups}).

This graphic novel is still going on nowadays with several other authors and artists, as well as three spinoff series. Two of these are centered around secondary characters (\textit{Kriss of Valnor} and \textit{Louve}), whereas the third focuses on the youth of Thorgal and Aaricia youth (\textit{Young Thorgal}). As of November 2021, a total of $63$ volumes were published ($39$ in the main series and $24$ in the spinoff ones), resulting in more than $15$ million books sold. This makes \textit{Thorgal} a bestseller on the French-speaking market~\cite{Potet2018}. In the Supplementary Material, Tables~\ref{tab:Volumes} and~\ref{tab:NarrArcs} list all the volumes and the $23$ narrative arcs constituting the series, respectively. \textit{Thorgal} is also the object of academic work in human sciences~\cite{Desfontaine2018,Thiry2019}, was adapted into a video game by Cryo Interactive~\cite{Parker2003} and there is an ongoing project of TV series adaptation~\cite{Meza2018} by Academy Award winner Florian Henckel von Donnersmarck.

We chose to study \textit{Thorgal} not only because of its cultural and commercial significance, but also because only two bandes dessinées series have been studied in the complex network literature as of yet (cf. Table~\ref{tab:SurveyList}), and this work is the occasion to expand this corpus. Moreover, both of these series are relatively short, whereas \textit{Thorgal} develops a story involving hundreds of characters over tens of volumes. Another interesting property of the series is the way it handles time. 
In \textit{Thorgal}, characters get old as the story unrolls, their relationships evolve, some important characters die while others are born. The eponymous character is a teenager in the first volume, whereas in the latest one he is over forty and a father of three, with an elder son that left the parental home.

Based on the literature, we identified two research questions, that we want to answer by analyzing the character network extracted from the series. 


\paragraph{RQ1} \textit{How similar is Thorgal's story to Icelandic sagas?} Thiry~\cite{Thiry2019} stresses that the \textit{Thorgal} series is deeply rooted in the \textit{Íslendingasögur} (\textit{Sagas of Icelanders}), hybridizing it with more modern narrative elements. Such sagas have been thoroughly studied in the complex network literature~\cite{MacCarron2013}, therefore we can leverage these results to answer this question.

\paragraph{RQ2} \textit{What is the position of women in \textit{Thorgal}?} Desfontaine \textit{et al}.~\cite{Desfontaine2018} argue that Thorgal can be seen as a model of soft masculinity, involving a more balanced relationship between men and women. To explore this point, we propose to characterize the position of female characters in the character network of the series.



\subsection{Dataset Properties}
\label{sec:DataProp}
A single annotator identified the characters, interactions, and scene boundaries for the whole series, as explained in Section~\ref{sec:ExtractionMethodsAnnotation}. In order to assess the reliability of this annotation, three independent annotators did the same work on 4 randomly picked volumes. The agreement between these annotators and the main one is strong, with micro-average $F$-measure scores of $0.97$  for scene boundaries and of $0.94$ for character identification (cf. Section~\ref{sec:AddDataAnnot} of the Supplementary Material for more technical details). 

In addition, we also annotated the biological sex of the characters: \textit{Female}, \textit{Male}, \textit{Mixed} (for groups of characters), or \textit{Unknown}. Each volume follows the \textit{48CC} bandes dessinées standard, i.e. a 48-page hardcover color book (of which 46 pages are actually drawn). It takes approximately one hour to annotate such a book. Our dataset\footnote{DOI:~\href{https://doi.org/10.5281/zenodo.6395874}{\texttt{10.5281/zenodo.6395875}}}, including the raw data and the extracted networks, as well as our source code\footnote{\url{https://github.com/CompNet/NaNet/releases/tag/v1.0.1}} are both available online.

The $63$ volumes constituting our dataset contain $21{,}259$ panels distributed over $2{,}925$ pages. They involve a total of $1{,}480$ characters interacting in $4{,}622$ scenes. The average volume and scene include $34.2$ and $3.1$ characters, respectively. One could assume that longer scenes tend to involve more characters, but this is not the case: the correlation between both quantities is rather weak ($\rho_k = 0.18$). 

\begin{table}[htb!]
    \caption{Most frequent characters of the corpus, with their number of occurrences expressed in volumes, pages, scenes and panels. 
    The last row corresponds to the average values over all characters.}
    \label{tab:FrequentChars}
    \begin{tabular}{p{5.8cm} r r r r}
        \toprule
        \textbf{Character} & \textbf{Volumes} & \textbf{Pages} & \textbf{Scenes} & \textbf{Panels} \\
        \colrule
        Thorgal        & 54   & 1,552   & 1,752   & 9,852   \\
        Aaricia        & 47   &   733   &   816   & 4,375   \\
        Jolan          & 32   &   672   &   805   & 4,147   \\
        Kriss          & 24   &   573   &   683   & 3,518   \\
        Louve          & 26   &   481   &   561   & 3,010   \\
        Muff           & 15   &   137   &   152   &   761   \\
        \colrule
        Average (All)  &  1.5 &     9.2 &     9.8 &    51.7 \\
        \botrule
    \end{tabular}
\end{table}

\begin{figure}[htb!]
    \centering
    \begin{tikzpicture}
        \node[anchor=south west,inner sep=0] (image) at (0,0) {
	        \begin{subfigure}[t]{0.49\textwidth}
                \includegraphics[height=5.85cm]{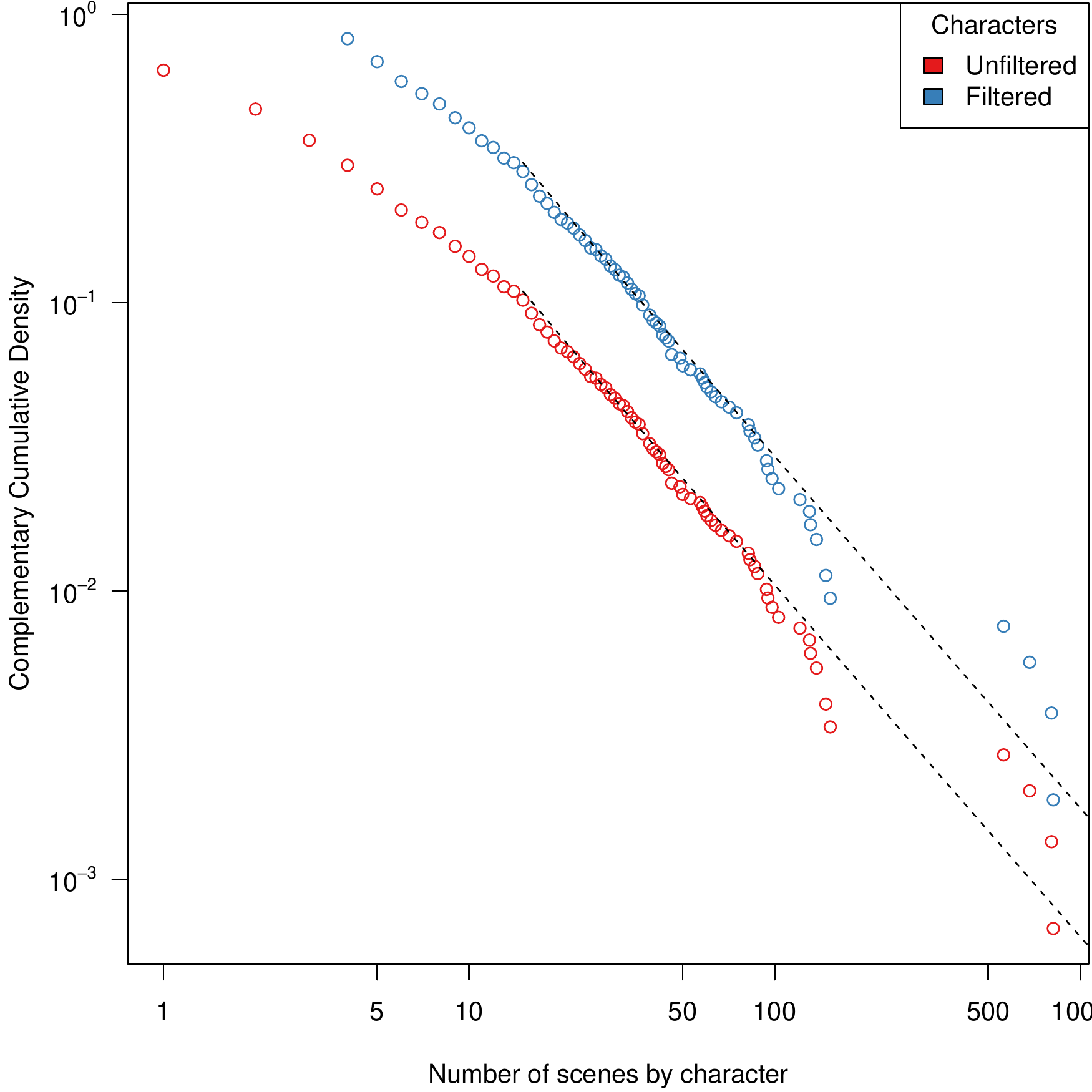}
                \phantomsubcaption\label{fig:SceneDistr}
            \end{subfigure}\hfill
            \begin{subfigure}[t]{0.49\textwidth}
	            \includegraphics[height=5.85cm]{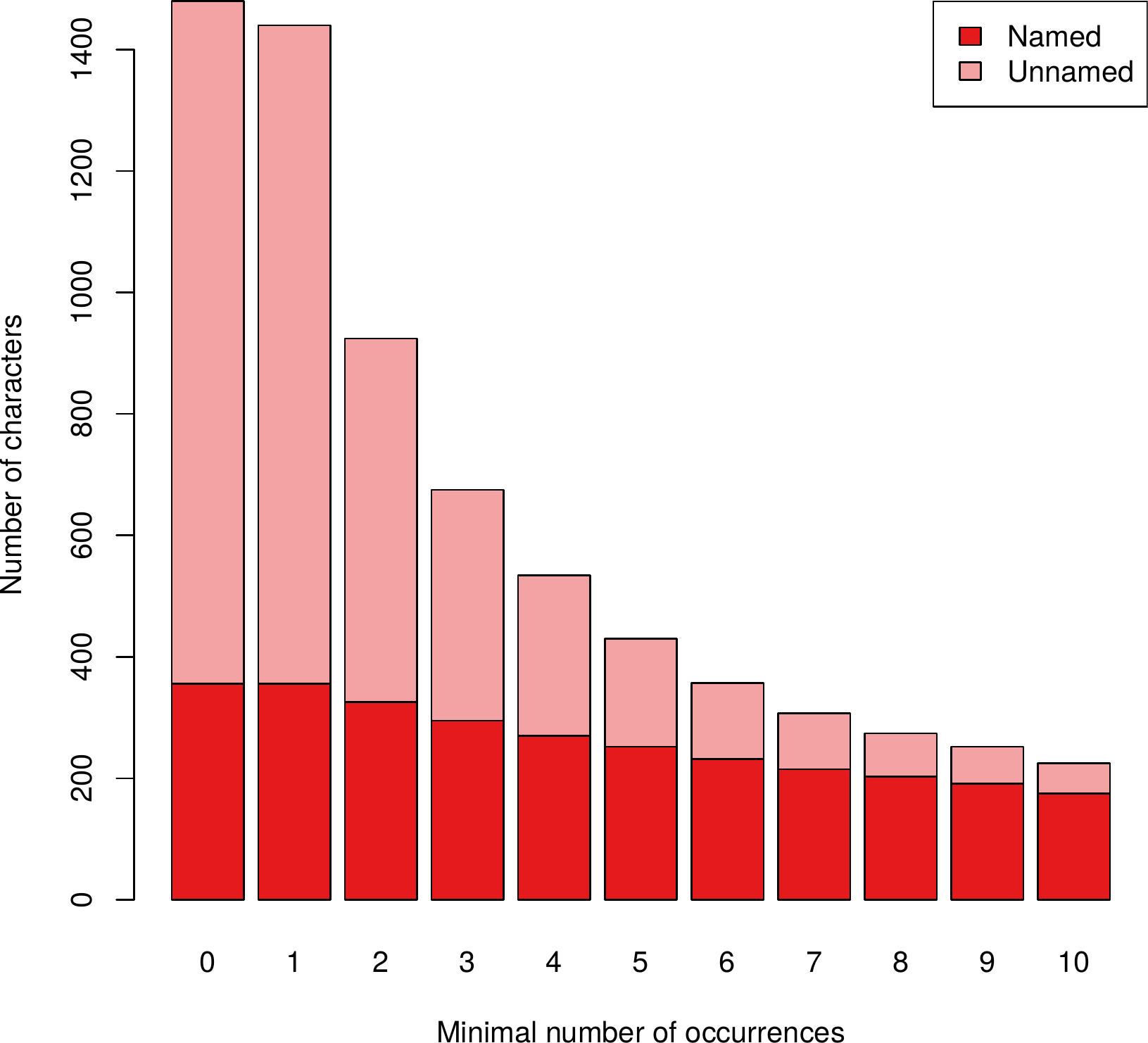}
                \phantomsubcaption\label{fig:CharNbr}
            \end{subfigure}
        };
        \node[anchor=west, circle, fill=white, inner sep=0.5pt] at (0.00,0.60) {\fontsize{8}{8}\selectfont{}\textbf{a)}};
        \node[anchor=west, inner sep=0.5pt] at (6.40,0.60) {\fontsize{8}{8}\selectfont{}\textbf{b)}};
    \end{tikzpicture}
    \vspace{-0.5cm}
    \caption{(a)~Complementary cumulative distribution of the number of scenes by character, for all characters (red) and after filtering them (blue). The dotted lines are power laws corresponding to the best fit. Both axes use a logarithmic scale. (b)~Number of characters as a function of the minimal threshold used to filter them based on their number of occurrences. The red part represents characters whose proper name is explicitly mentioned in the narrative, whereas the blue part represents the others. Figure available at \href{https://doi.org/10.5281/zenodo.6573491}{10.5281/zenodo.6573491} under CC-BY license.}
\end{figure}

The story is clearly dominated by Thorgal and his family, as illustrated by Table~\ref{tab:FrequentChars}, that shows the most frequent characters. Besides the main protagonist, Thorgal himself, which is present in 38\% of the scenes, these include his wife Aaricia (18\%); his eldest son Jolan (17\%, introduced in vol.6); the antagonist Kriss of Valnor (15\%, introduced in vol.9); his daughter Louve (12\%, introduced in vol.16) and Jolan's dog Muff (3\%, shown in Figure~\ref{fig:BdExAnimal}, introduced in vol.8 and retired in vol.25). Note that the spinoff series focus on some of these characters (Louve, Kriss, Aaricia), otherwise the imbalance would be even more marked in favor of Thorgal. By comparison, the average character appears in only 0.2\% of the scenes. The distribution of scenes by character, showed in red in Figure~\ref{fig:SceneDistr}, has a tail best fit by a power law ($x_{\min} = 15$), according to Clauset \textit{et al}.'s procedure~\cite{Clauset2009}\footnote{We always use Clauset \textit{et al}.'s~\cite{Clauset2009} method to identify distributions in the rest of this article.}. This means that this strong imbalance between characters, regarding their occurrence over the narrative, is observed also for less-frequent characters. The same observation holds for other narrative units (pages, panels, and to a lesser extent, volumes): cf. Figure~\ref{fig:OtherOccDistr} in the Supplementary Material.

Finally, $67\%$ of the characters are males, $13\%$ are females, and the rest either correspond to collective characters mixing both sexes ($15\%$) or to entities whose sex could not be determined ($5\%$). It is worth noting that the last three groups are more frequent than males when considering characters appearing only once; however males are prevalent among characters appearing several times. This suggests that the recurrent characters of the series are essentially males, and that other character types make only quick appearances, in general. This constitutes the beginning of an answer to RQ2, but we will come back to this point when studying the network itself.

\subsection{Character Filtering}
\label{sec:DataFiltering}
When extracting character networks from works of fiction, many authors apply a filtering step in order to remove very minor characters, namely extras, whose contribution to the story is not deemed sufficient to be taken into account during the subsequent analysis~\cite{Labatut2019}. 

\paragraph{Criteria}
For this purpose, they use mainly three criteria. First, the most widespread is a minimal threshold over the number of occurrences. For instance, Elson \textit{et al}.~\cite{Elson2010} remove characters that appear three times or fewer in a novel, or that amount to $1\%$ or less of all occurrences; Suen \textit{et al}.~\cite{Suen2013} ignore characters that speak fewer than five times over a play or movie; and the \textit{Marvel Chronology Project} dataset used by Alberich \textit{et al}.~\cite{Alberich2002} and Gleiser~\cite{Gleiser2007} is built by retaining only super-heroes and super-villains, provided they appear in at least two issues. 

The second criterion is to keep only characters interacting with a sufficient number of others characters, which amounts to setting a minimal threshold over the vertex degree. For instance, when dealing with novels, Zhang \textit{et al}.~\cite{Zhang2021o} retain only characters interacting with at least six other persons. This criterion is more questionable though, as the literature exhibits examples of low-degree characters that are central according to other topological measures~\cite{Rochat2014a}, and therefore of great importance for the story. Imagine, for instance, a character interacting with only two persons, each one belonging to an otherwise separate part of the narrative. 

Finally, the third criterion is the explicit mention of the character's proper name. All approaches relying only on Named Entity Recognition to extract character networks from novels implicitly assume that unnamed characters should be ignored. The \textit{Marvel Chronology Project} considers that an entity is a character if it has a proper super-hero or super-villain name, or a full regular name (both first and last names). However, as mentioned by Grener \textit{et al}.~\cite{Grener2017}, even unnamed characters can hold functionally and structurally significant positions in the narrative.

\begin{table}[htb!]
    \centering
    \caption{Effect of both criteria used to filter characters: number of occurrences (\textit{Occurrences}) and number of distinct interaction partners (\textit{Interactions}).}
    \label{tab:CharFiltering}
    \begin{tabular}{r r r r}
        \cline{2-4}\noalign{\smallskip}
         & ~~~$\mathbf{Occurrences} \leq 3$ & ~~~$\mathbf{Occurrences} > 3$ & \textbf{~~~~Total} \\
        \noalign{\smallskip}\hline\noalign{\smallskip}
        ~~~~$\mathbf{Interactions} \leq 1$ & 221 & 1 & 222 \\
        $\mathbf{Interactions} > 1$ & 724 & 534 & 1,258 \\
        \noalign{\smallskip}\hline\noalign{\smallskip}
        \textbf{Total} & 945 & 535 & 1,480 \\
        \noalign{\smallskip}\hline
    \end{tabular}
\end{table}

\paragraph{Our Approach}
In this work, in order to avoid discarding important characters, we use the number of interaction partners (i.e. degree) only to remove leaves and isolates from our networks. Put differently, we keep vertices with a degree of at least two. We do not use the presence of name as a criterion, as certain important characters of \textit{Thorgal} do not have a proper name. For instance, the so-called Guardian of the Keys appears in $10$ volumes, each time in a critical role, and volume 17 is even titled \textit{The Guardian of the Keys}. On the contrary, the names of certain extras are revealed incidentally (e.g. a random soldier hailed by his boss during an action scene). We leverage this information though, in order to set up the most appropriate threshold over the minimal number of occurrences, through the elbow method (cf. Figure~\ref{fig:CharNbr}). 

According to our observations, it seems reasonable to remove all characters with three occurrences or fewer. Table~\ref{tab:CharFiltering} summarizes how both criteria affect the filtered characters. It is worth observing that, for the selected threshold values, the occurrence criterion almost completely subsumes the degree one. Put differently, apart from one individual, all characters interacting with a single person (or less) also participate in three scenes or fewer. There are $534$ characters remaining after this process, which amounts to $37\%$ of the initial character set. Among them, $270$ are named, which represents $77\%$ of the named characters of the initial character set. They constitute a giant component of $530$ characters, whereas the rest are distributed over two very small components. These are constituted of secondary characters that are never showed interacting with others, so we discard them to obtain a connected graph. Note that, in order to assess the effect of this filtering step, we perform our network analysis on both unfiltered and filtered networks. Like its unfiltered counterpart, the distribution of scenes by filtered characters, shown in blue in Figure~\ref{fig:SceneDistr}, has a tail best fit by a power law ($x_{\min} = 15$), according to Clauset \textit{et al}.'s method~\cite{Clauset2009}.

\section{Network Analysis}
\label{sec:DescriptiveAnalysis}
\begin{figure*}[htb!]
    \centering
    \includegraphics[trim={15cm 20cm 18cm 16cm}, clip, width=1\linewidth]{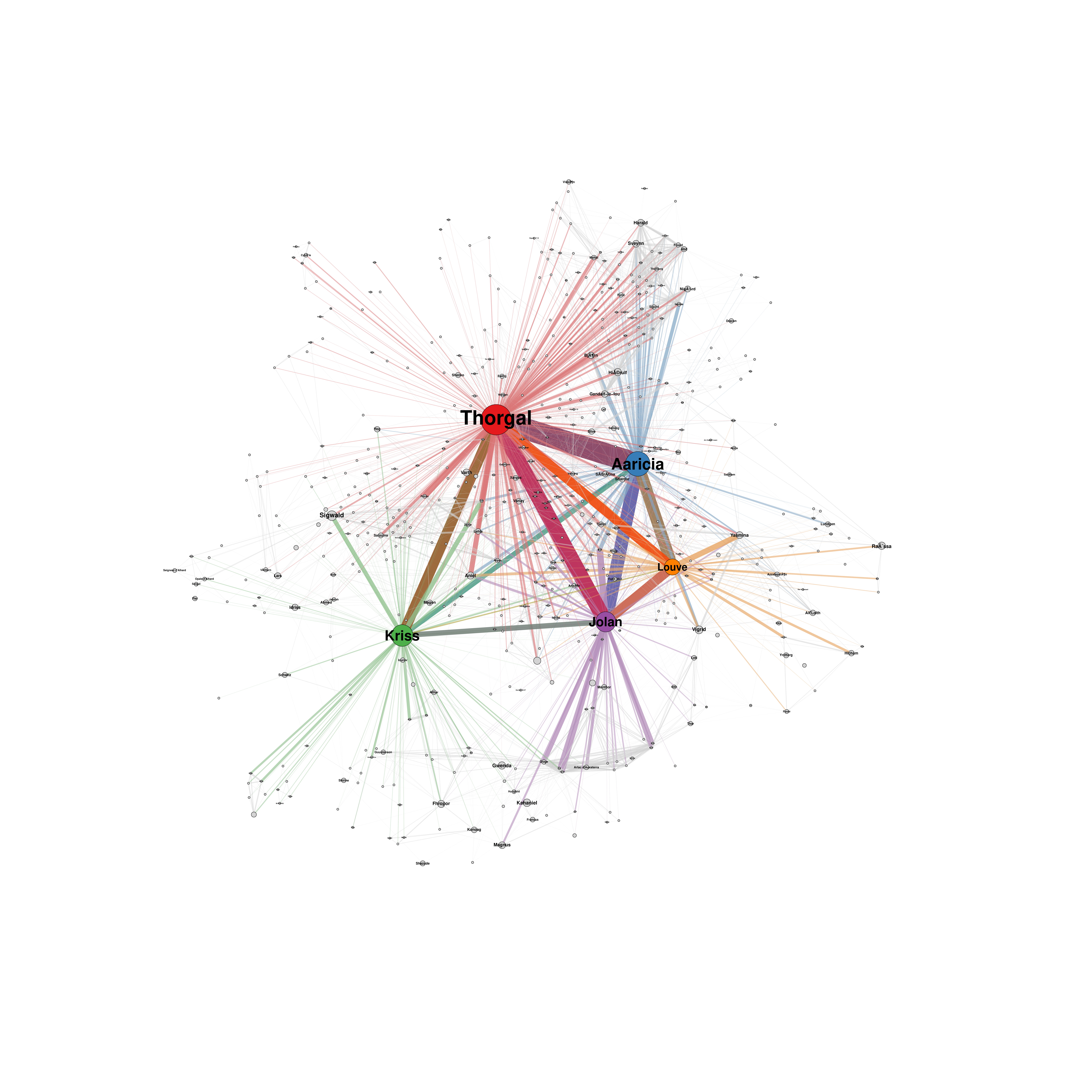}
    \caption{Filtered network extracted from the whole \textit{Thorgal} series. Vertex size is a function of betweenness, and edge width is a function of the number of co-occurrences. The five most frequent characters (in terms of scenes) are shown in a specific color: Thorgal (red), his wife Aaricia (blue), their elder son Jolan (purple), their daughter Louve (Orange), and the antagonist Kriss of Valnor (green). The other characters are shown in gray. The unfiltered network is shown in Figure~\ref{fig:UnfilteredNet} (Supplementary Material). Figure available at \href{https://doi.org/10.5281/zenodo.6573491}{10.5281/zenodo.6573491} under CC-BY license.}
    \label{fig:FilteredNet}
\end{figure*}

In this section, we proceed with the descriptive analysis of our character networks. Figure~\ref{fig:FilteredNet} shows the filtered network extracted from the whole \textit{Thorgal} series; whereas the unfiltered version is shown in Figure~\ref{fig:UnfilteredNet} (Supplementary Material). The top part of Table~\ref{tab:MainTopoProps} contains the main topological properties of both networks. As expected, since it focuses on the most important characters, the filtered network is much denser (see $\delta$, $\langle k \rangle$) and has smaller hubs ($k_{\max}$). Its local transitivity ($C$, a.k.a. clustering coefficient) is higher, which indicates that main characters tend to form more triangles. Indeed, many of the removed characters are leaves and therefore not part of triangles, which can also explain the lower diameter ($d_{\max}$) and average distance ($\langle d \rangle$). Interestingly, filtering does not affect seriously degree assortativity ($\rho_k$); however, as we will see later, it affects degree distribution.

\begin{table}[htb!]
    \setlength{\tabcolsep}{5pt}
    \centering
    \caption{Main topological properties of the considered networks: numbers of vertices ($n$) and edges ($m$), density ($\delta$), average degree ($\langle k \rangle$), maximal degree ($k_{\max}$), degree assortativity ($\rho_k$), average distance ($\langle d \rangle$), diameter or maximal distance ($d_{\max}$), and average local transitivity ($\langle C \rangle$). The top part corresponds to the unfiltered and filtered networks extracted from \textit{Thorgal}, and the bottom part to two \textit{Íslendingasögur} networks discussed in Section~\ref{sec:RqIceland}.}
    \label{tab:MainTopoProps}
    \begin{tabular}{p{2.3cm} r r r r r r r r r}
        \hline\noalign{\smallskip}
        \textbf{Network} & $n$ & $m$ & $\delta$ & $\langle k \rangle$ & $k_{\max}$ & $\rho_k$ & $\langle d \rangle$ & $d_{\max}$ & $\langle C \rangle$ \\
        \noalign{\smallskip}\hline\noalign{\smallskip}
        Unfiltered & $1{,}453$ & $6{,}579$ & $0.006$ & $9.06$ & $676$ & $-0.17$ & $2.78$ & $7$ & $0.74$ \\
        Filtered & $524$ & $3{,}702$ & $0.027$ & $14.13$ & $358$ & $-0.18$ & $2.32$ & $5$ & $0.83$ \\
        \noalign{\smallskip}\hline\noalign{\smallskip}
        \textit{18 Sagas}~\cite{MacCarron2013} & $1{,}549$ & $4{,}266$ & $0.004$ & $5.51$ & $83$ & $0.07$ & $4.30$ & $19$ & $0.50$ \\
        \textit{Egils Saga}~\cite{MacCarron2013} & $293$ & $769$ & $0.018$ & $5.25$ & $59$ & $-0.07$ & $4.20$ & $12$ & $0.60$ \\
        \noalign{\smallskip}\hline
    \end{tabular}
\end{table}

In the following, we discuss in more detail the main topological properties of our networks, successively focusing on aspects related to degree (Section~\ref{sec:DescAnalyDegree}), distance and transitivity (Section~\ref{sec:DescAnalyDist}), and centrality (Section~\ref{sec:DescAnalyCentr}).
Additionally, in the Supplementary Material (Section~\ref{sec:AddAnalyNetTypes}), we also leverage Rochat \& Triclot's typology~\cite{Rochat2017} (mentioned in Section~\ref{sec:RelatedWork}) to study the structure of Thorgal's narrative arc networks.

\subsection{Degree and Scale-Free Property}
\label{sec:DescAnalyDegree}

\paragraph{Average Degree}
One way to assess the average degree is to compare it with the value expected for a random model of comparable size. Alberich \textit{et al}.~\cite{Alberich2002} argue that their comic book network is a collaboration network, obtained through the projection of a bipartite graph containing two types of vertices (characters vs. comic book issues), and that the selected random model used to study the resulting unipartite network must therefore take this property into account. Like them, we adopt the model proposed by Newman \textit{et al}.~\cite{Newman2001c} for this purpose. We first extract a bipartite network from our data, which contains character and scene vertices. We then compute the degree of both types of vertices, and use it for the generation of random bipartite networks possessing the same size and degree distribution. We project these bipartite networks into their character dimension, and use the resulting unipartite network as a random model for comparison with our character network. 

The expected average degree is $19.57$ for the unfiltered network and $37.19$ for the filtered one (cf. Table~\ref{tab:SmallWorldStats}). In both cases, this is largely above the values measured for our networks, by factors of $2.16$ and $2.63$, respectively. This corresponds to what is observed by Newman \textit{et al}.~\cite{Newman2001c} on real-world collaboration networks. By comparison, Alberich \textit{et al}.~\cite{Alberich2002} get an even larger factor of $3.4$ on their \textit{Marvel} data, and see this as the result of writers forcing \textit{Marvel} characters to collaborate more often with the same characters than people would in real life. They consider this as a mark of the artificiality of the \textit{Marvel} network. By comparison, in this aspect, the \textit{Thorgal} network is closer to a real-world social network. This could be due to the fact that \textit{Thorgal} contain much fewer recurrent characters: only $13\%$ of the characters appear in several volumes. The main characters obviously interact a lot with one another, but they also frequently go on solo adventures, during which they interact with a number of non-recurrent characters.

\begin{figure}[htb!]
    \centering
    \begin{tikzpicture}
        \node[anchor=south west,inner sep=0] (image) at (0,0) {
	        \begin{subfigure}[t]{0.49\textwidth}
                \includegraphics[width=1\textwidth]{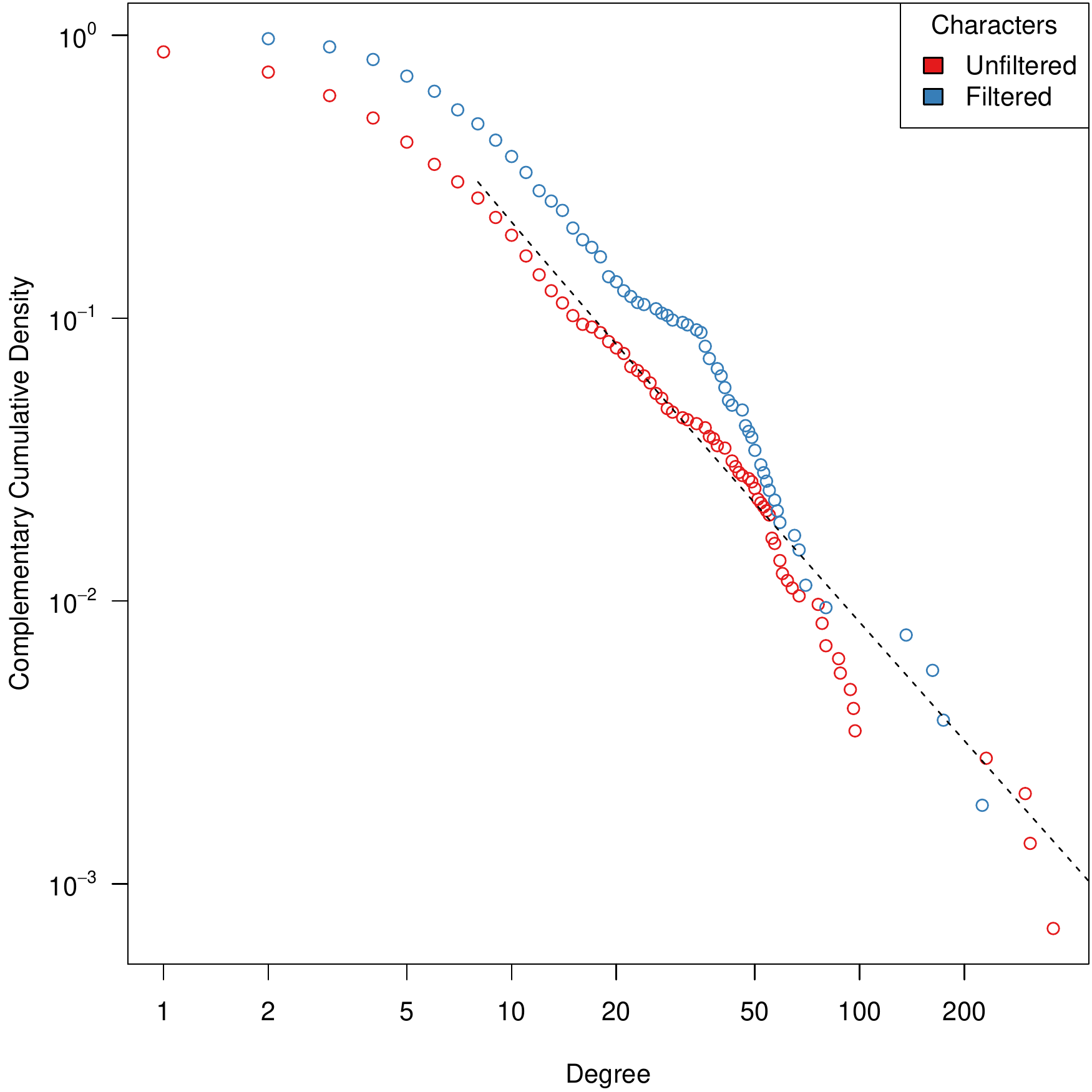}
                \phantomsubcaption\label{fig:DegreeUnfiltDistr}
            \end{subfigure}~~
	        \begin{subfigure}[t]{0.49\textwidth}
                \includegraphics[width=1\textwidth]{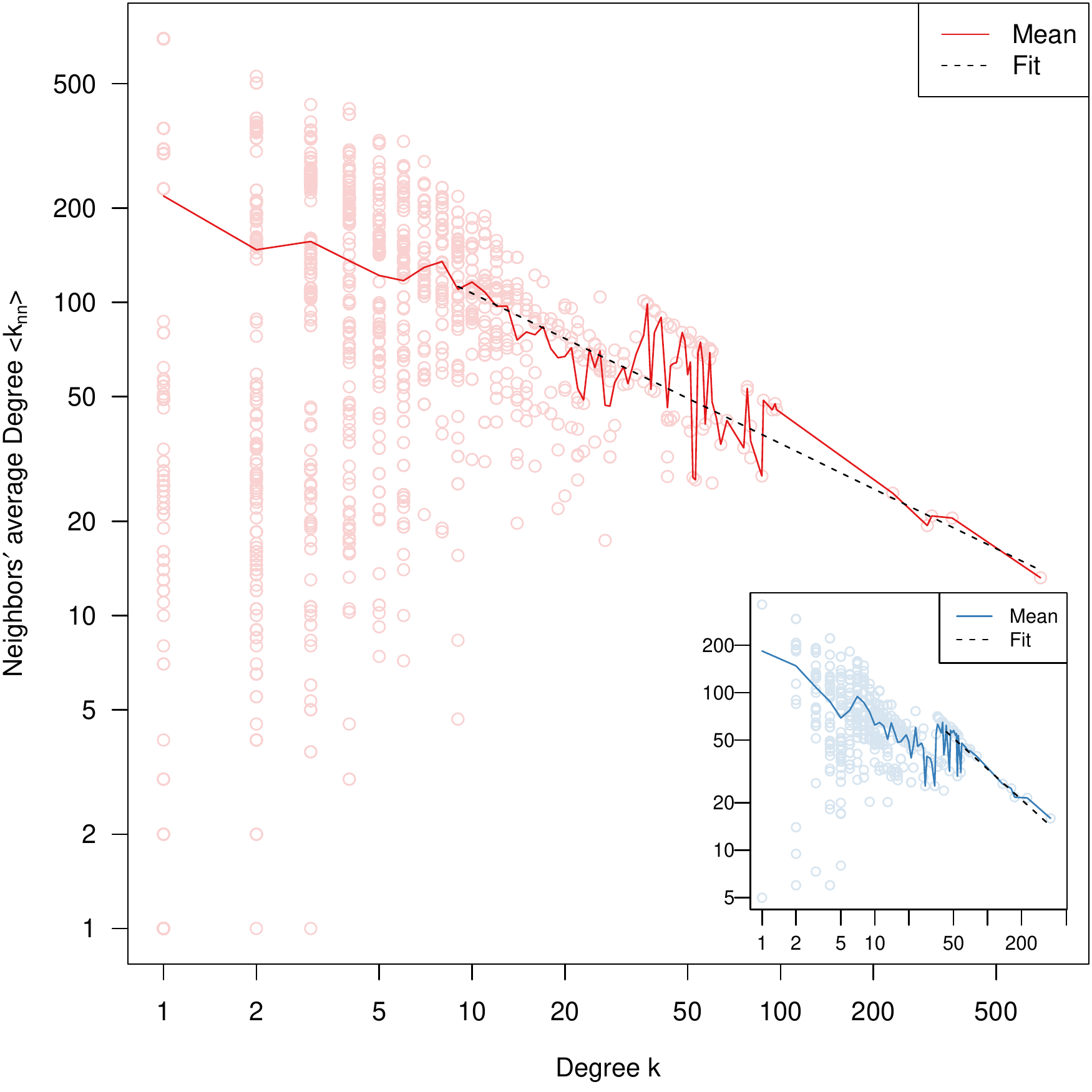}
                \phantomsubcaption\label{fig:DegreeUnfiltNeigh}
            \end{subfigure}
        };
        \node[anchor=west, circle, fill=white, inner sep=0.5pt] at (0.00,0.60) {\fontsize{8}{8}\selectfont{}\textbf{a)}};
        \node[anchor=west, inner sep=0.5pt] at (6.60,0.60) {\fontsize{8}{8}\selectfont{}\textbf{b)}};
    \end{tikzpicture}
    \vspace{-0.9cm}
    \caption{(a)~Degree distribution in the unfiltered (red) and filtered (blue) networks. The dotted line shows the power law best fitting the former. (b)~Average degree of the neighbors as a function of vertex degree, for the unfiltered (red) and filtered (blue) networks. The dotted line represents the power law best fitting the function. In both plots, the axes use a logarithmic scale. Figure available at \href{https://doi.org/10.5281/zenodo.6573491}{10.5281/zenodo.6573491} under CC-BY license.}
\end{figure}

\paragraph{Scale-Free Property}
Figure~\ref{fig:DegreeUnfiltDistr} shows the degree distribution observed in the unfiltered and filtered networks. The tail of the former is best fit by a power law ($x_{\min}=8$) with exponent $\alpha = 2.45$, which is in the $[2;3]$ range identified by Barabási~\cite{Barabasi2015} as characteristic of \textit{scale-free} networks. This is consistent with the presence of many lesser characters acquainted to only a few other persons, and of a few hubs, i.e. very important characters whose number of interacting partners is larger by several order of magnitude. Many real-world networks exhibit the same scale-free property, including the Web, or protein interaction networks, and more interestingly collaboration networks~\cite{Newman2003b}. Such networks are expected to contain large hubs, which in turn affect the average distance~\cite{Barabasi2015} (cf. Section~\ref{sec:DescAnalyDist}). By comparison, the exponent estimated by Alberich \textit{et al}.~\cite{Alberich2002} for the \textit{Marvel} network is smaller than $2$, which is uncommon in real-world networks. Barabási calls this the \textit{anomalous regime}: the hubs of such networks grow faster than the network itself, and $\langle d \rangle$ does not depend on $n$. Alberich \textit{et al}. interprets this as an excessive importance given to the main \textit{Marvel} characters, relative to real-world networks. A power law-distributed degree is often the result of a \textit{preferential attachment} process, as described by Barabási \& Albert~\cite{Barabasi1999a}. In Section~\ref{sec:AddAnalyPrefAtt}, we discuss this aspect by considering the \textit{preferential attachment rate} of a vertex as a function of its degree $k$.

The \textit{Thorgal} network \textit{seems} realistic with respect to degree distribution. However, it is worth stressing that, for a real-world social network, being scale-free reflects a high heterogeneity in the way social interactions are distributed over the considered population. In the case of a fictional network though, this heterogeneity concerns interactions that are \textit{shown in the narrative}. Maybe certain extras or minor characters have a rich social life that the author chose not to show because this does not help to move the story forward: these characters can be considered as possibly incompletely described. Compared to the unfiltered network, the filtered one is narrowed to the characters whose social interactions are the focus of the story, and can be assumed to be more completely described. Its degree is not power law -distributed, and none of the tested alternative distributions are good fits either. From this perspective, the \textit{Thorgal} network does not appear to be socially very realistic. Moreover, it appears that filtering the lesser characters affects the degree distribution to such an extent that it looses its heavy tail. Interestingly, a variety of degree distributions is observed in the character network literature, besides the power law: truncated power law~\cite{MacCarron2013}, exponential law~\cite{MacCarron2012}, log-normal law~\cite{Yose2016}. As the filtering procedures are also quite different from one author to the other, we hypothesize that the nature of this procedure may affect the degree distribution of the filtered network.

One could assume that the hubs correspond to the most frequent characters discussed in Section~\ref{sec:DataProp}, and more generally that the degree of a vertex depends on the number of occurrences of the corresponding character. This is partly correct indeed, as Spearman's rank correlation coefficient between degree and number of occurrences (expressed in scenes) returns a high value of $r_s = 0.75$. Figure~\ref{fig:DegreeVsScenes} shows that the degree is a non-linear increasing function of the number of scenes, an observation also made by Rochat regarding Rousseau's autobiographical novel \textit{Les confessions}~\cite{Rochat2014a}. The match is not perfect, though. On the one hand, it is possible for a character to appear frequently in the narrative while interacting always with the same few characters, in which case his degree will be small. This is the case of Muff, Jolan's dog: he is the sixth most frequent character, but he always follows the family. His degree is consequently much lower than for the other members of the family, which often leave for their own adventures. On the other hand, one character could appear in a few scenes in which he interacts with many characters. His degree would be high despite his low frequency. This is the case of the three Elders of Aran, a collective character that essentially appears in a single volume, and consequently has the 170\textsuperscript{th} frequency among all characters. These Elders are the organizers of a tournament involving many characters, though, and thus get the 22\textsuperscript{nd} higher degree.

\paragraph{Disassortativity}
According to Table~\ref{tab:MainTopoProps}, both networks are slightly disassortative ($\rho_k$), meaning that hubs tend to connect to low degree vertices. Figure~\ref{fig:DegreeUnfiltNeigh} shows how the degree averaged over the neighbors of a vertex evolves as a decreasing function of this vertex degree. In the unfiltered network, the tail of this function is best fit by a power law with exponent $\gamma = 0.31$ ($x_{\min}=9$), represented by the dotted line in the figure. The same computation over a degree-preserving randomized version of the network leads to the same result, which means that the observed disassortativity is \textit{structural}~\cite{Barabasi2015}. It is caused by the network being scale-free: its hubs are interconnected by fewer edges than required for the network to be neutral (i.e. neither assortative or disassortative). The same observations hold for the filtered network (inset), with $\gamma = 0.37$. 

By comparison, real-world collaboration networks are generally assortative~\cite{Newman2003b,Barabasi2015}, i.e. vertices tend to connect to other vertices of similar degree. We can therefore conclude that this aspect of the \textit{Thorgal} network is not realistic, reflecting its artificial nature. As shown by Gleiser~\cite{Gleiser2007}, the \textit{Marvel} network exhibits an even higher level of disassortativity ($\gamma = 0.52$). This could be a feature of fictional character networks, as the same observation was made for other works of fiction, spanning different types of media, e.g. myths~\cite{MacCarron2012}, novels~\cite{Zhang2014q,Rochat2014a} and TV series~\cite{Liu2017d}. This could be related to the principle of \textit{Chekhov's Gun}, 
which states that the narrative should only mention elements necessary to tell the story. As a consequence, narratives generally overlook relationships between minor characters.

\subsection{Distance, Transitivity, and Small-World Property}
\label{sec:DescAnalyDist}
The unfiltered network contains a giant component that includes 94\% of the nodes, as in most real-world collaboration networks~\cite{Newman2001c}. By comparison, the filtered network is connected by construction. As mentioned before, the filtered network is slightly more compact (see $\langle d \rangle$, $d_{\max}$ in Table~\ref{tab:MainTopoProps}), and its transitivity (a.k.a. clustering coefficient, $\langle C \rangle$) is higher. Both observations can be explained by the removal of many leaves ($222$ out of $1,480$ vertices, according to Table~\ref{tab:CharFiltering}) during filtering, resulting in a denser and more tightly connected network.

\paragraph{Small-World Property.}
The small-world property was originally defined by Watts \& Strogatz~\cite{Watts1998} by comparing the average distance and the average local transitivity of the considered network to those of two reference networks: a random model and a lattice, both of comparable size and density. A small-world network possesses 1) a small average distance, i.e. of the same order as the random network's and much smaller than the lattice's; and 2) a large average local transitivity, i.e. close to the lattice's and much larger than the random network's. In their seminal work, the random model selected by Watts \& Strogatz is the well-known Erdős–Rényi model~\cite{Erdos1960}, and they use a regular ring lattice. The Erdős–Rényi does not preserve the degree distribution of the original network, which is an essential property of the network, as discussed in Section~\ref{sec:DescAnalyDegree}. For this reason, other authors~\cite{Telesford2011} propose to use the Configuration model~\cite{Molloy1995} instead. For the same reason, we do not use a regular ring lattice, but a ring lattice with the same degree distribution as in the original graph (cf. Section~\ref{sec:AddAnalyLattice} of the Supplementary Material). Finally, we also consider the bipartite model described in Section~\ref{sec:DescAnalyDegree}.

\begin{table*}[htb!]
    \small
    \caption{Average degree ($\langle k \rangle$), maximal degree ($k_{\max}$), degree assortativity ($\rho_k$), average distance ($\langle d \rangle$), diameter ($d_{\max}$) and average local transitivity ($\langle C \rangle$) obtained for both versions of the \textit{Thorgal} network and for the selected models.}
    \label{tab:SmallWorldStats}
    \setlength{\tabcolsep}{5pt}
    \begin{tabular}{l p{3.3cm} r r r r r r}
        \cline{2-8}\noalign{\smallskip}
        & \textbf{Model} & $\langle k \rangle$ & $k_{\max}$ & $\rho_k$ & $\langle d \rangle$ & $d_{\max}$ & $\langle C \rangle$ \\
        \noalign{\smallskip}\hline\noalign{\smallskip}
        \textbf{Unfiltered} & \textit{Thorgal} Network & $9.06$ & $676.00$ & $-0.17$ & $2.78$ & $7.00$ & $0.737$ \\
        \noalign{\smallskip}\cline{2-8}\noalign{\smallskip}
        & Bipartite Model & $19.57$ & $932.51$ & $-0.23$ & $2.35$ & $5.33$ & $0.641$ \\
        & Erdős–Rényi Model & $9.06$ & $20.33$ & $0.00$ & $3.56$ & $6.07$ & $0.006$ \\
        & Configuration Model & $9.07$ & $697.00$ & $0.00$ & $3.07$ & $7.33$ & $0.113$ \\
        & Ring Lattice & $9.08$ & $10.00$ & $1.00$ & $136.87$ & $275.00$ & $0.577$ \\
        \noalign{\smallskip}\hline\hline\noalign{\smallskip}
        \textbf{Filtered} & \textit{Thorgal} Network & $14.13$ & $358.00$ & $-0.18$ & $2.32$ & $5.00$ & $0.829$ \\
        \noalign{\smallskip}\cline{2-8}\noalign{\smallskip}
        & Bipartite Model & $37.20$ & $495.68$ & $-0.25$ & $1.94$ & $3.02$ & $0.561$ \\
        & Erdős–Rényi Model & $14.13$ & $26.57$ & $0.00$ & $2.66$ & $4.03$ & $0.027$ \\
        & Configuration Model & $14.12$ & $364.00$ & $0.00$ & $2.57$ & $4.96$ & $0.159$ \\
        & Ring Lattice & $14.10$ & $18.00$ & $0.90$ & $33.39$ & $66.00$ & $0.643$ \\
        \noalign{\smallskip}\hline
    \end{tabular}
\end{table*}

Table~\ref{tab:SmallWorldStats} summarizes the statistics obtained for the original networks and the considered models. For both the unfiltered and filtered networks, the average distance is of the same order of magnitude as for all considered random models, and much smaller than for the lattice. On the contrary, their transitivity is much larger than the random models', and even than the lattice's. Based on Watts \& Strogatz's definition~\cite{Watts1998}, we can conclude that the network is small-world. Newman \textit{et al}.~\cite{Newman2001c} observe that real-world collaborative networks tend to have a transitivity roughly twice as large as for their bipartite model. For the \textit{Thorgal} networks, this factor is $1.15$ (unfiltered) and $1.48$ (filtered), which puts them somewhere between the real-world and random cases. By comparison, the transitivity of the \textit{Marvel} network is twice as large as the bipartite model's. However Alberich \textit{et al}. do not consider it as small-world, because its transitivity is small in absolute terms~\cite{Alberich2002}. 

\begin{figure}[htb!]
    \centering
    \begin{tikzpicture}
        \node[anchor=south west,inner sep=0] (image) at (0,0) {
	        \begin{subfigure}[t]{0.49\textwidth}
                 \includegraphics[width=1\textwidth]{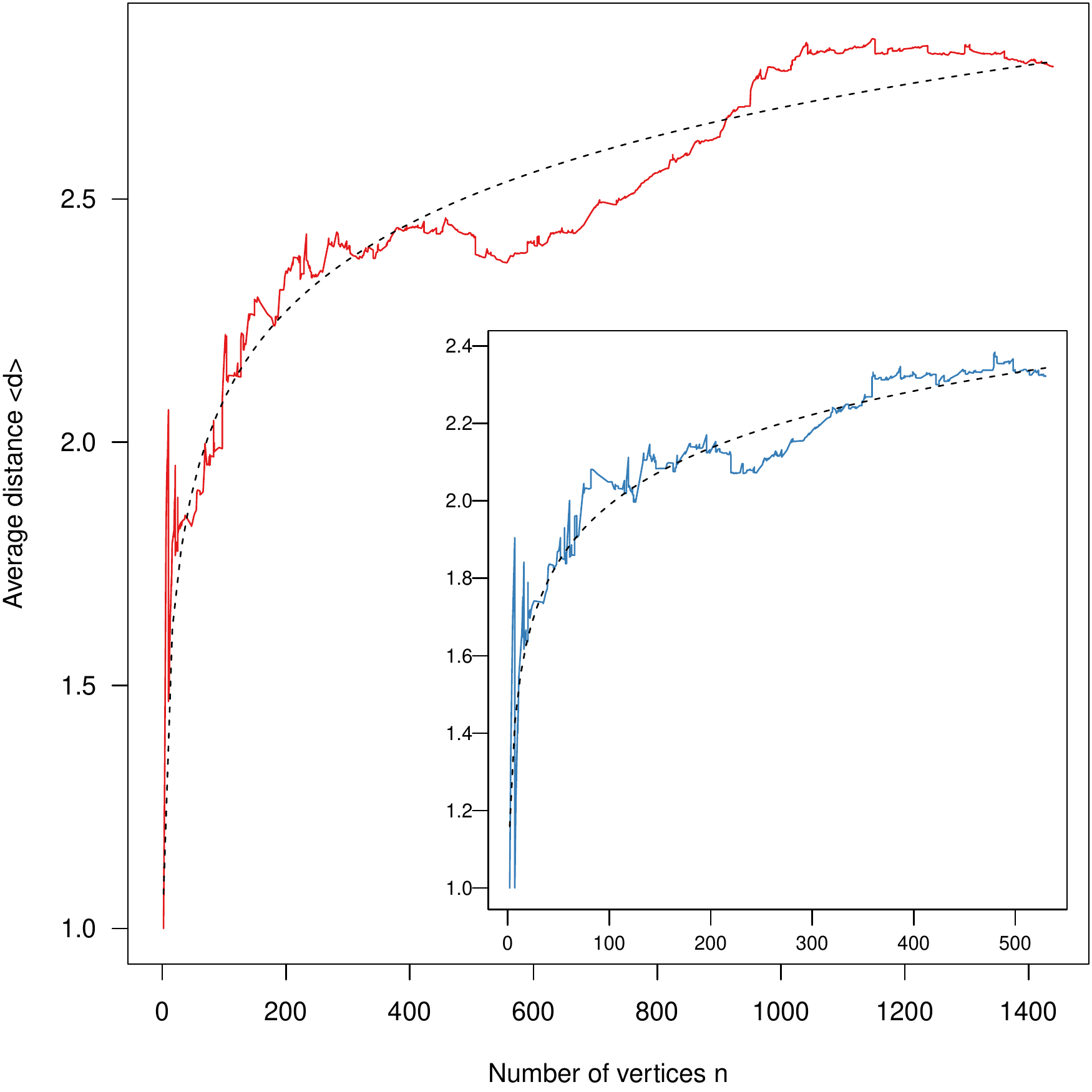}
                \phantomsubcaption\label{fig:DistEvol}
            \end{subfigure}~~
	        \begin{subfigure}[t]{0.49\textwidth}
                \includegraphics[width=1\textwidth]{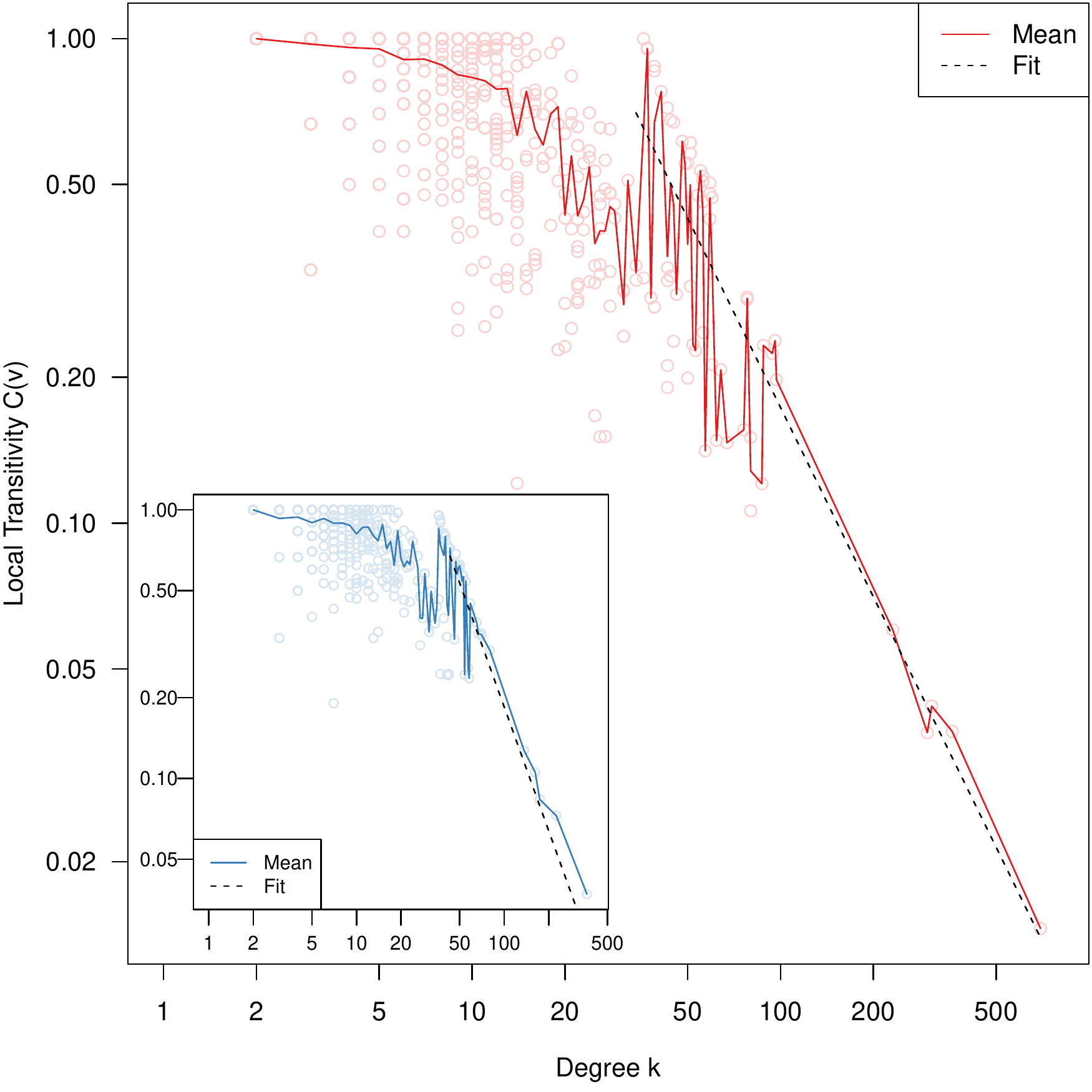}
                \phantomsubcaption\label{fig:TransVsDeg}
            \end{subfigure}
        };
        \node[anchor=west, circle, fill=white, inner sep=0.5pt] at (0.00,0.60) {\fontsize{8}{8}\selectfont{}\textbf{a)}};
        \node[anchor=west, inner sep=0.5pt] at (6.60,0.60) {\fontsize{8}{8}\selectfont{}\textbf{b)}};
    \end{tikzpicture}
    \vspace{-0.9cm}
    \caption{(a)~Evolution of the average distance as a function of the number of vertices in the unfiltered (red) and filtered (blue) networks. Each time step corresponds to one scene. The dotted lines represent the best logarithm fits. (b)~Local transitivity as a function of the degree, in the unfiltered (red) and filtered (blue) networks, using a logarithmic scale on both axes. The dotted lines represent the best power law fits. Figure available at \href{https://doi.org/10.5281/zenodo.6573491}{10.5281/zenodo.6573491} under CC-BY license.}
\end{figure}

The small-world property can alternatively be defined as a logarithmic dependence between the average distance and the network size and average degree~\cite{Watts1998,Barabasi2015}: $\langle d \rangle \sim \ln n / \ln \langle k \rangle$. The scale-free nature of the unfiltered network even indicates that it should be ultra-small-world, i.e. $\langle d \rangle \sim \ln \ln n$~\cite{Cohen2003b,Barabasi2015}. As a way of testing this assumption, we propose to extract a series of \textit{cumulative} networks, using one scene as the time step. Each network in this series models the characters and interactions occurring in the scene of interest, as well as those of the previous scenes. Figure~\ref{fig:DistEvol} shows how their average distance evolves over the sequence, as a function of the number of vertices. This function is well fit by a logarithm (represented by a dotted line in the figure), thereby confirming the small-world property. This holds for both unfiltered and filtered versions of the network. 


The average distance and transitivity of the \textit{Thorgal} networks are revealing of the story told in the series. \textit{Thorgal} is a rather episodic series, meaning it can be broken down into a few loosely connected narrative arcs (cf. Table~\ref{tab:NarrArcs}, Supplementary Material). The large average transitivity reflects the presence of relatively separated cliquish subgroups matching the social groups involved in these episodes. The small average distance indicates that the story is compact~\cite{Labatut2019}, which means that there is a certain narrative proximity between all the characters, even if these are very different and appear in distant moments of the story. This shows that, despite its episodic nature, the series is so centralized around Thorgal and its family and friends, that these efficiently connect very remote parts of the character space.

\paragraph{Hierarchical Structure}
Figure~\ref{fig:TransVsDeg} shows the average transitivity as a function of the degree.  In both the unfiltered and filtered networks, the tail of this decreasing function is best fit by a power law with exponent $0.42$ ($x_{\min}=39$) and $1.53$ ($x_{\min}=44$), respectively. This behavior is similar to various real-world networks, including collaboration networks~\cite{Ravasz2003,Barabasi2015}, and indicates the presence of a hierarchical structure. Gleiser shows that it is also the case of the \textit{Marvel} network~\cite{Gleiser2007}. Vertices with small degree tend to have a higher transitivity, forming tightly-knit clusters, whereas hubs tend to link these small structures and make the network connected at a higher level.

However, it is worth stressing that the dispersion observed in terms of transitivity is very different for \textit{Thorgal} and \textit{Marvel}, compared to other networks. Indeed, it mainly concerns small degree nodes, whereas in the cases shown in the literature, hubs tend to exhibit the wider range of transitivity~\cite{Ravasz2003,Barabasi2015}. This consistently low transitivity among the main characters suggests that the structure is more pronounced at the top of the hierarchy. By comparison, in real-world networks, top vertices tend to collaborate with individuals which are not as consistently disconnected from each other. 

In the \textit{Marvel} network, Gleiser~\cite{Gleiser2007} observes that the small clusters are organized around a few super-heroes surrounded by super-villains. Central super-heroes from different clusters are connected, forming higher level bonds, whereas this is not true for super-villains. Gleiser assumes that this is due to the \textit{Comics Authority Code} (CAC), a set of rules applying to comics. It prevents authors from having villains playing leading roles, and requires that good triumphs. Consequently, villains are not likely to band together, whereas hero collaboration allows making the story more interesting, even though the end is known in advance. In \textit{Thorgal}, we observe clusters of $5$--$20$ characters roughly matching specific narrative arcs or one-shot volumes. An interesting difference with the \textit{Marvel} network is that some of these groups are connected through antagonists (e.g. Kriss of Valnor, Varth). This can be explained by the fact that \textit{Thorgal} is not constrained by the CAC, and that its characters are much less Manichean: some of them hold an ambiguous position (e.g. Kriss of Valnor, who is both Thorgal's antagonist and lover).


\subsection{Vertex Centrality}
\label{sec:DescAnalyCentr}
We consider four standard centrality measures to characterize the position of individual vertices in the network, as these are widely use in the character network literature: degree, Eigenvector centrality~\cite{Bonacich1987}, betweenness~\cite{Freeman1978}, and closeness~\cite{Freeman1978}. 

\paragraph{Comparison Between Measures}
As mentioned before, high degree vertices represent characters interacting with many partners. A high betweenness typically denotes a broker, i.e. a character connecting otherwise well separated parts of the network, which in our case correspond to clusters matching distinct narrative arcs. The closeness allows describing the position of a character in terms of core vs. periphery, at the global and/or local levels. The Eigenvector centrality highlights hubs located in dense parts of the networks, containing other hubs. In addition to the power law distribution already described for the degree in Section~\ref{sec:DescAnalyDegree}, and similarly to Rochat's results for Rousseau's \textit{Les Confessions}~\cite{Rochat2014a}, we observe a long-tailed distribution for the Eigenvector centrality and betweenness, whereas the closeness has a more homogeneous distribution (i.e. normal-like). 

Plotting the measures against each other shows a general positive correlation between them, as shown by Figure~\ref{fig:CentrVsCentr} (Supplementary Material). This is confirmed when computing Spearman's rank correlation: closeness vs. Eigencentrality ($0.94$), degree vs. Eigencentrality ($0.78$), betweenness vs. degree ($0.70$), closeness vs. degree ($0.63$), betweenness vs. Eigencentrality ($0.50$), betweenness vs. closeness ($0.35$). Overall, our results exhibit a similar level of correlation compared to those obtained by Rochat for \textit{Les confessions}~\cite{Rochat2014a}. This relatively high correlation reflects the fact that most vertices get a similar (relative) centrality according to all measures~\cite{Rochat2014a}. But what interests us here are the vertices who do \textit{not}.

\begin{table*}[htb!]
    \caption{Ranks of the most central characters, according to the four considered centrality measures. All characters among the 12 most frequent or central characters (according to at least one measure) are represented.}
    \label{tab:CentralCharsDozen}
    \small	
    \begin{tabular}{p{2.0cm} r r r r r}
        \toprule
        \textbf{Character} & \textbf{Scene} & \textbf{Degree} & \textbf{Eigencentrality} & \textbf{Betweenness} & \textbf{Closeness} \\
        & \textbf{Rank} & \textbf{Rank} & \textbf{Rank} & \textbf{Rank} & \textbf{Rank} \\
        \colrule
        Thorgal  &   1 &   1 &   1 &  1 &   1 \\
        Aaricia  &   2 &   2 &   2 &  3 &   2 \\
        Jolan    &   3 &   3 &   3 &  4 &   3 \\
        Kriss    &   4 &   4 &   5 &  2 &   4 \\
        Louve    &   5 &   5 &   4 &  5 &   5 \\
        \colrule
        Muff     &   6 &   8 &   7 & 28 &   8 \\
        Sigwald  &   7 &  10 &  41 &  6 &  60 \\
        Petrov   &   7 &  11 &  44 & 22 &  48 \\
        Yasmina  &   9 &  24 &  47 & 10 &  51 \\
        Aniel    &  10 &  16 &  43 & 14 &  24 \\
        Tjall    &  11 &   7 &   6 & 39 &   6 \\
        Lehla    &  12 &   9 &   9 & 16 &   9 \\
        Vigrid   &  13 &  18 &  46 &  7 &  49 \\
        Argun    &  15 &  13 &  12 & 18 &  10 \\
        Gandalf  &  16 &   6 &   8 &  8 &   7 \\
        Hierulf  &  21 &  11 &  38 & 11 &  40 \\
        Darek    &  21 &  13 &  10 & 30 &  14 \\
        Variay   &  45 &  21 &  13 & 36 &  11 \\
        Gwenda   &  67 &  79 & 248 &  9 &  55 \\
        Veronar  & 125 &  19 &  14 & 53 &  12 \\
        Elders   & 169 &  21 &  11 & 96 &  13 \\
        Kahaniel & 184 & 164 & 864 & 12 & 782 \\
        \botrule
    \end{tabular}
\end{table*}

\paragraph{Typology of secondary characters}
Table~\ref{tab:CentralCharsDozen} shows the \textit{ranks} of the most frequent and central characters, according to their centrality. It includes all characters which are among the 12 most central characters according to at least one measure. The 5 most central characters are always the same, almost in the same order, whatever the considered centrality measure. In addition, these correspond to the most frequent characters (identified using specific colors in Figure~\ref{fig:FilteredNet}): Thorgal and his family. However, the measures disagree regarding the other characters, allowing us to distinguish four different cases. 

First, six characters have a much lower betweenness rank, compared to the other measures. Among them, Muff, Tjall, Darek and Lehla are recurrent sidekicks, that support the main protagonists for extended periods of time; whereas Veronar and the Elders are antagonists that appear only in a single volume. In both cases, they are well-connected locally, but at least one of their neighbors has a better position (generally: one of the main characters) and claims most of the shortest paths, hence the low betweenness. 

Second, quite similarly to the main protagonists, two characters have approximately the same rank for all measures. Argun is a friend of Thorgal's family, that follows them in their adventures, and Gandalf is one of their main antagonists. The main difference with our first category is a higher betweenness rank, which can be explained by a stronger recurrence of these characters: they appear in more narrative arcs, which makes them better connected to more distinct parts of the network.

Third, and contrarily to the first category, four characters have a much better betweenness rank, compared to the other measures. They connect some of the main characters to relatively remote parts of the network, corresponding to certain narrative arcs, but are not very well connected locally. Vigrid is a lesser god that interacts a lot with Thorgal and Aaricia in separate adventures, connecting them to major gods. Yasmina is a friend of Louve and has an important broker role in her spinoff. Gwenda and Kahaniel are recent antagonists connecting the main series to some narrative arcs developed in the spinoffs. They are both ranked poorly for all measure except betweenness.

Fourth, in addition to a better betweenness rank, a few characters also have a better degree rank. This means that they are at the same time secondary hubs and brokers. Their low Eigencentrality and closeness ranks indicate that their surroundings are not tightly connected, though. The remaining characters lie in between these somewhat extreme four categories.



\section{Research Questions}
\label{sec:ResearchQuestions}
In this section, we tackle the research questions described in Section~\ref{sec:DataSeries}. In Section~\ref{sec:RqIceland}, we compare the structure of \textit{Thorgal} and \textit{The Icelanders Saga}. In Section~\ref{sec:RqSex}, we study the position of women in \textit{Thorgal}.

\subsection{Comparison with the \textit{Íslendingasögur}}
\label{sec:RqIceland}

In this section, we focus on RQ1, i.e. the hypothesized similarity between Thorgal and the \textit{Íslendingasögur}, or \textit{Sagas of Icelanders}. It is a corpus of medieval Icelandic literature telling the story of the early settlers of Iceland. It is based on events that took place between the 9\textsuperscript{th} and 11\textsuperscript{th} centuries, but also includes supernatural elements. Some of these stories focus on a single adventurer, while others follow whole families over several generations.

As mentioned in Section~\ref{sec:DataSeries}, Thiry argues that \textit{Thorgal} can be seen as a hybridization between the \textit{Íslendingasögur} and more modern narrative elements~\cite{Thiry2019}. On the one hand, like this Icelandic corpus, the \textit{Thorgal} series is a collection of adventures that generally focus on the main character, but also tell the story of his family. Moreover, they combine realistic and supernatural elements derived from the Norse mythology. On the other hand, Thiry notes that Thorgal is an anti-mythological hero fighting to take control over his destiny, and distinguishing himself from the traditional Viking figure~\cite{Thiry2019}. Moreover, the series references other mythological sources (e.g. Greek tragedies), as well as fantasy, science-fiction and surrealist tropes.

In~\cite{MacCarron2013}, Mac Carron \& Kenna extract several character networks based on 18 sagas of the \textit{Íslendingasögur}, and study their structures in order to assess the level of historicity of the narrative. For this purpose, they use standard topological measures to describe the networks and compare them to values obtained in the literature for fictional and real-world social networks. Their assumption is that if the \textit{Íslendingasögur} networks reflect a realistic societal structure, they should be similar to the latter. We proceed similarly to compare \textit{Thorgal} to the \textit{Íslendingasögur}.

\paragraph{Aggregation of 18 Sagas}
We first compare the \textit{Íslendingasögur} network built upon the whole selection of 18 texts, and the unfiltered version of the \textit{Thorgal} network. As shown in Table~\ref{tab:MainTopoProps}, they roughly contain the same numbers of characters, which allows comparing their characteristics directly. This reveals several similarities. Both are built around a giant component, corresponding to 99\% of the characters for the former and 94\% for the latter. They both are small-world, exhibiting a high average local transitivity and a small average distance (cf. Section~\ref{sec:DescAnalyDist}). They both are scale-free, following power law distributions with exponents in the range $[2;3]$ ($\alpha = 2.99$ and $2.45$, respectively) (cf. Section~\ref{sec:DescAnalyDegree}). Both are also robust to random attacks but fragile to targeted attacks (cf. Section~\ref{sec:AddAnalyCentr}, in the Supplementary Material).

But there are also noticeable differences. The \textit{Thorgal} network is $50\%$ denser. Its hubs are much larger, Thorgal possessing a degree more than $8$ times that of the largest Icelandic hub. It is clearly disassortative, whereas the \textit{Íslendingasögur} is slightly assortative ($\rho_k = -0.17$ vs. $0.07$). The Thorgal network is more compact, with substantially smaller average distance ($\langle d \rangle = 2.78$ vs. $4.30$) and diameter ($d_{\max} = 7$ vs. $19$), and higher transitivity ($\langle C \rangle = 0.74$ vs. $0.50$). Overall, this depicts a network that is not as centered on a tight group of characters as \textit{Thorgal}, and whose edge distribution better matches real-world social networks. This realism is precisely what makes the \textit{Íslendingasögur} special~\cite{MacCarron2013,Thiry2019}, and we can only conclude that \textit{Thorgal} differs significantly from this Icelandic saga, at least from the perspective of the character network.

\paragraph{Egils Saga}
However, this difference could be explained by the fact that the considered Islandic network is the aggregation of 18 stories which, although sharing many characters, focus on different individuals or families. By comparison, \textit{Thorgal} has a single focus, on the main character and his direct family (half of the $20$ most frequent characters are direct relatives). Therefore, we now turn to the network of a single Icelandic story. Among those considered by Mac Carron \& Kenna~\cite{MacCarron2013}, we select \textit{Egils Saga}, because it is the most similar to the story told in \textit{Thorgal}. First, it is at the same time a \textit{family saga} (telling the story of a family) and a \textit{poet's saga} (focusing on the adventures of a protagonist). Second, it is not set in a single location, as the protagonist travels a lot. Third, it contains more supernatural elements than the other 17 sagas. We compare it with the filtered \textit{Thorgal} network, as they contain a similar number of vertices.

As before, both networks have a giant component and are small-world. However, the filtered \textit{Thorgal} network is not scale-free, whereas \textit{Egils Saga} degree follows a log-normal distribution~\cite{MacCarron2014}, and is therefore heavy-tailed. As before, the \textit{Thorgal} network is $50\%$ denser and has larger hubs, it is more compact in terms of average distance and diameter, and has a higher transitivity. However, unlike the aggregated Icelandic network, \textit{Egils Saga} is slightly disassortative ($\rho_k = -0.07$), which makes it similar to \textit{Thorgal}, on this point. According to Mac Carron \& Kenna, this is characteristic of sagas centering on one or a few characters, by opposition to a larger society~\cite{MacCarron2013}. Finally, and as before, both networks are robust to random attacks but fragile to targeted attacks.

\paragraph{Concluding Remarks}
According to Thiry, \textit{Thorgal} retains mainly three aspects of the Icelandic saga: 1) the story follows an extraordinary man and his family (also extraordinary), including his ancestors and children; 2) it depicts the Viking daily life, customs, and system of belief with a certain historical accuracy; and 3) it uses the Norse mythology as a background, but also as a driving narrative force. Certain structural similarities exhibited by both \textit{Thorgal} networks with respect to the \textit{Íslendingasögur} seem to support this observation: similar numbers of characters, small-world and scale-free structures suggesting realistic social interactions, strong focus on Thorgal and his family, high transitivity revealing an episodic story, supernatural entities with central positions (e.g. Vigrid, Kahaniel). However, there are also clear differences, in particular the \textit{Thorgal} networks are denser and more compact, with a stronger degree disassortativity, which reveals a higher focus on the main characters. In~\cite{MacCarron2012}, Mac Carron \& Kenna make similar observations when comparing networks extracted from mythological narratives and from other works of fiction, including novels, plays, and the \textit{Marvel} corpus. These differences could illustrate the hybridization noted by Thiry: recycling of foreign myths (Greek, Abrahamic, but also contemporary), more frequent occurrence of supernatural events through fantasy and science-fiction subplots, and Thorgal acting as a (post)modern hero by refusing the fate decided for him by the Norse gods.



\subsection{Position of Women}
\label{sec:RqSex}

One characteristic of the \textit{Thorgal} series is that its eponymous protagonist does not comply with the Viking value system of his adoptive clan. Instead, he goes by his own rules, which are at odds with those of his time, but consistent with the modern standards of the series authors'~\cite{Desfontaine2018}. As a teenager, he is an outcast in his village, and is consequently not trained as a warrior but as a skald (a Scandinavian poet): his male identity is not built upon brute force and hegemony, but rather on empathy and humanism, and on more balanced social gender relations. This is not to say that Thorgal does not exhibit traditional virile features that characterize mythological heroes (athleticism, courage, pugnacity, righteousness, etc.), but rather that those are mitigated by his personal values, resulting in what Desfontaine \textit{et al}.~\cite{Desfontaine2018} call a \textit{soft masculinity}. This aspect is highlighted by the fact that the main female characters are strong and independent women and girls. In particular, his main antagonist is Kriss of Valnor, which challenges him on most levels, and is treated on an equal foot not only by Thorgal himself, but more generally by the other characters.

\begin{figure*}[htb!]
    \centering
    \includegraphics[trim={15cm 20cm 18cm 16cm}, clip, width=1\linewidth]{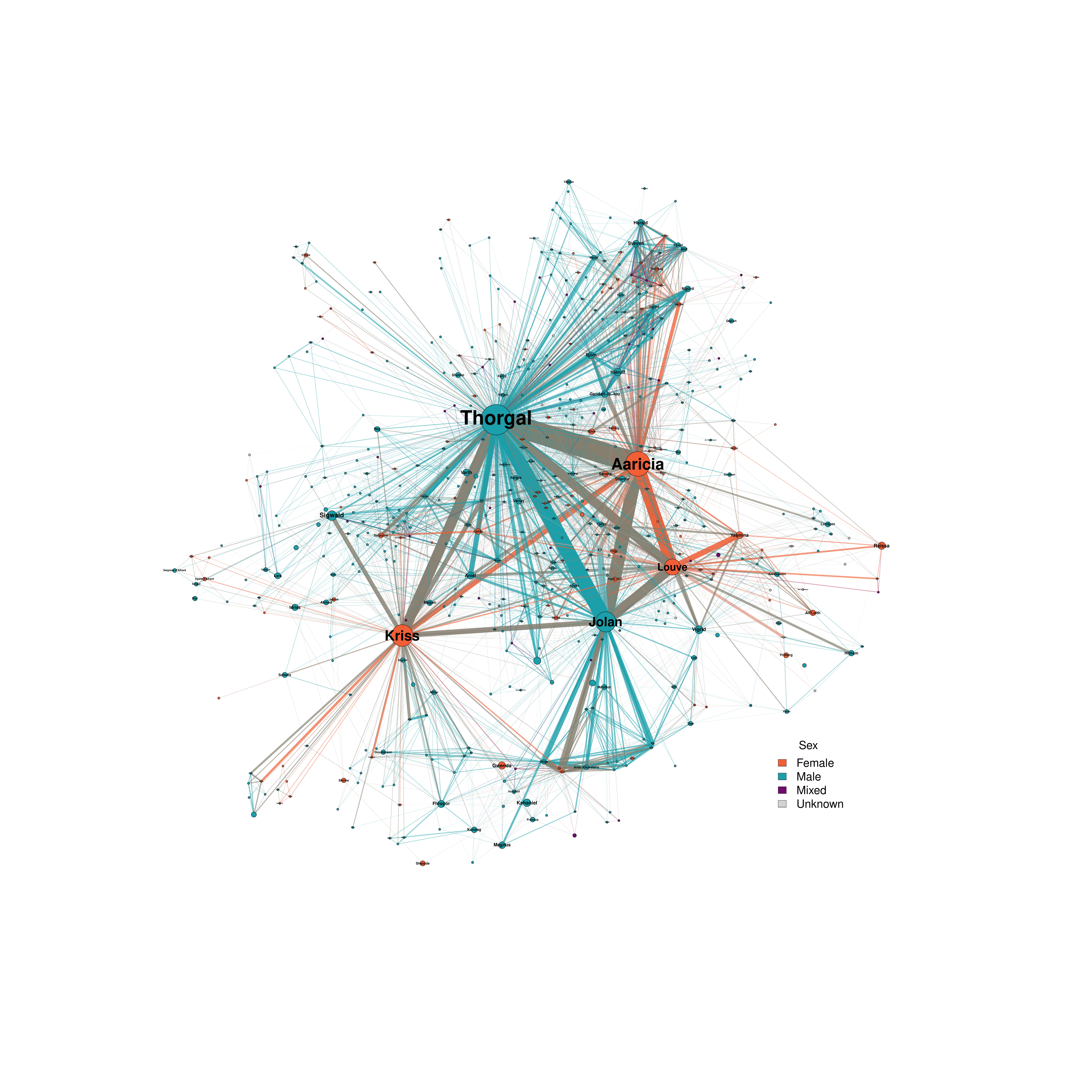}
    \caption{Sex of the characters in the filtered network, using the layout of Figure~\ref{fig:FilteredNet}. Figure available at \href{https://doi.org/10.5281/zenodo.6573491}{10.5281/zenodo.6573491} under CC-BY license.}
    \label{fig:FilteredNetSex}
\end{figure*}

\paragraph{Individuals}
This section focuses on RQ2, studying these aspects of \textit{Thorgal} by analyzing the position of female characters in the network. Figure~\ref{fig:FilteredNetSex} shows the filtered network using colors to represent the characters' sex. It appears clearly that males are prevalent. Among the unfiltered characters whose sex could be identified, 84\% are males and 16\% are females. After filtering, the proportion of males decreases to 79\% for 21\% females, which indicates that the proportion of women is larger when focusing on important characters. By comparison, there are $22\%$ women among \textit{Marvel} characters, 
with also an improvement when focusing on important characters: $31\%$~\cite{Hickey2014}. However, it is worth stressing that only super-heroes and super-villains are taken into account in these statistics, whereas our \textit{Thorgal} dataset contains all characters appearing in the narrative.


When focusing on the 10 most central characters, we find $4$ or $5$ females, depending on the considered centrality measure. In~\cite{Kagan2019}, Kagan \textit{et al}. study gender bias in movies spanning a period of one century, through the analysis of their character networks. By comparison with \textit{Thorgal}, they find a lower average of $33\%$ women among the 10 most central roles.
Note that their dataset includes only \textit{named} characters. They get a slightly lower proportion when considering the 5 most central roles. However, in the case of \textit{Thorgal}, 3 out of the 5 most central characters are always female, whatever the considered measure. Moreover, as explained in Section~\ref{sec:DescAnalyCentr}, these top 5 characters are clearly more central than the rest of the vertices. Thorgal is  unambiguously the main protagonist of the series, but $60\%$ of the core characters are females. In his study of \textit{Marvel} comics, Gleiser~\cite{Gleiser2007} observes that \textit{all} the central characters are males. Moreover, women do not act as bridges between parts of the network, as also observed for super-villains. In \textit{Thorgal}, the high betweenness of certain female characters shows that some women hold this intermediary position. On the same note, if in \textit{Marvel} comics women tend to be heroes and not villains~\cite{Hickey2014}, in \textit{Thorgal} the main antagonist (and fourth most frequent character) is a woman, Kriss of Valnor.

We compare the full distributions of the four centrality measures studied in Section~\ref{sec:DescAnalyCentr}, relative to character sex (cf. Figure~\ref{fig:CentrDistribSex}). We use the permutation test for integer-valued measures and the Kolmogorov–Smirnov test for real-valued measures. We find significant differences ($p<0.05$) for the degree and Eigencentrality, but not for the betweenness and closeness. This indicates that despite the strong male prevalence, female characters manage to retain important positions in terms of long-range centrality. In short, women tend to be less central than men for a separate episode, but they are as important when considering the whole narrative.

\paragraph{Relationships}
Regarding the relationships, in the unfiltered network, the majority of edges model full male interactions ($51\%$), followed by mixed ($43\%$) and full female ($6\%$) interactions. The proportions are very similar for the filtered network, with slightly more female and mixed edges ($7\%$). This is of course due to the large sex imbalance observed before between the characters. However, if the edges were randomly distributed, we would get 70\% of male, 27\% of mixed and 3\% of female edges: we can thus conclude that the network structure favors interactions involving females, more than expected by chance. We compute the density of the male- and female-induced subgraphs. The unfiltered and filtered female subgraphs are respectively three and two times denser than their male counterparts. This confirms that the narrative develops proportionally more connections between female than male characters. The sex-based assortativity measure is almost zero in both cases though, which shows that this property is hidden by the wide prevalence of male characters.

Following the method used by Kagan \textit{et al}.~\cite{Kagan2019}, we enumerate the triangles constituting the network, distinguishing the vertices only in terms of sex. Interestingly, in both the unfiltered and filtered networks, the most frequent type of triangle is FMM (one female and two male characters, $44\%$). Triangles of type MMM are only second ($33\%$), followed by FFM ($20\%$) and FFF ($3\%$) women. By comparison, in their corpus of movies, Kagan \textit{et al} obtain very similar proportions, with an important difference: MMM triangles are the most frequent ($41\%$), and FMM ones only come second ($36\%$). In both cases, $77\%$ of the triangles contain more men than women. Kagan \textit{et al}. notice some differences between movie genres: romantic movies tend to contain an increased number of triangles with a majority of women ($29\%$ FFM and $6\%$ FFF), whereas it is the opposite for war movies ($25\%$ FMM and $64\%$ MMM). This is interesting, because \textit{Thorgal} contains a number of fight scenes, and some characters participate in battles and even lead a war. One would therefore expect a much larger number of MMM triangles. There are love stories in \textit{Thorgal}, but only a few, especially between the protagonist and his wife: this does not seem to be enough to compensate for the war scenes. The Viking background of the narrative implies that women participate in warfare, which could be a better explanation of this peculiarity.

\paragraph{Gender Equality Testing}
The Bechdel test is a popular method to test whether a narrative represents women fairly. It has three criteria: 1) there must be at least two named female characters; 2) they must talk to each other; and 3) about another topic than men. Out of the $4,622$ scenes constituting our dataset, $441$ (10\%) involve only women. Among them, $200$ have more than one female character, which talk to each other in $172$ cases. After a manual assessment of these scenes, it turns out that $87$ ($51\%$) comply with the third criterion of the Bechdel test.


However, as noted by Kagan \textit{et al}.~\cite{Kagan2019}, the Bechdel test can be considered as too lenient, as a single interaction is enough for a narrative to pass, independently of the rest of the story. Instead, they propose a quantitative approach relying on a gender-based comparison of the numbers of interactions. They use the notion of \textit{volume} of a subgraph, which is the sum of the degrees over the vertices constituting this subgraph. Their \textit{Gender Degree Ratio} is the ratio of the male to female volumes. They consider that a situation of gender equality is described by a ratio between $0.8$ and $1.2$. In the case of \textit{Thorgal}, the ratio is $0.32$, so the series is clearly out of this interval. By comparison, Kagan \textit{et al}. get an average ratio of $0.6$ for their corpus of movies, of which only $12\%$ pass this test.

One could argue that the degree-based volume does not represent the number of \textit{punctual} interactions though, but rather the number of interaction \textit{partners}. This is likely to hide situations where recurrent female characters have many interactions over time. We propose to compute the \textit{strength}-based volume instead. Vertex strength is the total weight of the edges attached to the vertex of interest. In our case, we use the number of distinct interaction between two characters as weight. The ratio obtained for \textit{Thorgal} is $0.40$: this is higher than before, which indicates that this series matches the case mentioned above, but still far from  Kagan \textit{et al}.'s minimal threshold of $0.8$.

\paragraph{Concluding Remarks}
The soft masculinity of Thorgal is not obvious at first sight in the network structure, when only considering sex distribution among the characters. There is a large male prevalence, on par with other works of fiction such as the \textit{Marvel} comics and popular movies. However, focusing on the most important characters reveals that women indeed hold a different position in this network. They are prevalent in the top 5 characters, and hold central roles, unlike in the other considered works of fiction. In terms of relationships, the \textit{Thorgal} network similarly exhibits much more full male edges, but full female edges are more frequent than explained by the male vertex prevalence. Its triangles tend to involve more females than in other works of fiction. However, if \textit{Thorgal} passes the Bechdel test, its gender degree ratio does not meet the criteria proposed by Kagan \textit{et al}.~\cite{Kagan2019}, which are fulfilled by many of the movies they study. This could be due to them discarding unnamed characters, though.


\section{Conclusion}
\label{sec:Conclusion}
In this article, we proposed a new corpus describing the characters of the graphic novel \textit{Thorgal}, and the interactions occurring between them in the course of the narrative. We then leveraged this dataset to extract the series character network. We conducted a descriptive analysis which showed that it exhibits certain realistic properties (i.e. features also present in real-world social networks): small-worldness, scale-freeness, hierarchical structure, and sensitivity to attacks. By comparison, the comics-based \textit{Marvel} network studied by Alberich \textit{et al}.~\cite{Alberich2002} is less realistic (less transitive, not scale-free, robust to targeted attacks). However, the \textit{Thorgal} network is also disassortative, a trait generally found in fictional character networks. Moreover, we argue that it exhibits only the \textit{appearance} of realism: the structure of the network does not necessarily represent realistic social interactions, but rather the outcome of the creative process behind the unfolding of the story, which happens to possess some of the properties observed when modeling real-world relationships.

During our descriptive analysis, we compared two versions of our network: unfiltered (all characters) vs. filtered (important characters only). Filtering is generally performed quite systematically (and implicitly) in the character networks literature, and we wanted to assess its effect. It turns out it essentially affects the degree distribution, resulting in a loss of the scale free property. Interestingly, fictional character networks are often \textit{not} scale free~\cite{MacCarron2012}: according to our results, this could be explained not by the intrinsic nature of works of fiction, but rather by the way the networks are extracted and, more importantly, filtered.

In addition, we leveraged our character network to answer two research questions related to the series. In summary, we first showed that the similarity between \textit{Thorgal} and the \textit{Sagas of Icelanders} observed in~\cite{Thiry2019} does not transpire in the network structure in a meaningful way, especially when considering the characteristics of other fictional narratives. Second, we showed that the notion of soft masculinity discussed in~\cite{Desfontaine2018} seems to affect the network structure, both intrinsically and when compared to other works of fiction.

Our work can be extended in several ways. We plan to expand the dataset by annotating the character interactions in order to include directionality and polarity: this will allow us to extract directed and signed networks, and therefore to apply additional tools to complement our analysis. Another direct extension is to use the narrative smoothing method presented in~\cite{Bost2016} to extract a dynamic version of our network and study its evolution. We also want to conduct a comparative study of character networks extracted from works belonging to the three main graphic novel traditions, namely North American, European, and East Asian.


\section*{Acknowledgments}
I first thank Christian Labatut, who introduced me to \textit{Thorgal} and comics in general, more than 30 years ago. I also thank Arthur Amalvy, Noé Cécillon, and Elise Labatut, whose annotation work helped assess the quality of the main corpus.

\section*{Credits}
The French version of \textit{Thorgal} is published by Le Lombard, and the English version by Cinebook Ltd. The panels shown in this article are taken from the following books, all authored by Jean Van Hamme (writer) and Grzegorz Rosiński (artist):
\begin{itemize}
    \item Figure~\ref{fig:BdExAnimal}: \textit{Thorgal} vol.16 ``Wolfcub'', p.18, 1990.
    \item Figure~\ref{fig:BdExUnilateral}: \textit{Thorgal} vol.14 ``Aaricia'', p.41, 1989.
    \item Figure~\ref{fig:BdExNarration}: \textit{Thorgal} vol.20 ``The Brand of the Exiles'', p.9, 1995.
    \item Figure~\ref{fig:BdExImpers}: \textit{Thorgal} vol.17 ``The Guardian of the Keys'', p.28, 1991.
    \item Figure~\ref{fig:BdExEnters}: \textit{Thorgal} vol.20 ``The Brand of the Exiles'', p.16, 1995.
    \item Figure~\ref{fig:BdExLeaves}: \textit{Thorgal} vol.21 ``Ogotai's Crown'', p.33, 1995.
    \item Figure~\ref{fig:BdExEllipsis}: \textit{Thorgal} vol.18 ``The Sun Sword'', p.8, 1992.
    \item Figure~\ref{fig:BdExUnconsc}: \textit{Thorgal} vol.27 ``The Barbarian'', p.48, 2002.
    \item Figure~\ref{fig:BdExInvisible}: \textit{Thorgal} vol.28 ``Kriss of Valnor'', p.19, 2004.
    \item Figure~\ref{fig:BdExDistant}: \textit{Thorgal} vol.18 ``The Sun Sword'', p.40, 1992.
    \item Figure~\ref{fig:BdExGroups}: \textit{Thorgal} vol.18 ``The Sun Sword'', p.48, 1992.
    \item Figure~\ref{fig:BdExMultiple}: \textit{Thorgal} vol.15 ``The Master of the Mountains'', p.36, 1989.
    \item Figure~\ref{fig:BdEx4}: \textit{Thorgal} vol.11 ``The Eyes of Tanatloc'', p.11--12, 1986.
\end{itemize}

\bibliographystyle{ws-acs}
\bibliography{Labatut2022.bib}

\newpage
\color{black!60!blue}
\renewcommand{\thesection}{SM\arabic{section}}
\setcounter{section}{0}
\renewcommand{\thetable}{SM\arabic{table}}
\setcounter{table}{0}
\renewcommand{\thefigure}{SM\arabic{figure}}
\setcounter{figure}{0}
\begin{center}
    \Huge{Supplementary Material} \\[0.3cm]
    \Large{Complex Network Analysis of a Graphic Novel: \\[0.0cm] The Case of the Bande Dessinée \textit{Thorgal}} \\[0.2cm]
    \large{Vincent Labatut}
\end{center}

\bigskip
This document contains additional resources and results completing those already shown in the main article. It also contains some results that could not be placed in the main text due to lack of space. 

\section{Corpus}
\label{sec:AddData}
This section provides some information regarding the titles and characteristics of the volumes and narrative arcs of \textit{Thorgal} (Section~\ref{sec:AddDataSeries}). It also compares the Thorgal corpus with other comic novel-based datasets (Section~\ref{sec:AddDataCompar}). 

\subsection{\textit{Thorgal} Series}
\label{sec:AddDataSeries}
Table~\ref{tab:Volumes} lists the volumes constituting the \textit{Thorgal} series, including the main series and its spinoffs. They are shown by order of publication date. The volumes of the main series are simply numbered, whereas the spinoffs are prefixed with a letter: \textit{L} (\textit{Louve}), \textit{K} (\textit{Kriss of Valnor}), and \textit{J} (\textit{Young Thorgal}, or \textit{La jeunesse de Thorgal} in French).

\begin{table*}[htb!]
    \color{black!60!blue}
    \caption{\color{black!60!blue} List of Thorgal volumes, by order of publication date.}
    \label{tab:Volumes}
    \centering
    \footnotesize
    \begin{tabular}{r l l}
        \hline\noalign{\smallskip}
        \textbf{Nbr} & \textbf{Volume title} \\
        \noalign{\smallskip}\hline\noalign{\smallskip}
        01 & The Betrayed Sorceress \\
        02 & The Island of Frozen Seas \\
        03 & The Three Elders of Aran \\
        04 & The Black Galley \\
        05 & Beyond the Shadows \\
        06 & The Fall of Brek Zarith \\
        07 & Child of the Stars \\
        08 & Alinoe \\
        09 & The Archers \\
        10 & The Land of Qa \\
        11 & The Eyes of Tanatloc \\
        12 & City of the Lost God \\
        13 & Between Earth and Sun \\
        14 & Aaricia \\
        15 & The Master of the Mountains \\
        16 & Wolf Cub \\
        17 & The Guardian of the Keys \\
        18 & The Sun Sword \\
        19 & The Invisible Fortress \\
        20 & The Brand of the Exiles \\
        21 & Ogotai's Crown \\
        22 & Giants \\
        23 & The Cage \\
        24 & Arachnea \\
        25 & The Blue Plague \\
        26 & The Kingdom Beneath the Sand \\
        27 & The Barbarian \\
        28 & Kriss of Valnor \\
        29 & The Sacrifice \\
        30 & I, Jolan \\
        31 & Thor's Shield \\
        32 & The Battle of Asgard \\
        \noalign{\smallskip}\hline
    \end{tabular}
    \hspace{0.25cm}
    \begin{tabular}{r l l}
        \hline\noalign{\smallskip}
        \textbf{Nbr} & \textbf{Volume title} \\
        \noalign{\smallskip}\hline\noalign{\smallskip}
        K1 & I Forget Nothing \\
        33 & The Blade Ship \\
        L1 & Raissa \\
        K2 & The Sentence of the Valkyries \\
        K3 & Fit for a Queen \\
        L2 & The Severed Hand of the God Tyr \\
        J1 & The Three Minkelsonn Sisters \\
        L3 & The Kingdom of Chaos \\
        K4 & Alliances \\
        34 & Kah-Aniel \\
        J2 & The Eye of Odin \\
        L4 & Crow \\
        K5 & Red like Raheborg \\
        L5 & Skald \\
        J3 & Runa \\
        K6 & The Island of Lost Children \\
        J4 & Berserkers \\
        L6 & Queen of the Black Elves \\
        35 & The Scarlet Fire \\
        L7 & Nidhogg \\
        J5 & Slivia \\
        K7 & The Time Mountain \\
        J6 & The Ice Drakkar \\
        K8 & The Master of Justice \\
        36 & Aniel \\
        J7 & The Blue Tooth \\
        37 & The Hermit of Skellingar \\
        J8 & The Two Bastards \\
        38 & The Selkie \\
        J9 & The Tears of Hel \\
        39 & Neokora \\
        & & \\
        \noalign{\smallskip}\hline
    \end{tabular}
\end{table*}

Table~\ref{tab:NarrArcs} lists the narrative arcs of the \textit{Thorgal} series, according to the official website\footnote{\url{http://www.thorgal.com/}}, also by order of publication. Each arc is typically constituted of 2--3 volumes.

\begin{table}[htb!]
    \color{black!60!blue}
    \caption{\color{black!60!blue} Narrative arcs of the \textit{Thorgal} series, according to the official website. The last column indicates the type of network according to Rochat \& Triclot's nomenclature~\cite{Rochat2017}: \textit{kernel} (central group of characters), \textit{unicentric} (a single main character), \textit{polycentric} (several distinct centers), \textit{acentric} (no clear center).}
    \label{tab:NarrArcs}
    \centering
    \begin{tabular}{r p{6.3cm} l l}
        \hline\noalign{\smallskip}
        \textbf{Number} & \textbf{Arc title} & \textbf{Volumes} & \textbf{Type} \\
        \noalign{\smallskip}\hline\noalign{\smallskip}
        1 & The Queen of Frozen Seas & 01--02 & Unicentric \\
        2 & The Cursed Village & 03 & Acentric \\
        3 & The Masters of Brek Zarith & 04--06 & Kernel \\
        4 & The Origins & 07, 14 & Polycentric \\
        5 & The Dreamed Child & 08 & Acentric \\
        6 & The Great Country & 09--13 & Polycentric \\
        7 & To the North & 15--17 & Polycentric \\
        8 & Shaïgan-the-Merciless & 18--23 & Polycentric \\
        9 & The Wandering Viking & 24--26 & Kernel \\
        10 & The Last Trip & 27--29 & Polycentric \\
        11 & The Successor & 30--32 & Polycentric \\
        12 & The Valkyries' Court & K1--K2 & Unicentric \\
        13 & The Red Mages & 33--36 & Polycentric \\
        14 & The Fallen Mage & L1--L3 & Unicentric \\
        15 & North-Levant & K3--K5 & Polycentric \\
        16 & The Sald's Song & J1--J2 & Polycentric \\
        17 & The Black She-Wolf & L4--L5 & Polycentric \\
        18 & The Beast-Warriors & J3--J4 & Acentric \\
        19 & Aniel & K6--K8 & Polycentric \\
        20 & The Underground Kingdoms & L6--L7 & Kernel \\
        21 & The Betrayed Sorceress & J5--J6 & Polycentric \\
        22 & The Kingdom of the Danes & J7--J9 & Unicentric \\
        23 & New Horizons & 37--39 & Polycentric \\
        \noalign{\smallskip}\hline
    \end{tabular}
\end{table}

\begin{figure*}[htb!]
    \centering
    \begin{tikzpicture}
        \node[anchor=south west,inner sep=0] (image) at (0,0) {
	        \begin{subfigure}[t]{0.49\textwidth}
                 \includegraphics[height=6.13cm]{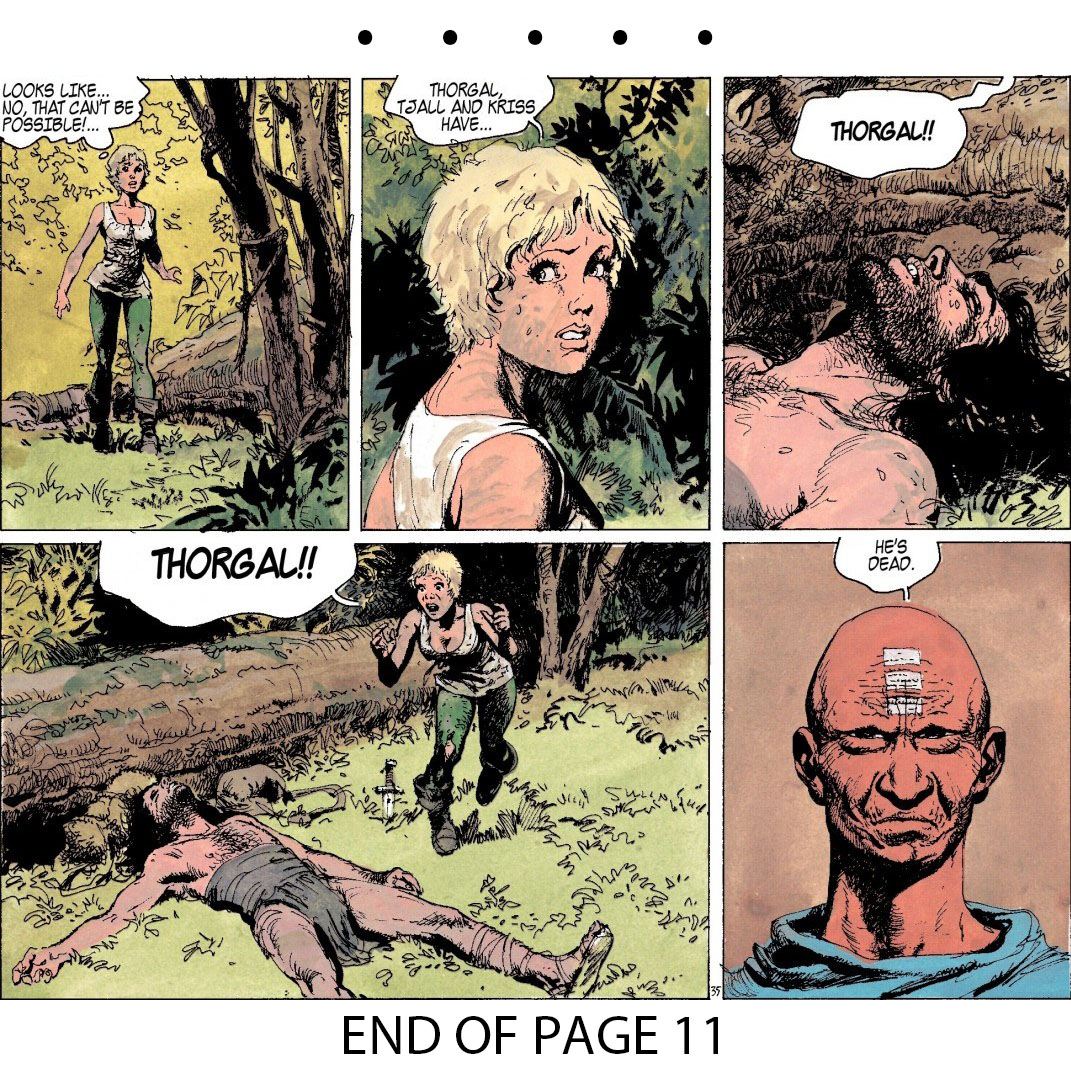}
                \phantomsubcaption\label{fig:BdExSuspStart}
            \end{subfigure}~~
	        \begin{subfigure}[t]{0.49\textwidth}
                \includegraphics[height=6.13cm]{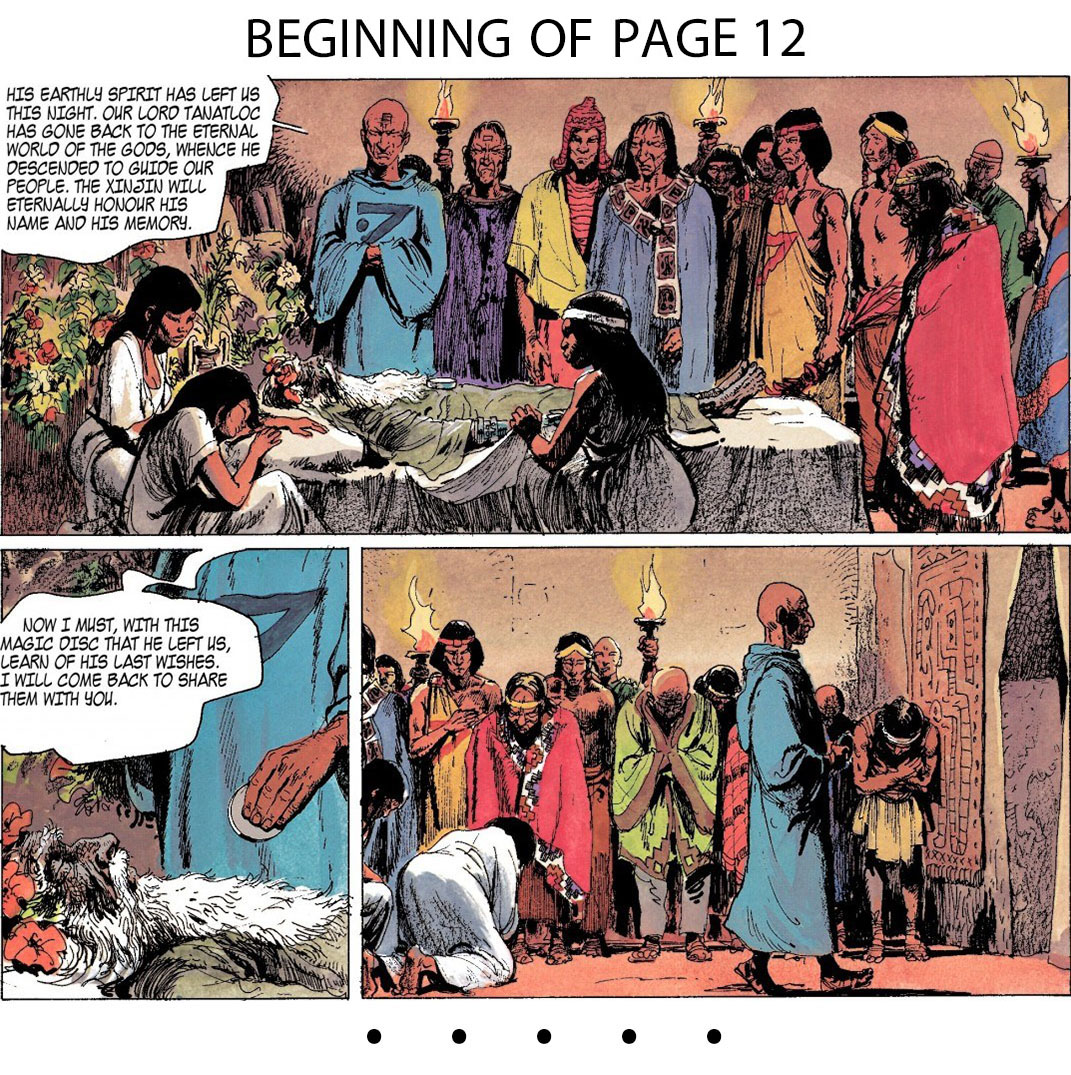}
                \phantomsubcaption\label{fig:BdExSuspEnd}
            \end{subfigure}
        };
        \node[anchor=west, circle, fill=white, inner sep=0.5pt] at (0.08,0.75) {\fontsize{8}{8}\selectfont{}\textbf{c)}};
        \node[anchor=west, circle, fill=white, inner sep=0.5pt] at (6.65,0.75) {\fontsize{8}{8}\selectfont{}\textbf{d)}};
    \end{tikzpicture}
    \vspace{-0.5cm}
    \caption{\color{black!60!blue} Example of the way artists use page layout to build up suspense. (a)~Aaricia finds Thorgal unconscious, and as he was previously very sick, the reader assumes that he may be dead. The last panel is part of another scene, happening in a completely different place. (b)~The next page reveals who is actually dead: Variay announces the death of Tanatloc.}
    \label{fig:BdEx4}
\end{figure*}

Figure~\ref{fig:BdEx4} complements our explanations from Section~\ref{sec:ExtractionMethodsGraph} by illustrating how artists use page layout to control the pace of the narrative.

\begin{figure*}[htb!]
    \centering
    \begin{tikzpicture}
        \node[anchor=south west,inner sep=0] (image) at (0,0) {
	        \begin{subfigure}[t]{0.32\textwidth}
                \includegraphics[width=1\textwidth]{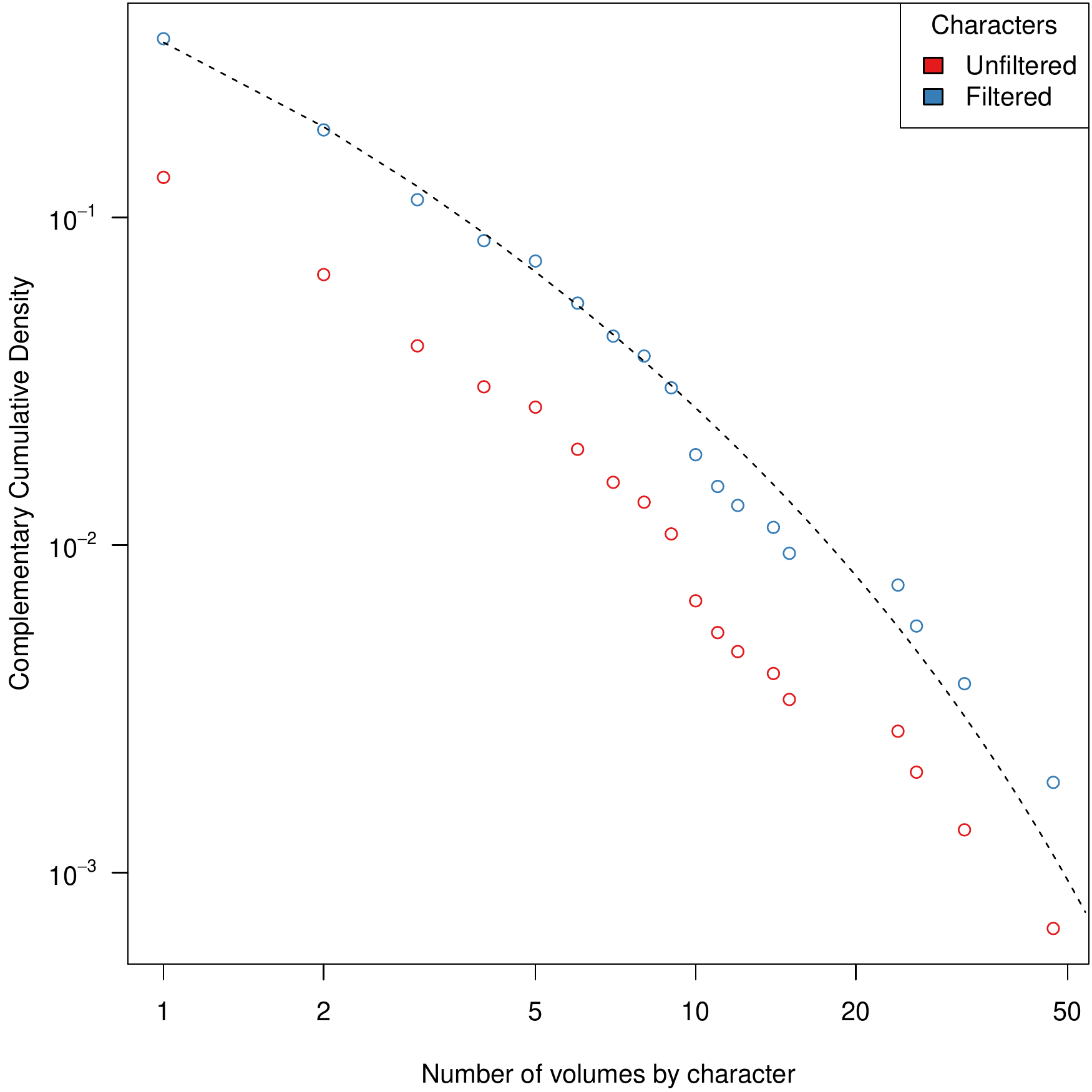}
                \phantomsubcaption\label{fig:VolDistr}
            \end{subfigure}~
	        \begin{subfigure}[t]{0.32\textwidth}
                \includegraphics[width=1\textwidth]{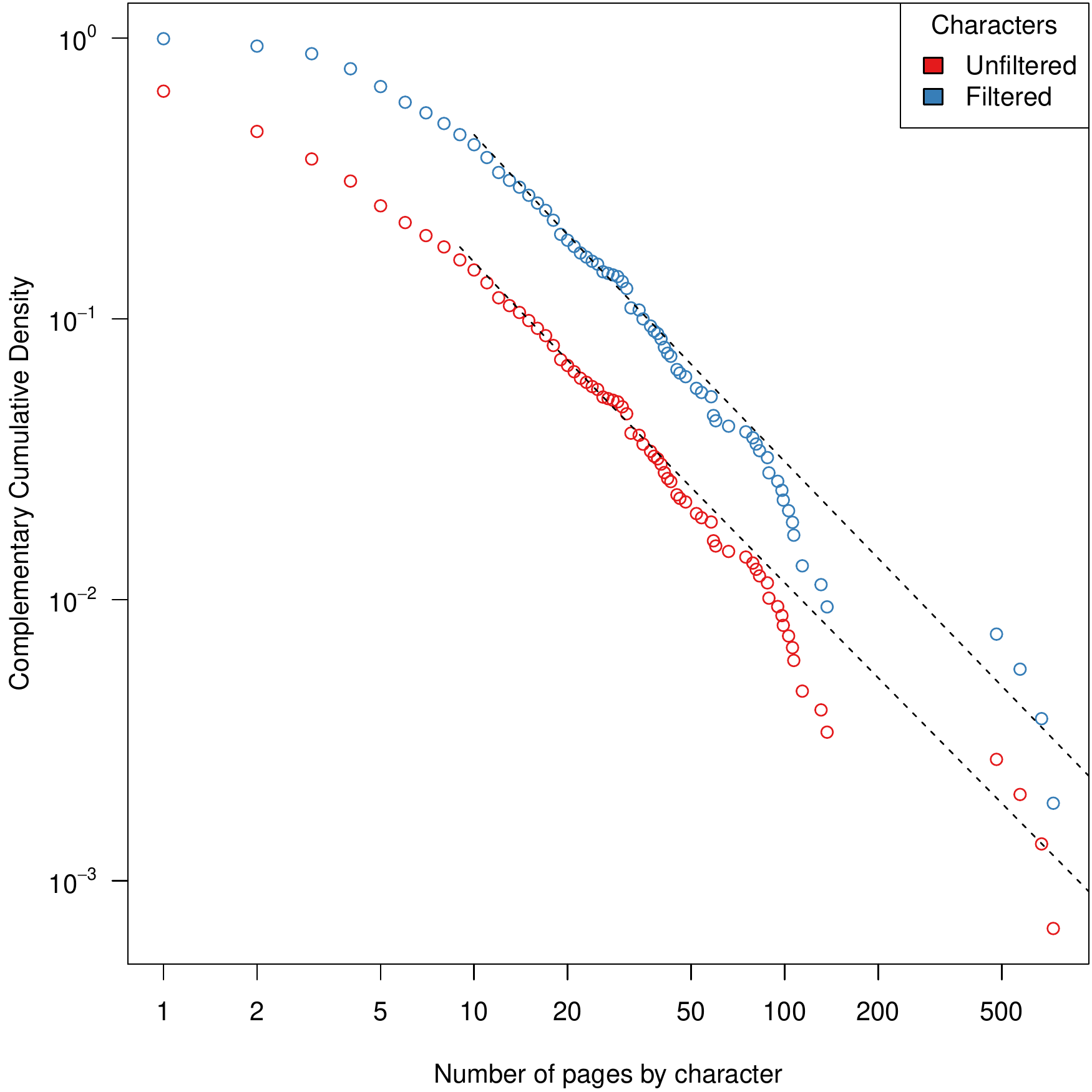}
                \phantomsubcaption\label{fig:PageDistr}
            \end{subfigure}~
	        \begin{subfigure}[t]{0.32\textwidth}
                \includegraphics[width=1\textwidth]{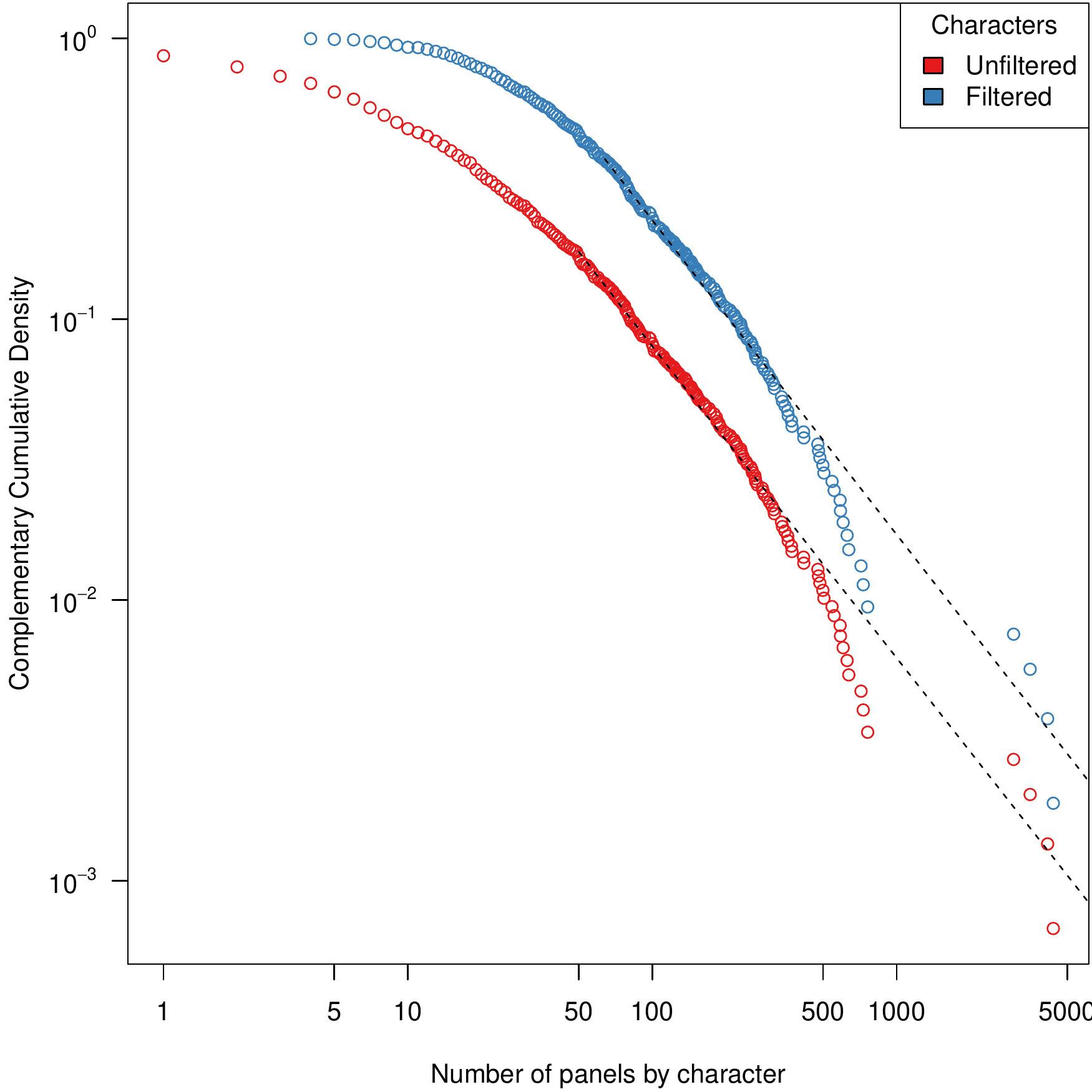}
                \phantomsubcaption\label{fig:PanelDistr}
            \end{subfigure}
        };
        \node[anchor=west] at (0.00,0.50) {\fontsize{8}{8}\selectfont{}\textbf{a)}};
        \node[anchor=west] at (4.20,0.50) {\fontsize{8}{8}\selectfont{}\textbf{b)}};
        \node[anchor=west] at (8.50,0.50) {\fontsize{8}{8}\selectfont{}\textbf{c)}};
    \end{tikzpicture}
    \vspace{-0.7cm}
    \caption{\color{black!60!blue} Complementary cumulative distribution of the number of (a)~volumes, (b)~pages, and (c)~panels by character, for all characters (red) and after filtering them (blue), in \textit{Thorgal}. The dotted lines represent the best fits, when available (see text). Figure available at \href{https://doi.org/10.5281/zenodo.6573491}{10.5281/zenodo.6573491} under CC-BY license.}
    \label{fig:OtherOccDistr}
\end{figure*}

As explained in Section~\ref{sec:DataProp} and shown in Figure~\ref{fig:SceneDistr}, the distribution of scenes by character in \textit{Thorgal} is best fit by a power law, both for unfiltered and filtered characters. For the sake of exhaustiveness, Figure~\ref{fig:OtherOccDistr} shows the distributions of volumes, pages and panels by character, for both the unfiltered and filtered versions of our dataset. The volume distribution (Figure~\ref{fig:VolDistr}) is best fit by a truncated power law for filtered characters, whereas no good fit is detected for unfiltered characters. The other distributions (Figures~\ref{fig:PageDistr} and~\ref{fig:PanelDistr}) all follow power laws. 


\subsection{Annotation}
\label{sec:AddDataAnnot}
As explained in Section~\ref{sec:DataProp}, we assess the quality of the main annotation by comparing it with the annotations produced by three additional annotators, on four randomly chosen volumes: \textit{Thorgal} vol.6, 16 and 20, and \textit{Kriss of Valnor} vol.1. In order to conduct this comparison, we consider the main annotation as our ground truth, and leverage the Precision, Recall, and $F$-measure separately for scene boundary detection and for character identification. Note that for the former task, these measures are very conservative, as they do not involve any tolerance regarding the position of the boundaries~\cite{Ercolessi2011}. 

A scene boundary correctly located is a true positive (TP). A scene boundary incorrectly defined by a secondary annotator is a false positive (FP). A scene boundary missed by a second annotator is a false negative). For characters identification, we first match each scene detected by a secondary annotator to a scene of the main annotation. For this purpose, we select the ground truth scene with the largest intersection, in terms of panels. Then, a character detected by a secondary annotator is a true positive if it is also present in the ground truth. It is a false positive if it is not present in the ground truth. On the contrary, a character that is not detected by the secondary annotator while it appears in the ground truth is a false negative.

Based on the numbers of TPs, FPs, and FNs, we can compute the Precision and Recall, using their standard definitions: 
\begin{align}
    Pre &= TP / (TP + FP) \\
    Rec &= TP / (TP + FN).
\end{align}
The $F$-measure is the harmonic mean of the Precision and Recall, i.e.
\begin{equation}
    F = 2 \frac{Pre \cdot Rec}{Pre + Rec}.
\end{equation}
These three definitions hold for both tasks (scene boundary detection and character identification).

Here, we consider the \textit{micro-average} versions of these measures, which means that the TP, FP and FN numbers are computed over all annotators. Tables~\ref{tab:AnnotationAgreementSc} and~\ref{tab:AnnotationAgreementChar} show the scores obtained for each considered volume separately, as well as overall (for all four volumes at once). Table~\ref{tab:AnnotationAgreementSc} focuses on the scene detection task, and Table~\ref{tab:AnnotationAgreementChar} on character identification. 

\begin{table*}[htb!]
    \color{black!60!blue}
    \caption{\color{black!60!blue} Precision, Recall and $F$-measure (all three micro-average) obtained by the three secondary annotators on the scene boundary detection task, for each one of the $4$ considered volumes, and overall.}
    \label{tab:AnnotationAgreementSc}
    \small
    \begin{tabular}{p{4.4cm} r r r r r r r}
        \toprule
        \textbf{Volume} & \textbf{$TP$} & \textbf{$FP$} & \textbf{$FN$} & \textbf{~~~~~~~~$Pre$} & \textbf{$Rec$} & \textbf{~~~~~~~~$F$} \\
        \colrule
        \textit{Thorgal} vol.6  & $189$ &  $8$ &  $7$ & $0.959$ & $0.964$ & $0.962$ \\
        \textit{Thorgal} vol.16 & $221$ &  $8$ & $16$ & $0.965$ & $0.932$ & $0.948$ \\
        \textit{Thorgal} vol.20 & $319$ &  $5$ &  $8$ & $0.985$ & $0.976$ & $0.980$ \\
        \textit{Kriss of Valnor} vol.1 & $313$ &  $8$ & $12$ & $0.975$ & $0.963$ & $0.969$ \\
        \colrule
        \textbf{Overall} & $1{,}042$ & $29$ & $43$ & $0.973$ & $0.960$ & $0.967$ \\
       \botrule
    \end{tabular}
\end{table*}

The agreement level is good, with an overall $F$-measure of $0.97$ and $0.94$, respectively. A manual inspection reveals that the disagreement between annotators are generally not due to errors, but rather to differences in the way they understand a scene. As explained in Section~\ref{sec:ExtractionMethodsAnnotation}, our annotation process requires identifying the characters involved in a scene, even when they are not shown explicitly in certain panels. This necessitates a good understanding of the story, and also to make some inference and assumptions. Sometimes, the annotators do not reach the same conclusion on these aspects. Another source of disagreement is the identification of extras and minor characters, which are sometimes mistaken with other similar characters.

\begin{table*}[htb!]
    \color{black!60!blue}
    \caption{\color{black!60!blue} Precision, Recall and $F$-measure (all three micro-average) obtained by the three secondary annotators on the character identification task, for each one of the $4$ considered volumes, and overall.}
    \label{tab:AnnotationAgreementChar}
    \small
    \begin{tabular}{p{4.4cm} r r r r r r r}
        \toprule
        \textbf{Volume} & \textbf{$TP$} & \textbf{$FP$} & \textbf{$FN$} & \textbf{~~~~~~~~$Pre$} & \textbf{$Rec$} & \textbf{~~~~~~~~$F$} \\
        \colrule
        \textit{Thorgal} vol.6  & $492$ & $31$ & $30$ & $0.941$ & $0.943$ & $0.942$ \\
        \textit{Thorgal} vol.16 & $595$ & $34$ & $31$ & $0.946$ & $0.950$ & $0.948$ \\
        \textit{Thorgal} vol.20 & $1{,}181$ & $85$ & $96$ & $0.933$ & $0.924$ & $0.929$ \\
        \textit{Kriss of Valnor} vol.1 & $779$ & $46$ & $26$ & $0.944$ & $0.968$ & $0.956$ \\
        \colrule
        \textbf{Overall} & $3{,}047$ & $196$ & $183$ & $0.940$ & $0.943$ & $0.941$ \\
       \botrule
    \end{tabular}
\end{table*}

\subsection{Comparison}
\label{sec:AddDataCompar}
It is worth comparing some characteristics of our dataset with those listed in Section~\ref{sec:RelatedWork}, whenever possible. For the European and Asian graphic novels studied by Rochat \& Triclot~\cite{Rochat2017}, we use the dataset provided in the article\footnote{\url{https://github.com/mtriclot/Belfort}} to compute the statistics ourselves. For the \textit{Marvel} comics, we leverage the information provided in Alberich \textit{et al}.'s article~\cite{Alberich2002,Gleiser2007}, and also use the dataset available online\footnote{\url{http://bioinfo.uib.es/~joemiro/marvel.html}}. Unfortunately, the Manga- and Webtoon-based datasets respectively constituted by Murakami \textit{et al}.~\cite{Murakami2011,Murakami2018,Murakami2020} and Lee \& Kim~\cite{Lee2020f} are not available and consequently could not be included in this comparison. Note that we focus on the filtered version of \textit{Thorgal}, as only the main characters are considered in the other datasets.

\begin{table*}[htb!]
    \color{black!60!blue}
    \caption{\color{black!60!blue} Statistics related to characters, volumes, and chapters, for \textit{Thorgal} (filtered version) and certain graphic novels listed in Section~\ref{sec:RelatedWork}.}
    \label{tab:CompDatasets}
    \footnotesize
    \begin{tabular}{p{2.2cm} r r r r r r}
        \toprule
        \textbf{Statistic} & \textbf{Thorgal} & \textbf{Marvel} & \textbf{Metabarons} & \textbf{Aldebaran} & \textbf{Akira} & \textbf{Gunnm} \\
        \colrule
        Characters &
            $530$ &  $6{,}486$ & $102$ & $30$ & $42$ & $37$ \\
        Volumes    &
            $63$ & $12{,}942$ &   $8$ &  $5$ &  $3$ &  $6$ \\
        Chapters   &
            $-$ &        $-$ &   $-$ &  $-$ & $28$ & $32$ \\
        \colrule
        Characters/Volume  &
            $15.25$ & $14.90$ & $27.00$ & $21.00$ & $21.33$ & $12.50$ \\
        Characters/Chapter &
            $-$    &  $-$    &  $-$    &  $-$    &  $9.00$ & $6.84$ \\
        \colrule
        Volumes/Character  &
            $2.25$ &  $7.47$ &  $2.20$ &  $3.50$ &  $1.56$ &  $2.50$ \\
        Chapters/Character &
            $-$    &  $-$    &  $-$    &  $-$    &  $6.15$ &  $7.30$ \\
        \botrule
    \end{tabular}
\end{table*}

Table~\ref{tab:CompDatasets} summarizes the main statistics of the mentioned graphic novels. It reflects some differences in the way these are published in America, Europe and Japan, from the perspective of their format and production pace. These differences are important in the context of our study, as they are known to affect various aspects of the graphic novels, in particular their narrative structure~\cite{Lefevre2000}. In the United States, the primary medium of dissemination is the publication of monthly issues of approximately 30 pages, which are generally not systematically collected in omnibus books~\cite{Couch2000}. Recent \textit{Marvel} issues tend to be 20 pages long~\cite{Cronin2017}. In France and Belgium, bandes dessinées used to be pre-published in specialized weekly magazines collecting various distinct series, before being released as hardcover albums. Nowadays, such pre-publication still exist, but is less frequent, and the main means of dissemination is the album, typically 46 pages long. \textit{Thorgal} and \textit{Aldebaran} follow this format, whereas \textit{Metabarons} is a bit longer, with 62-page volumes. In Japan, mangas are also pre-published in similar weekly or monthly magazines, under the form of 20-page long chapters, then collected in so-called tankobons~\cite{Lefevre2000}. The number of pages produced is larger than in Europe though, as mangas are generally produced by studios and not by individual authors. The \textit{Akira} and \textit{Gunnm} volumes contain approximately 175 pages, and their chapter 20--30 pages. Based on these lengths, manga chapters seem to better fit a comparison with American and European volumes, \textit{a priori}.

The compared datasets vastly differ in their numbers of volumes. The \textit{Marvel} dataset~\cite{Alberich2002} is the largest, as it describes the work of many people over forty years. By comparison, the main \textit{Thorgal} series was produced by two persons; this dataset is second in number of volumes only because it is complete. It is not the case of the datasets produced by Rochat \& Triclot~\cite{Rochat2017}: \textit{Metabarons} is part of the \textit{Incal} universe, \textit{Aldebaran} focuses only on the second arc of the larger series, and \textit{Akira} and \textit{Gunnm} contain only the first few volumes of their respective series. Despite these differences, the average number of volumes by character is rather similar for all datasets (2--4) except \textit{Marvel}. 
The average number of characters by volume is also relatively close for all series (15--20), except for \textit{Metabarons} ($27$). 
Finally, contrary to our initial assumption, the statistics obtained for the mangas when using chapters instead of volumes are the most different from the other datasets. This could be due to a more action-oriented way of staging the story, involving fewer characters in each chapter.

\section{Descriptive Analysis}
\label{sec:AddAnaly}
This section contains part of our descriptive analysis that could not fit in the main article. These include a study of the preferential attachment mechanism in \textit{Thorgal} (Section~\ref{sec:AddAnalyPrefAtt}); a description of the lattice-based generative model used in Section~\ref{sec:DescAnalyDist} to assess the small-world property (Section~\ref{sec:AddAnalyLattice}); some additional plots related to the four centrality measures considered in the article (Section~\ref{sec:AddAnalyCentr}); and a study of the narrative arcs of \textit{Thorgal} based on the graph typology proposed by Rochat \& Triclot in~\cite{Rochat2017} (Section~\ref{sec:AddAnalyNetTypes}).

\subsection{Preferential Attachment}
\label{sec:AddAnalyPrefAtt}
\begin{figure*}[htb!]
    \centering
    \begin{tikzpicture}
        \node[anchor=south west,inner sep=0] (image) at (0,0) {
	        \begin{subfigure}[t]{0.32\textwidth}
                \includegraphics[width=1\textwidth]{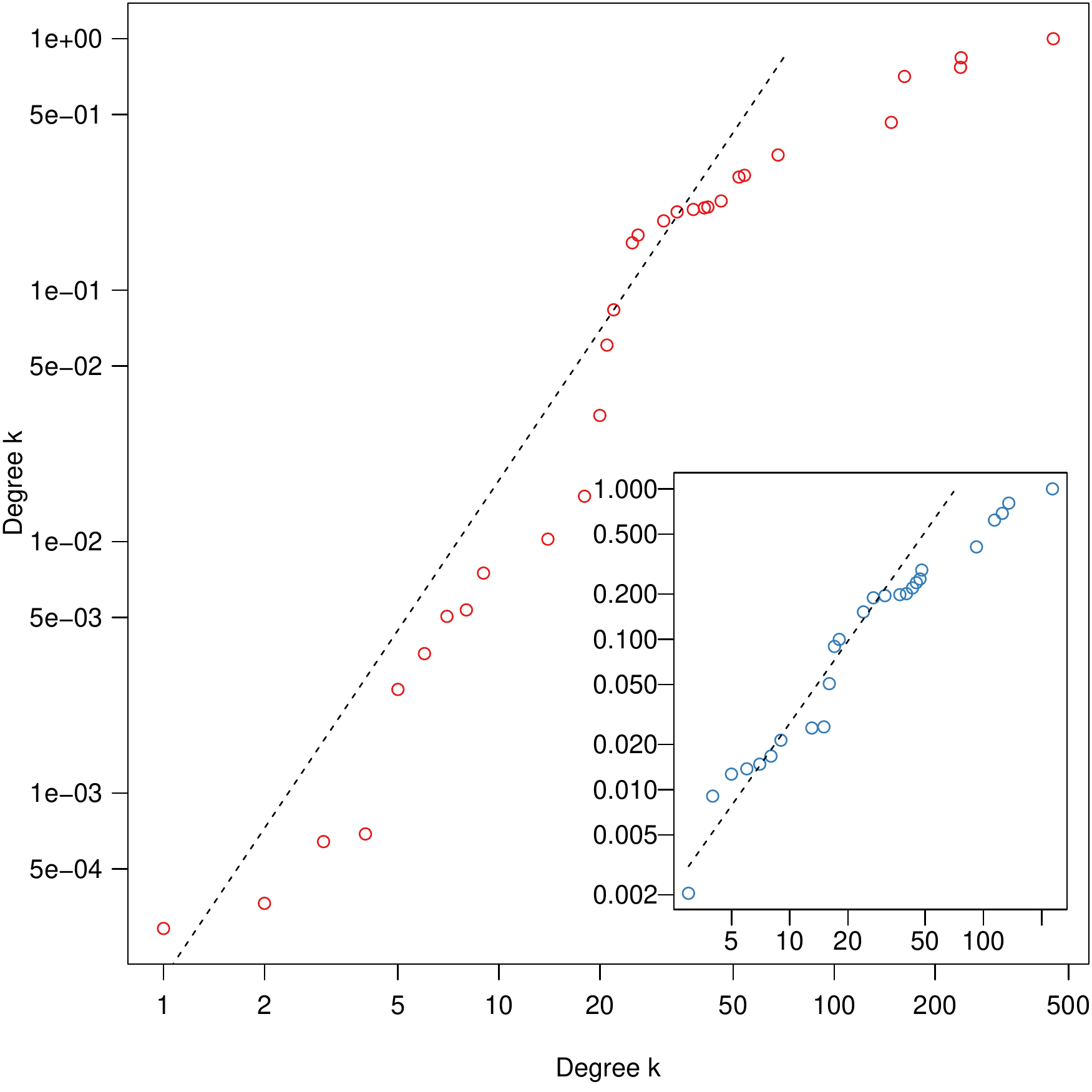}
                \phantomsubcaption\label{fig:DegreePrefAttAll}
            \end{subfigure}~
	        \begin{subfigure}[t]{0.32\textwidth}
                \includegraphics[width=1\textwidth]{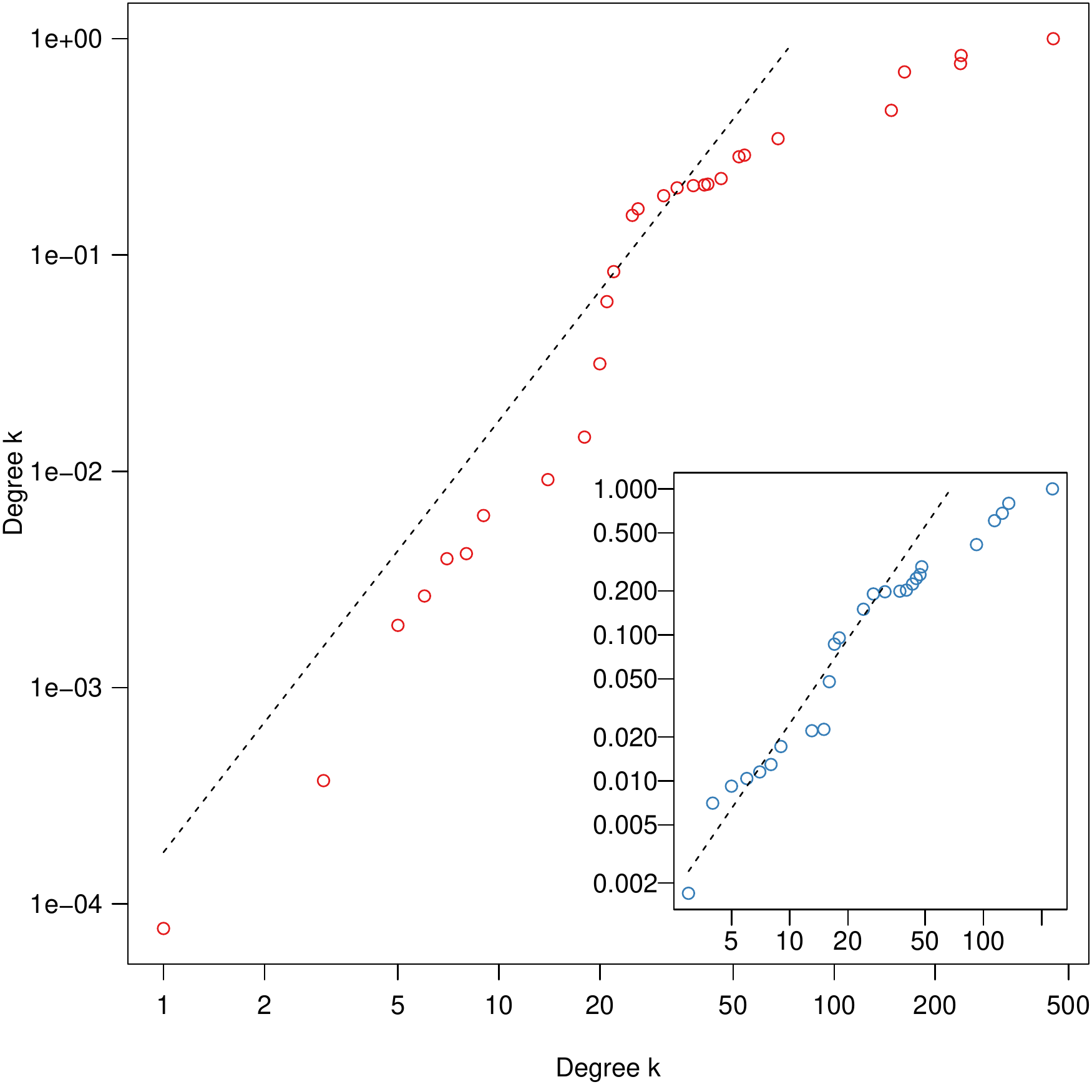}
                \phantomsubcaption\label{fig:DegreePrefAttExt}
            \end{subfigure}~
	        \begin{subfigure}[t]{0.32\textwidth}
                \includegraphics[width=1\textwidth]{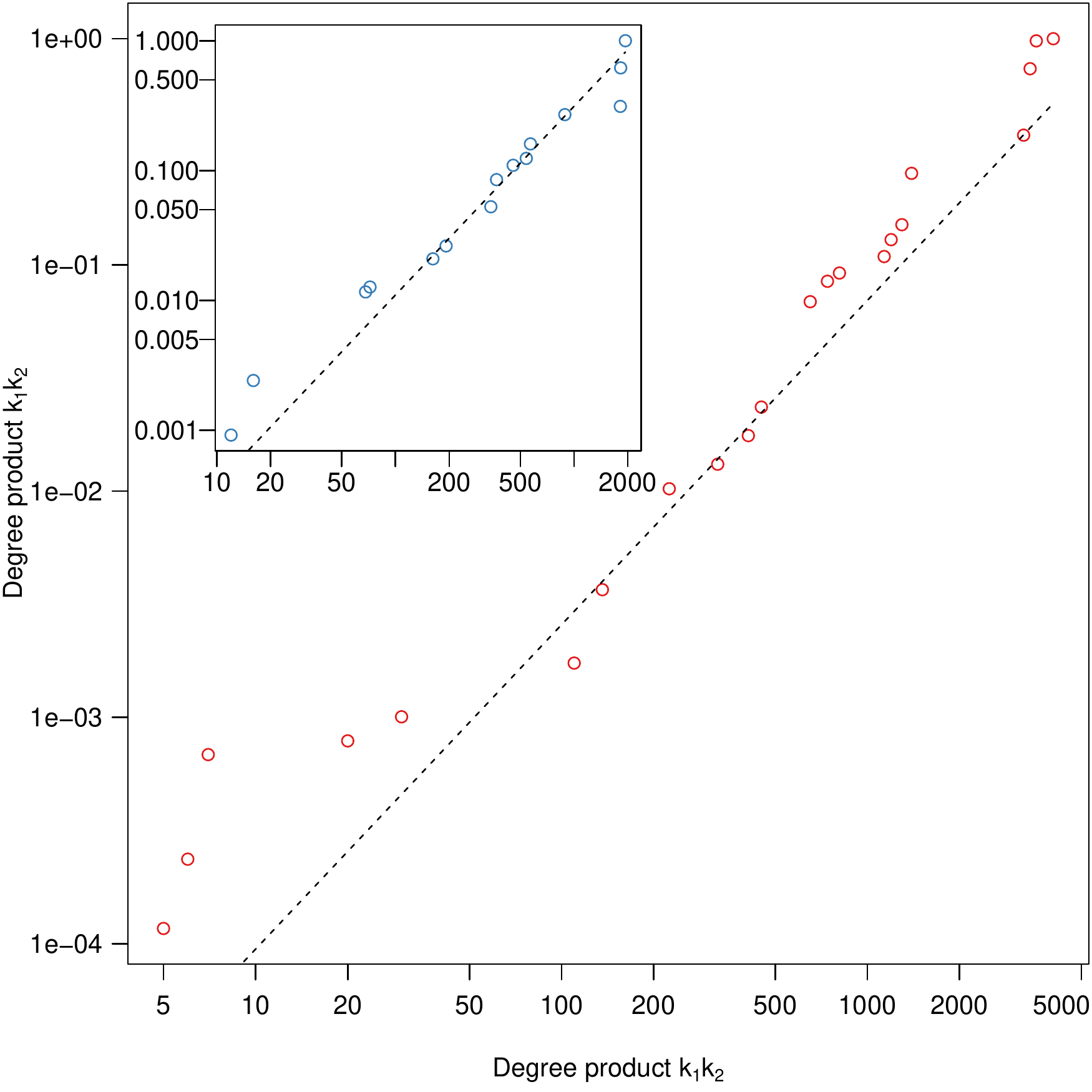}
                \phantomsubcaption\label{fig:DegreePrefAttInt}
            \end{subfigure}
        };
        \node[anchor=west] at (0.00,0.50) {\fontsize{8}{8}\selectfont{}\textbf{a)}};
        \node[anchor=west] at (4.20,0.50) {\fontsize{8}{8}\selectfont{}\textbf{b)}};
        \node[anchor=west] at (8.50,0.50) {\fontsize{8}{8}\selectfont{}\textbf{c)}};
    \end{tikzpicture}
    \vspace{-0.7cm}
    \caption{\color{black!60!blue} (a)~Cumulative preferential attachment of new vertices, as a function of the degree. (b)~Cumulative preferential attachment of new vertices, as a function of the degree. (b)~Cumulative preferential attachment of new internal edges, as a function of the degree product. The dotted lines show the power law best fitting the data. The unfiltered network is represented in red, and the filtered one in blue. In all three plots, the axes use a logarithmic scale. Figure available at \href{https://doi.org/10.5281/zenodo.6573491}{10.5281/zenodo.6573491} under CC-BY license.}
    \label{fig:DegreePrefAtt}
\end{figure*}

A power law-distributed degree is often the result of a \textit{preferential attachment} process, as described by Barabási \& Albert~\cite{Barabasi1999a}. It is possible to assess this property by estimating $\Pi(k)$, the preferential attachment rate of a vertex as a function of its degree $k$. In case of preferential attachment, this function is supposed to be linear in $k$, whereas if there is no preferential attachment, it should be independent of $k$. Following the method used in the literature~\cite{Jeong2003,Barabasi2015}, we do not study directly this function, but rather the \textit{cumulative} preferential attachment rate, noted $\kappa(k)$. Figure~\ref{fig:DegreePrefAttAll} shows an estimation of $\kappa(k)$ for both unfiltered and filtered networks. The left tail of the function is best fit by a power law with exponent $0.98$ for the unfiltered network and $0.72$ for the filtered one. The former displays an almost \textit{linear} preferential attachment, known to produce scale-free networks. The latter exhibits a \textit{sublinear} preferential attachment: this scaling regime is not sufficient to produce a power law degree distribution, which confirms our previous observation. Albeit apparently less frequent, this sublinear regime is also observed in real-world collaboration networks~\cite{Newman2001b,Jeong2003,Barabasi2015}, though.

In the context of a character network, preferential attachment means that new characters tend to interact more with existing characters that possess more acquaintances. In the case of a series like \textit{Thorgal}, the narrative typically focuses on a few main characters, and new characters are generally introduced \textit{because} they interact with them. An absence of preferential attachment would require the introduction of new characters related to minor characters only, which is very uncommon. The fact that preferential attachment is sublinear in the filtered network indicates that this observation is not as strong when discarding the most minor characters. There is no point of comparison for the preferential attachment rate in the character network literature, at least regarding other graphic novels. However, in novels, Waumans \textit{et al}.~\cite{Waumans2015} detect a linear dependence in the case of the \textit{Harry Potter} series. 


Figures~\ref{fig:DegreePrefAttExt} and~\ref{fig:DegreePrefAttInt} shows the external and internal cumulative preferential attachment rates, noted $\kappa(k)$ and $\kappa(k_1 k_2)$, respectively. The former (Figure~\ref{fig:DegreePrefAttExt}) reflects the probability for a new vertex to attach to an existing vertex, depending on its degree. The latter (Figure~\ref{fig:DegreePrefAttInt}) is related to the probability for a new edge to attach to two existing vertices, depending on the product of their degrees. Both functions are supposed to be linear in presence of preferential attachment~\cite{Jeong2003,Barabasi2015}. Compared to Figure~\ref{fig:DegreePrefAttAll}, these representations allow distinguishing between the effects of new edges connecting new vertices to existing ones (so-called \textit{external} edges) vs. new edges connecting two existing vertices (\textit{internal} edges). See~\cite{Jeong2003,Barabasi2015} for more methodological details. 

All functions are well fit by a power law, with approximately the same exponents for the unfiltered and filtered networks. The external rate is related linearly to the degree, with an exponent of $0.99$. The exponent is $0.44$ for the internal rate, which corresponds to a sublinear regime. Note that there are approximately $10$ times as many external than internal edges, on the considered period. Therefore, preferential attachment is stronger for edges connecting a new vertex to an existing one than for edges connecting two existing vertices. This indicates that new characters tend to interact first with important characters rather than minor ones, and then only develop connections with less important characters. 

\subsection{Lattice-Based Model}
\label{sec:AddAnalyLattice}
To obtain a lattice-based graph following a specified degree sequence, we proceed as follows. We start from a regular ring lattice, whose degree corresponds to the minimal value specified in the degree sequence. By construction, at this moment, all vertices have the same degree. 

Second, we randomly pick one of its vertex, which we note $v_1$, to become the largest hub of the final network. Its first left- and right-hand neighbors, spatially speaking, will consequently become the second and third largest hubs. Similarly, its second neighbors will become the fourth and fifth largest hubs, and so on. 

Third, we add edges between $v_1$ and its neighbors, starting from the spatially closest ones, until $v_1$ reaches the targeted degree. This will also increase the degree of the concerned neighbors. We proceed similarly with the second largest hub to be, and go on until all vertices have the desired degree.

\subsection{Centrality}
\label{sec:AddAnalyCentr}

\begin{figure*}[htb!]
    \centering
    \begin{tikzpicture}
        \node[anchor=south west,inner sep=0,align=left] (image) at (0,0) {
	        \begin{subfigure}[t]{0.49\textwidth}
                 \includegraphics[width=1\textwidth]{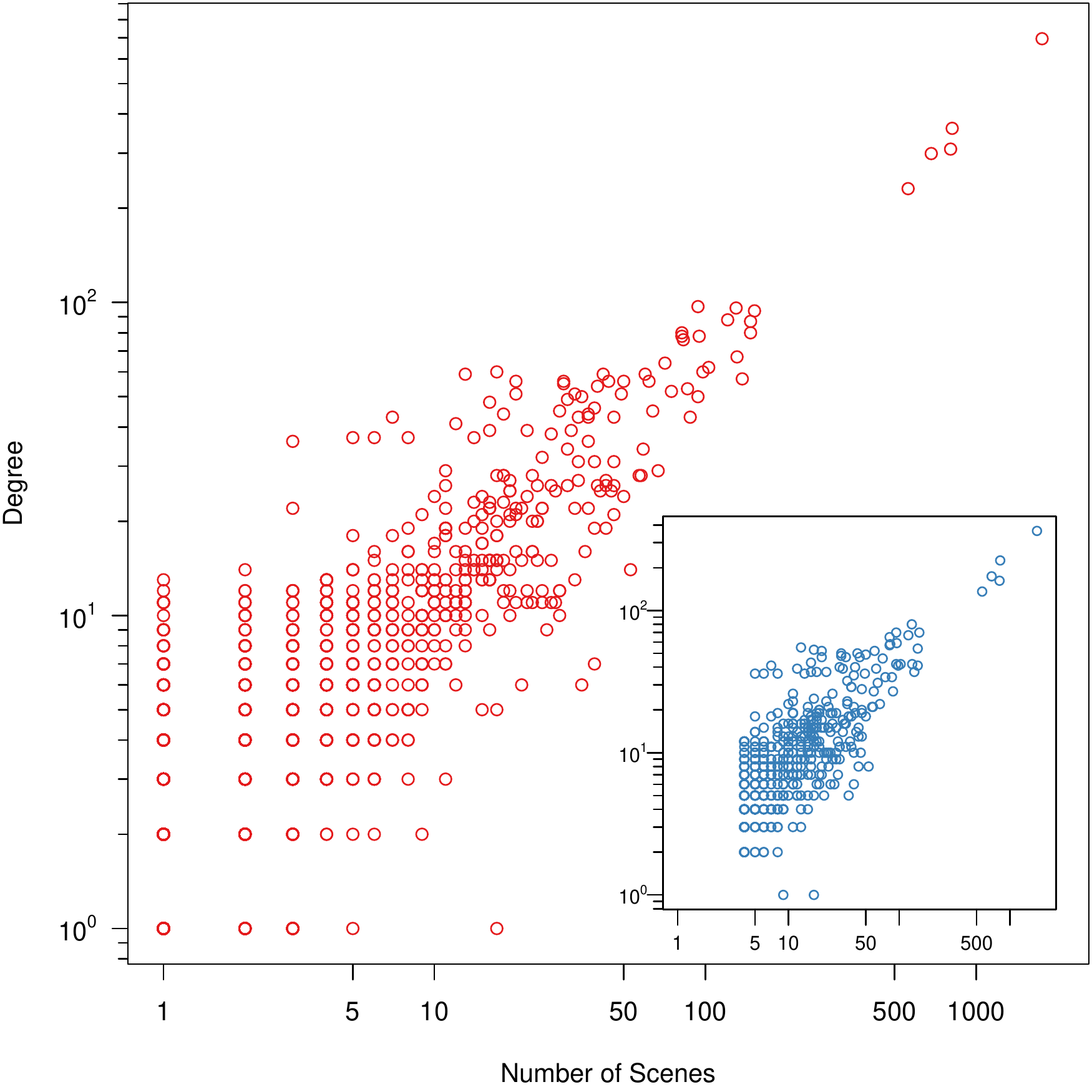}
                \phantomsubcaption\label{fig:DegreeVsScenes}
            \end{subfigure}~
	        \begin{subfigure}[t]{0.49\textwidth}
                \includegraphics[width=1\textwidth]{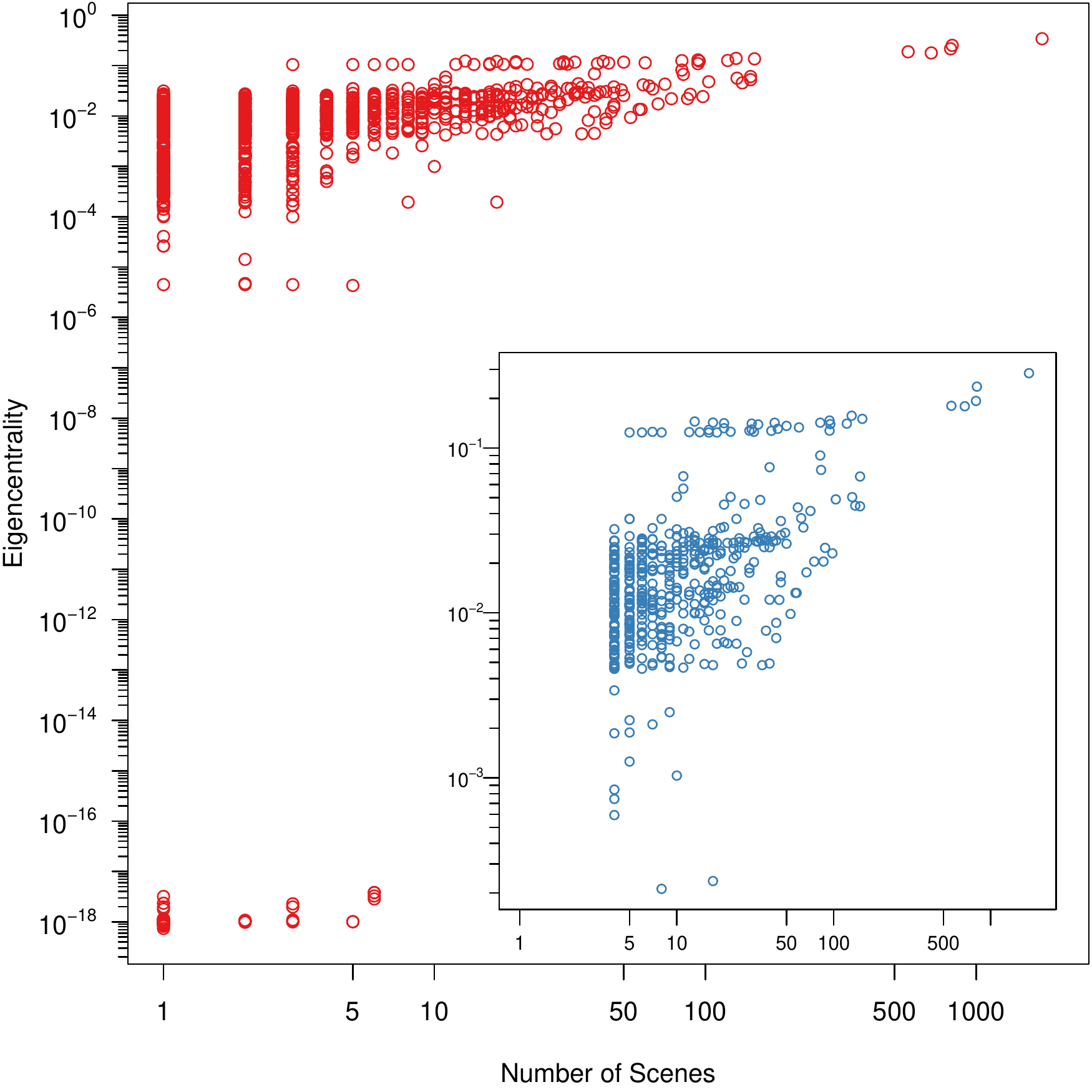}
                \phantomsubcaption\label{fig:EigenVsScenes}
            \end{subfigure}\\[-2mm]
	        \begin{subfigure}[t]{0.49\textwidth}
                 \includegraphics[width=1\textwidth]{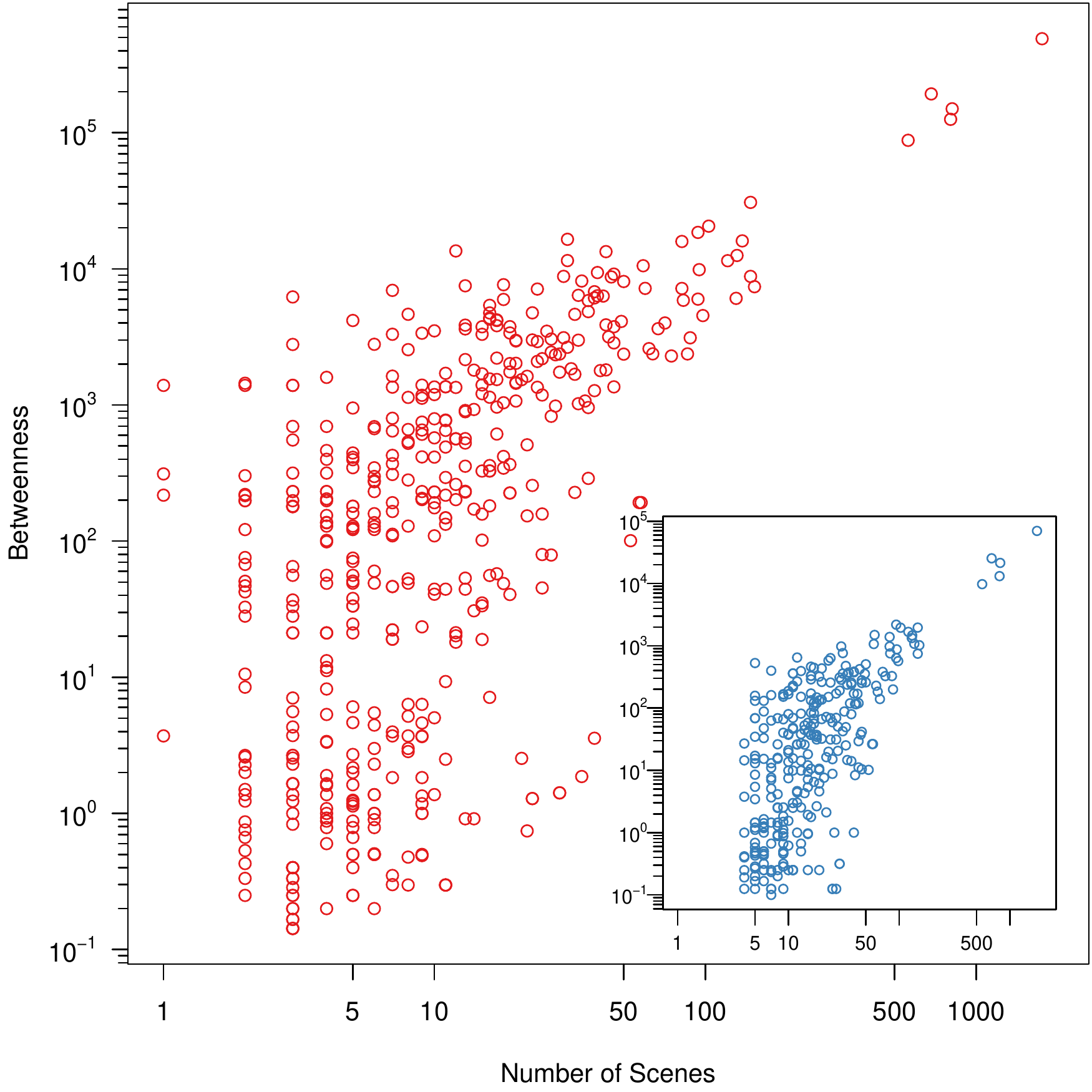}
                \phantomsubcaption\label{fig:BetwVsScenes}
            \end{subfigure}~
	        \begin{subfigure}[t]{0.49\textwidth}
                \includegraphics[width=1\textwidth]{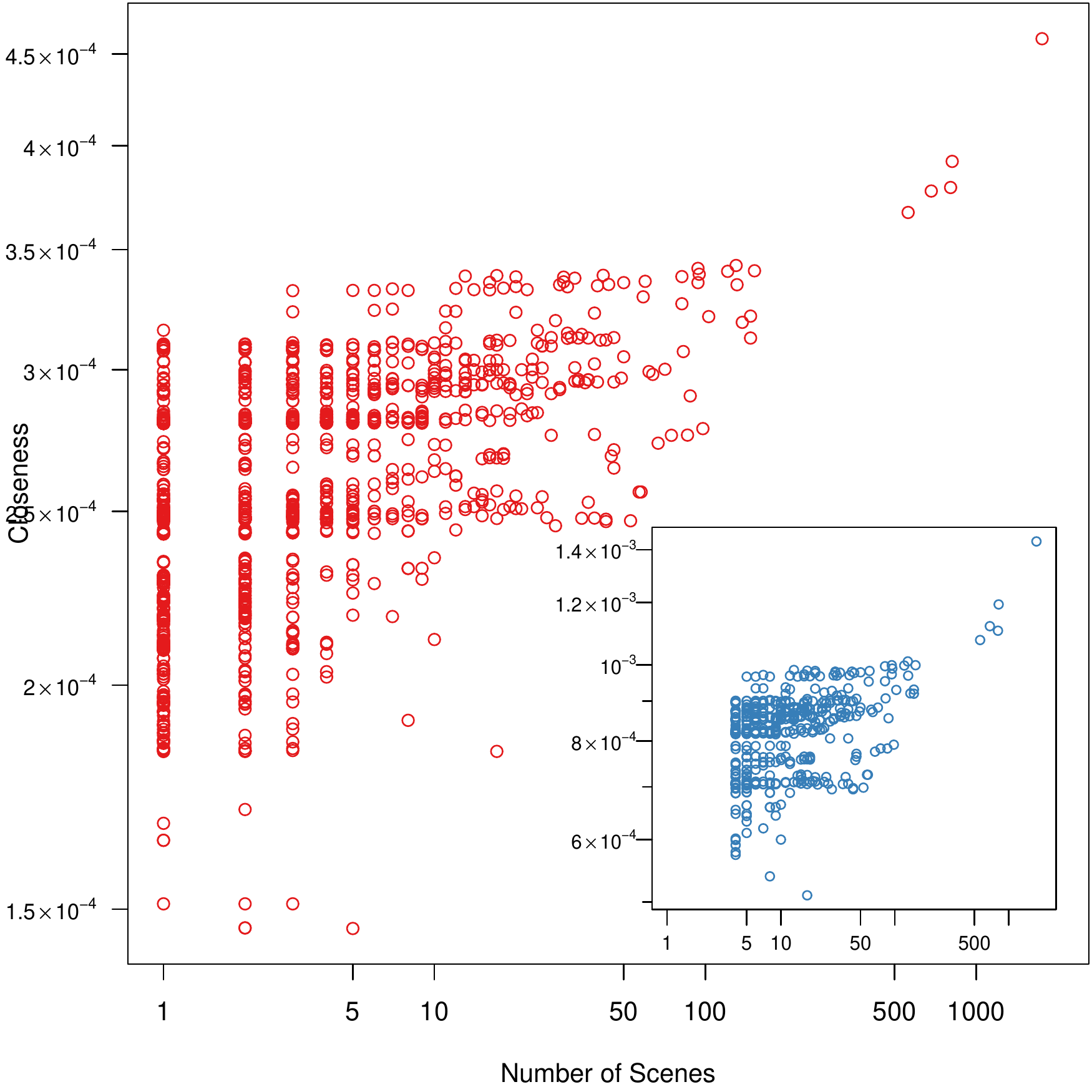}
                \phantomsubcaption\label{fig:CloseVsScenes}
            \end{subfigure}
        };
        \node[anchor=west] at (0.10,7.10) {\fontsize{8}{8}\selectfont{}\textbf{a)}};
        \node[anchor=west] at (6.60,7.10) {\fontsize{8}{8}\selectfont{}\textbf{b)}};
        \node[anchor=west] at (0.10,0.60) {\fontsize{8}{8}\selectfont{}\textbf{c)}};
        \node[anchor=west] at (6.60,0.60) {\fontsize{8}{8}\selectfont{}\textbf{d)}};
    \end{tikzpicture}
    \vspace{-0.8cm}
    \caption{\color{black!60!blue} Centrality of characters as a function of their number of occurrences expressed in scenes, for all characters (red) and after filtering them (blue). (a)~Degree. (b)~Eigenvector centrality. (c)~Betweenness. (d)~Closeness. All axes use a logarithmic scale. Figure available at \href{https://doi.org/10.5281/zenodo.6573491}{10.5281/zenodo.6573491} under CC-BY license.}
    \label{fig:CentrVsScenes}
\end{figure*}

\paragraph{Correlation Study}
Figure~\ref{fig:CentrVsScenes} shows character centrality as a function of their number of occurrences expressed in scenes. Note the logarithmic scales. It appears that centrality is an increasing function of the number of occurrences, whatever the considered centrality measure. Spearman's rank correlation coefficient is relatively high for all four measures vs. scenes. For the unfiltered network, we get $0.75$ (degree), $0.58$ (Eigencentrality), $0.73$ (betweenness), and $0.49$ (closeness). The same observation and similar correlation values hold when considering occurrences expressed in panels, pages and volumes (not shown). The dispersion is much larger for infrequent characters, whereas the $5$ most frequent characters are the most central according to all four considered measures.

The Eigenvector centrality computed for the unfiltered network exhibits a clear separation between the $85$ least central characters, whose centrality score is almost zero, and the rest of the population. They all correspond to filtered characters, which hints at using the Eigenvector centrality as the filtering criterion. However, the first non-minor character (Shazade) is the 113\textsuperscript{th} by order of increasing Eigencentrality, whereas we filtered around $1,000$ characters with the method described in Section~\ref{sec:DataFiltering}. We can conclude that this centrality measure alone is not sufficient to perform the filtering.

\begin{figure*}[htb!]
    \centering
    \begin{tikzpicture}
        \node[anchor=south west,inner sep=0,align=left] (image) at (0,0) {
	        \begin{subfigure}[t]{0.49\textwidth}
                 \includegraphics[width=1\textwidth]{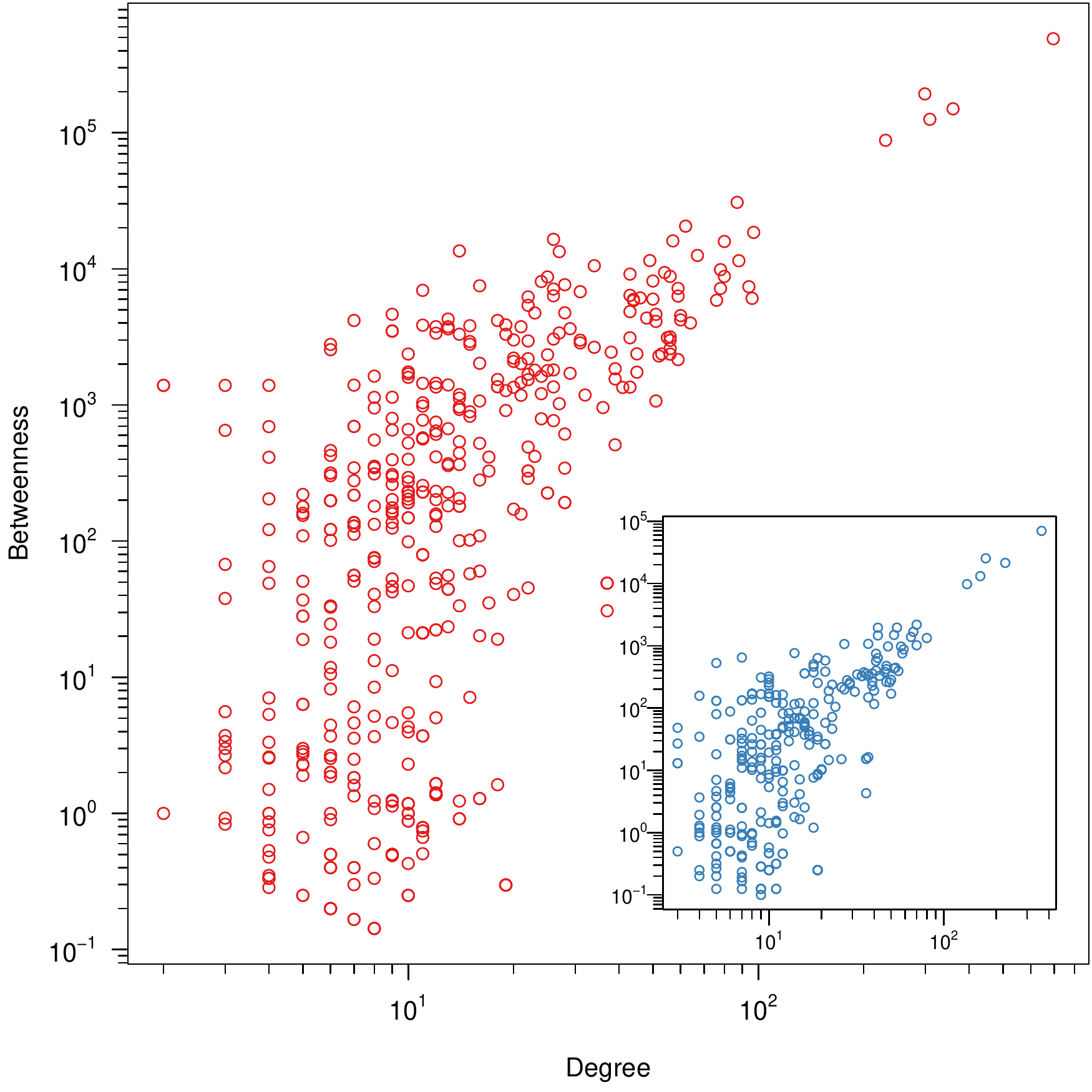}
                \phantomsubcaption\label{fig:DegreeVsBetw}
            \end{subfigure}~
	        \begin{subfigure}[t]{0.49\textwidth}
                \includegraphics[width=1\textwidth]{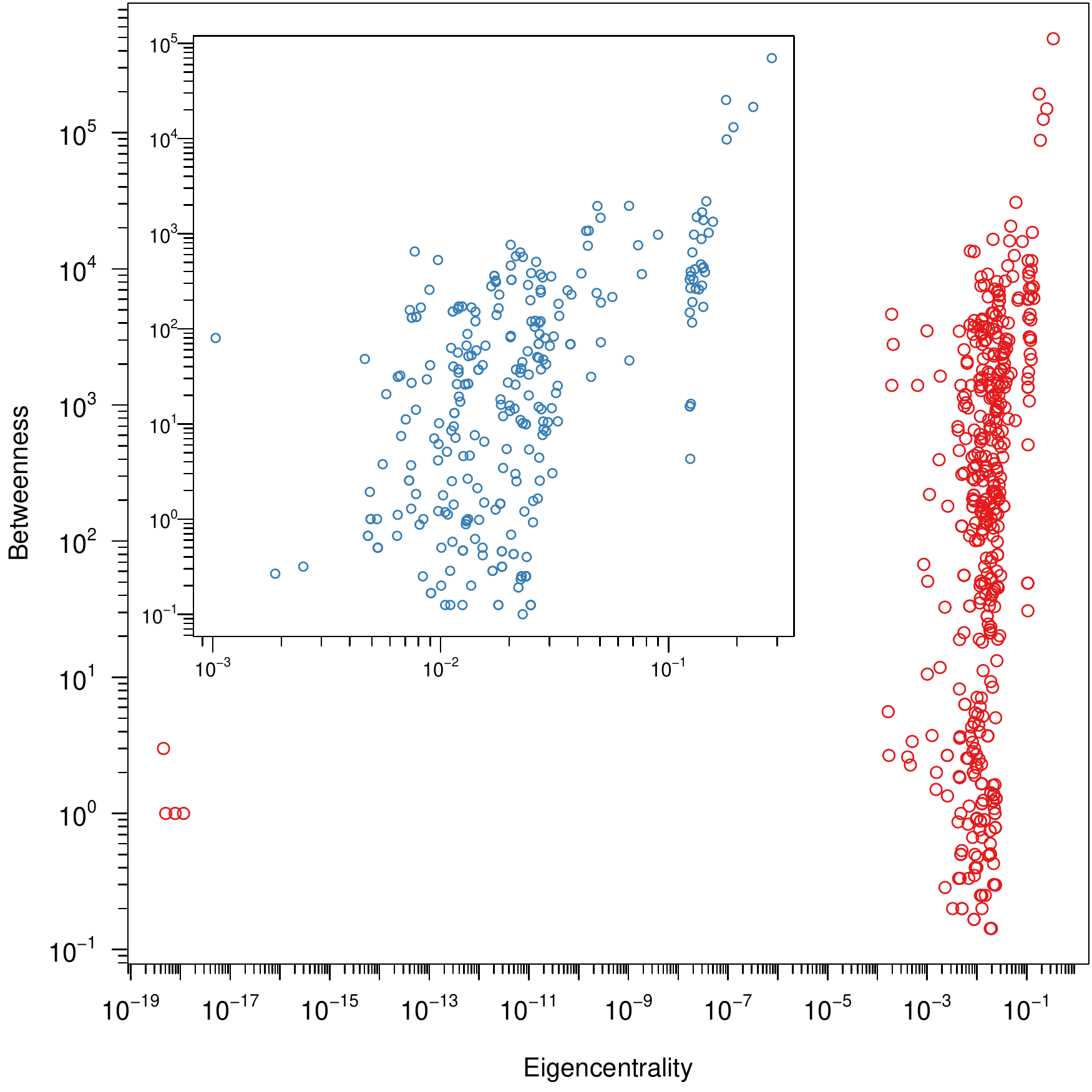}
                \phantomsubcaption\label{fig:EigenVsBetw}
            \end{subfigure}\\[-2mm]
	        \begin{subfigure}[t]{0.49\textwidth}
                 \includegraphics[width=1\textwidth]{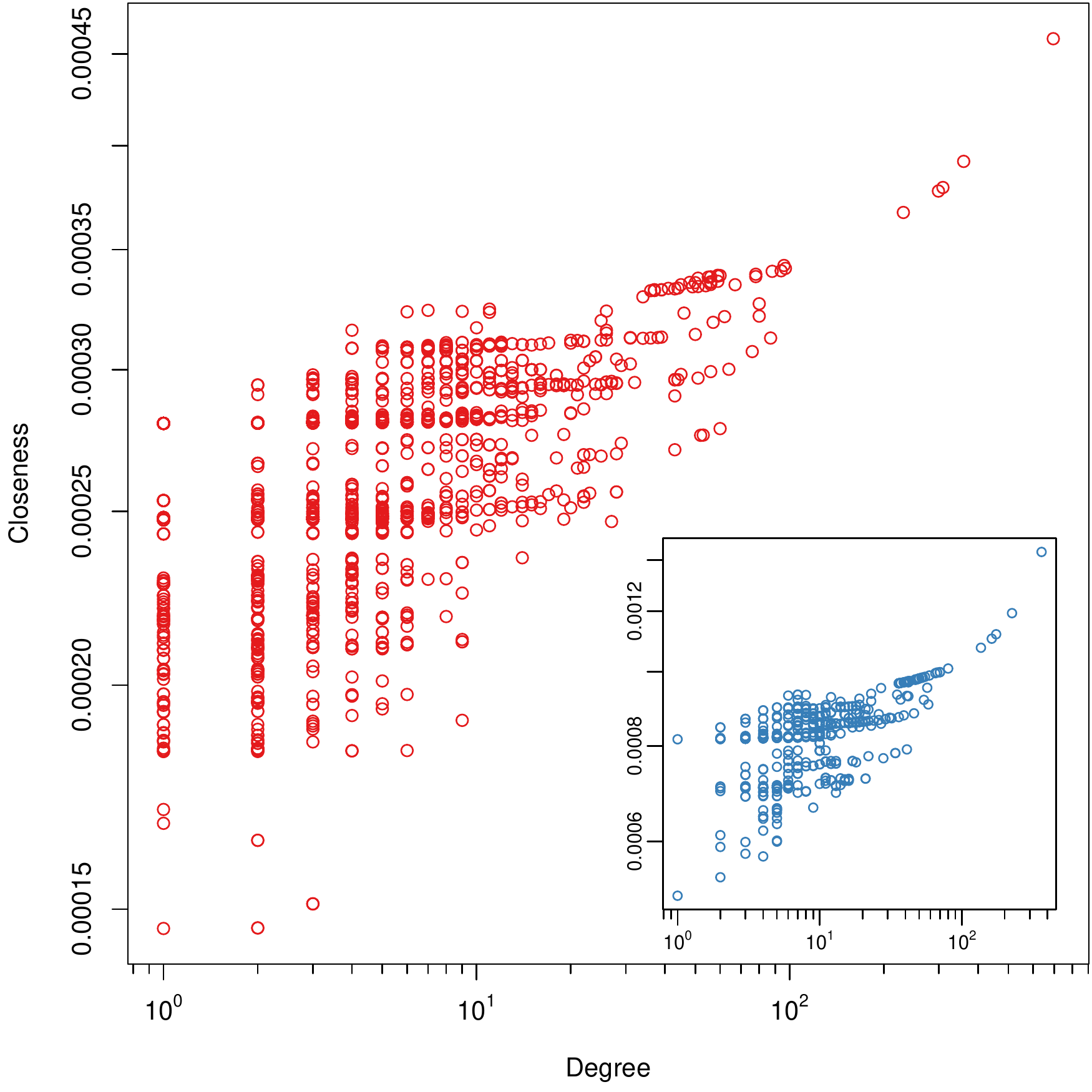}
                \phantomsubcaption\label{fig:DegreeVsClose}
            \end{subfigure}~
	        \begin{subfigure}[t]{0.49\textwidth}
                \includegraphics[width=1\textwidth]{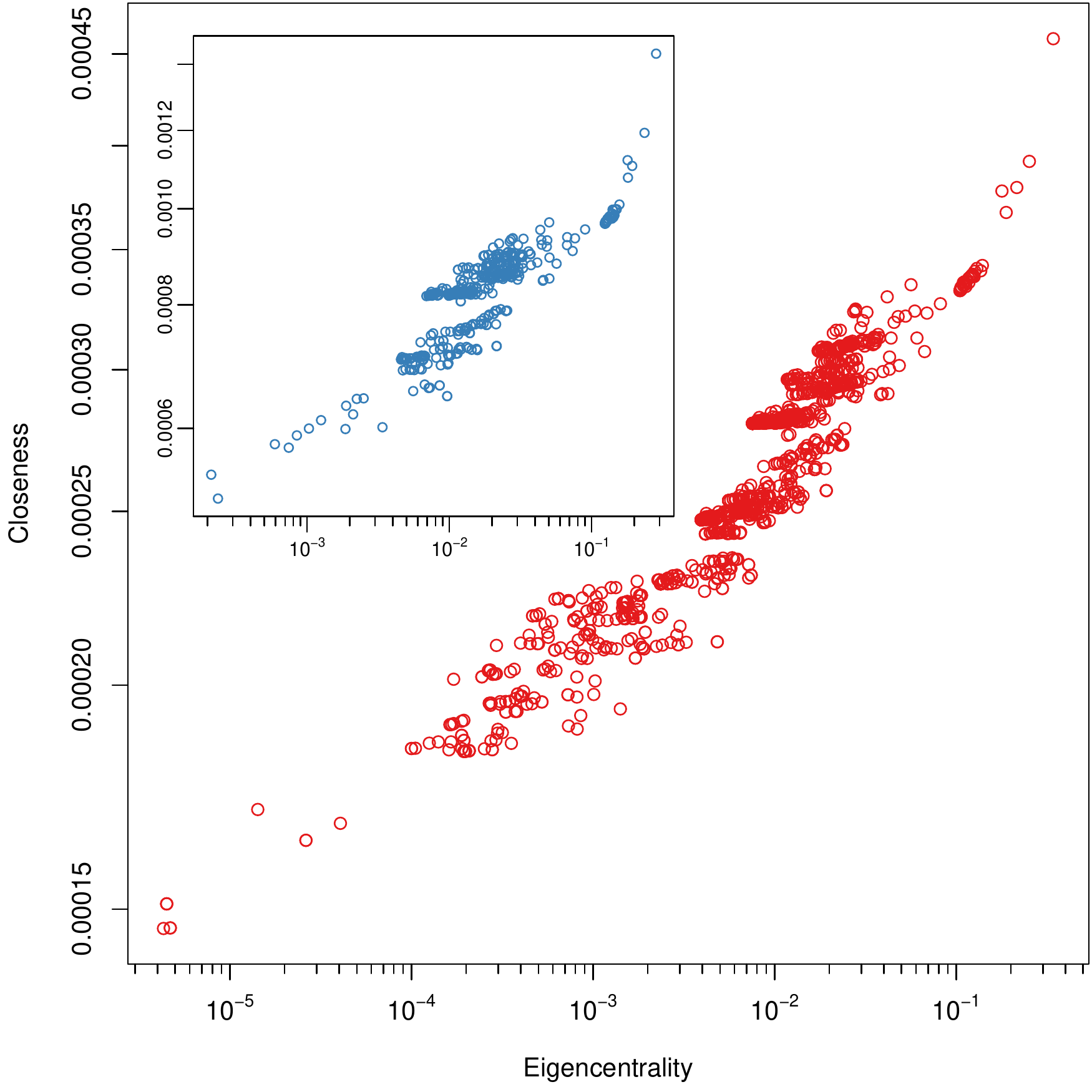}
                \phantomsubcaption\label{fig:EigenVsClose}
            \end{subfigure}\\[-2mm]
	        \begin{subfigure}[t]{0.49\textwidth}
                 \includegraphics[width=1\textwidth]{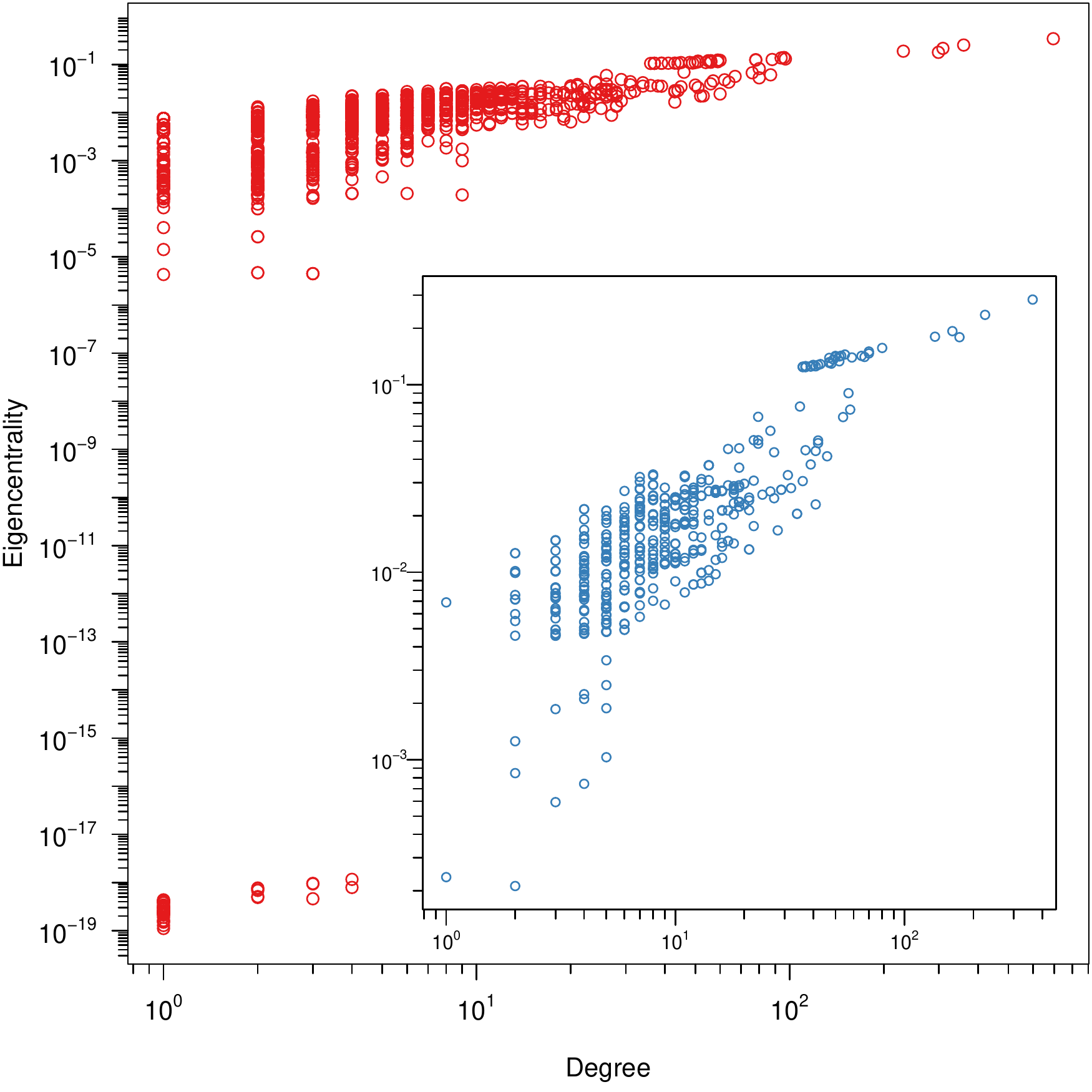}
                \phantomsubcaption\label{fig:DegreeVsEigen}
            \end{subfigure}~
	        \begin{subfigure}[t]{0.49\textwidth}
                \includegraphics[width=1\textwidth]{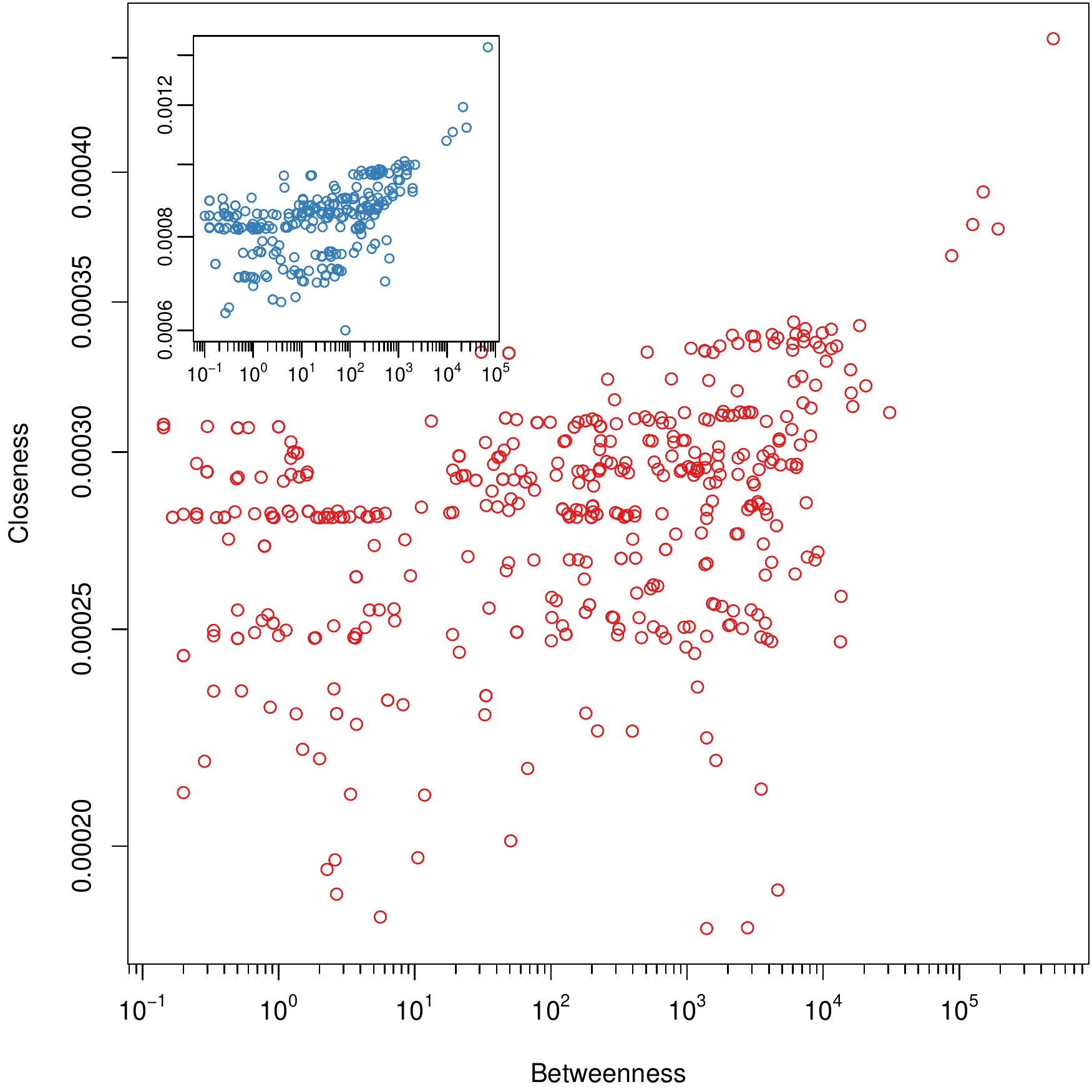}
                \phantomsubcaption\label{fig:BetwVsClose}
            \end{subfigure}
        };
        \node[anchor=west, circle, fill=white, inner sep=0.5pt] at (0.00,13.60) {\fontsize{8}{8}\selectfont{}\textbf{a)}};
        \node[anchor=west, inner sep=0.5pt] at (6.60,13.60) {\fontsize{8}{8}\selectfont{}\textbf{b)}};
        \node[anchor=west, circle, fill=white, inner sep=0.5pt] at (0.00,7.10) {\fontsize{8}{8}\selectfont{}\textbf{c)}};
        \node[anchor=west, inner sep=0.5pt] at (6.60,7.10) {\fontsize{8}{8}\selectfont{}\textbf{d)}};
        \node[anchor=west, circle, fill=white, inner sep=0.5pt] at (0.00,0.60) {\fontsize{8}{8}\selectfont{}\textbf{e)}};
        \node[anchor=west, inner sep=0.5pt] at (6.60,0.60) {\fontsize{8}{8}\selectfont{}\textbf{f)}};
    \end{tikzpicture}
    \vspace{-0.8cm}
    \caption{\color{black!60!blue} Comparison of four centrality measures (degree, Eigenvector centrality, betweenness and closeness), for all characters (red) and after filtering them (blue). All axes use a logarithmic scale. Figure available at \href{https://doi.org/10.5281/zenodo.6573491}{10.5281/zenodo.6573491} under CC-BY license.}
    \label{fig:CentrVsCentr}
\end{figure*}

Figure~\ref{fig:CentrVsCentr} allows assessing the association for each pair of considered centrality measures, in the unfiltered (red) and filtered (blue) networks. Visual assessment reveals that there is some form of non-linear relationship for all of them. Of course, that is a general observation, since some vertices of interest differ in how they are ranked by centrality measures, as discussed in Section~\ref{sec:DescAnalyCentr}.

\FloatBarrier

\paragraph{Robustness to Attacks}
One way to study whether the network structure relies on one or a few important vertices, is to study its robustness to targeted vs. random attacks. A targeted attack consists in removing the most central vertices~\cite{MacCarron2013} or edges~\cite{Gleiser2007} of the network, by opposition to randomly picking them. One assesses the network robustness by monitoring how the size of its largest component evolves as a function of the number of removed vertices or edges. In~\cite{MacCarron2012}, Mac Carron \& Kenna apply this approach while studying character networks extracted from myths, and comparing them to real-world and fictional social networks. The latter type corresponds to a selection of novels and plays, and also includes the \textit{Marvel} dataset of Alberich \textit{et al}.~\cite{Alberich2002}. Mac Carron \& Kenna identify robustness to targeted attacks as a sign of the network artificiality, in the sense that its structure relies too much on a few important characters to be realistic. All three considered types of social networks tend to be robust to random attacks, however real-world ones are vulnerable to targeted attacks whereas fictional ones are robust to these. The mythological networks studied in \cite{MacCarron2012} tend to be on the realistic side.

We adopt the same method to study \textit{Thorgal}. Both the unfiltered and filtered networks are robust to random attacks: removing 5\% of the vertices results in giant components containing 89\% and 100\% of the remaining vertices, respectively. On the contrary, they are fragile to targeted attacks, for all four considered centrality measures. The effect is noticeably stronger for the unfiltered network, which can be explained by its lower density. When considering degree and Eigencentrality, the largest component retains half the vertices for the unfiltered network, and three quarters for the filtered one. When considering betweenness and closeness, we get 5\% and 50\% of the vertices remaining in the unfiltered and filtered networks, respectively. We can conclude that the \textit{Thorgal} exhibits the characteristics of real-world social networks regarding these attacks.

\subsection{Graph Typology}
\label{sec:AddAnalyNetTypes}
As mentioned in Section~\ref{sec:RelatedWork}, when discussing the structure of the narratives constituting their science-fiction corpus, Rochat \& Triclot~\cite{Rochat2017} identify four main types of networks based on their overall structure. This typology is based on the relative position and interconnection of the main vs. secondary characters. 

\paragraph{Typology}
First, the \textit{kernel} network is built around a dense group of characters, dominated by a protagonist, and surrounded by some loosely connected periphery. In this type of story, the protagonist is typically assisted by some sidekicks, that follow him or her everywhere. The periphery can either be single characters or social groups, that are connected to the central group but disconnected from each other. 

Second, the \textit{unicentric} network can be seen as an extreme version of the previous type, in which the kernel is reduced to a single vertex constituting its clear center. It represents the story of an isolated character facing separated social groups. The protagonist may have sidekicks, but only punctually. 

Third, the \textit{polycentric} network is also built around an interconnected group of vertices, but each of these central characters acts as a gateway, providing exclusive access to one or several social groups. Such structure typically appears in case of \textit{hypodiegesis}, i.e. when a protagonist tells his story, which involves a separate social group, to some other main character; or when the novel is a \textit{fix-up} (a rework of some originally separate short stories). 

Fourth, the \textit{acentric} network has no recognizable center at all. This is caused either by a large number of characters with well distributed interactions, or on the contrary when the focus is put on a small number of characters, without any periphery. The former case is typical of serial, long-term stories, that have the time to develop the relationships between secondary characters. The latter is characteristic of behind-closed-doors narratives, in which all characters are important because they are few.

\begin{figure*}[htb!]
    \centering
    \begin{tikzpicture}
        \node[anchor=south west,inner sep=0,align=left] (image) at (0,0) {
	        \begin{subfigure}[t]{0.49\textwidth}
                 \includegraphics[height=5.6cm]{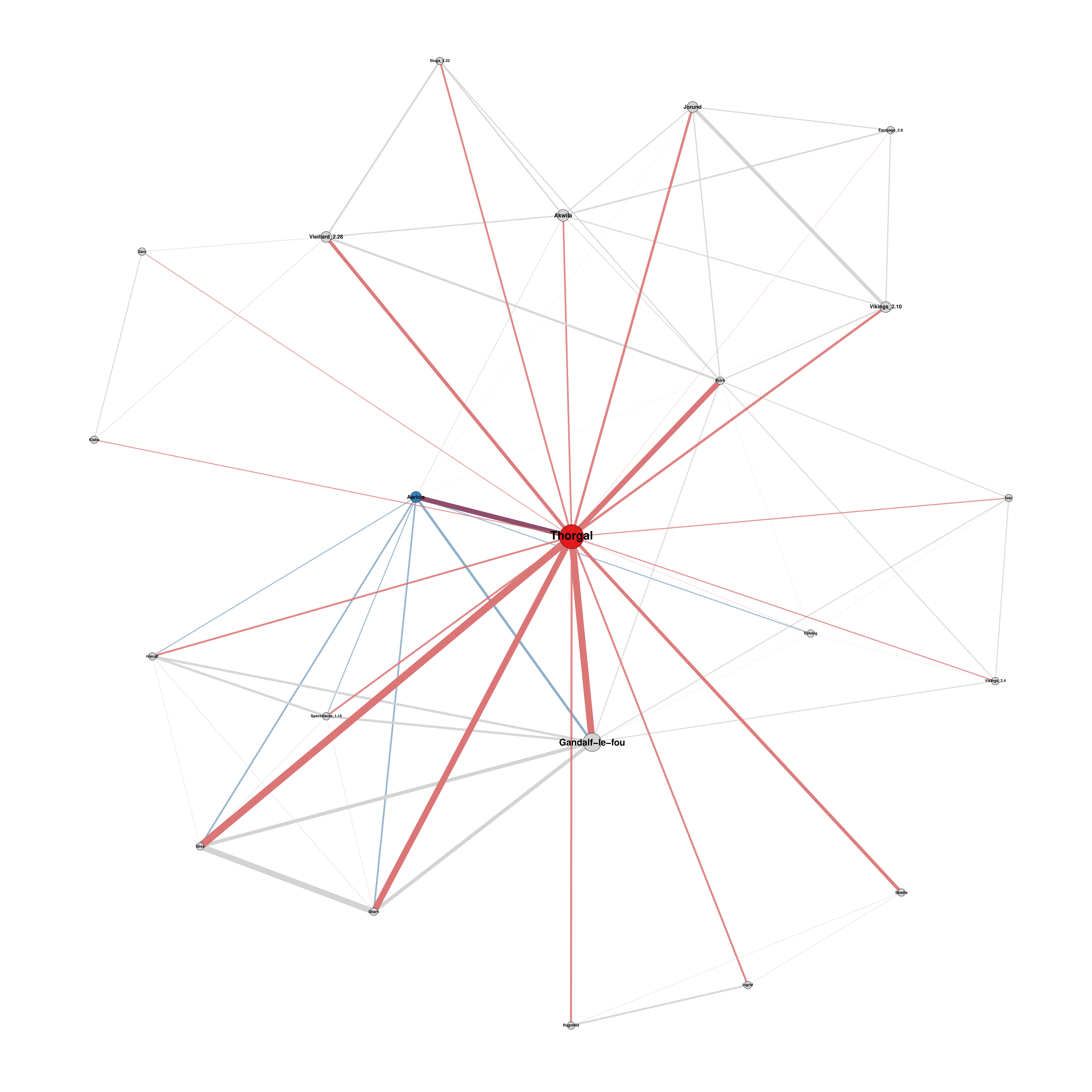}
                \phantomsubcaption\label{fig:NetTypeExArc1}
            \end{subfigure}~~
	        \begin{subfigure}[t]{0.49\textwidth}
                \includegraphics[height=5.6cm]{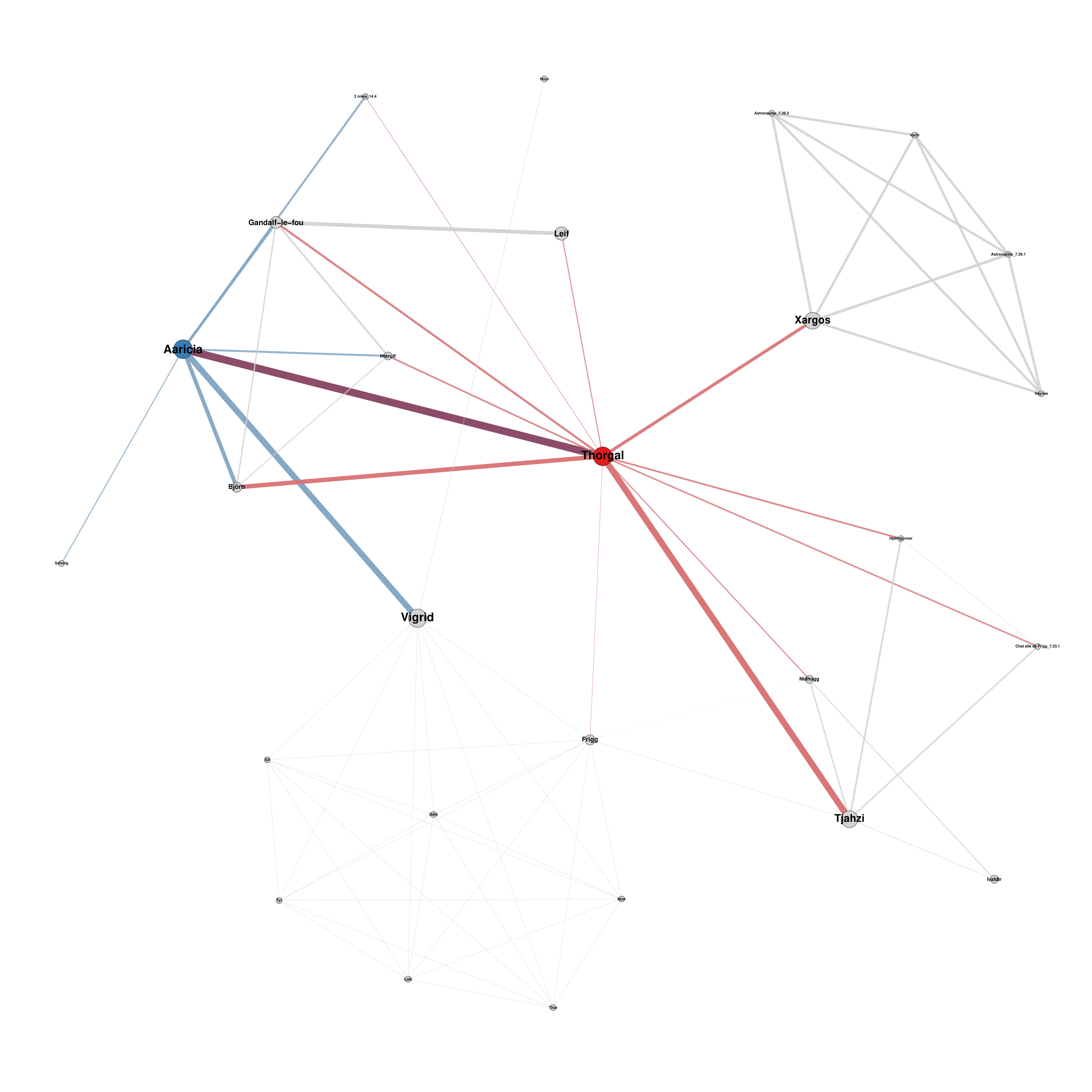}
                \phantomsubcaption\label{fig:NetTypeExArc4}
            \end{subfigure}\\[-4mm]
	        \begin{subfigure}[t]{0.49\textwidth}
                 \includegraphics[height=5.6cm]{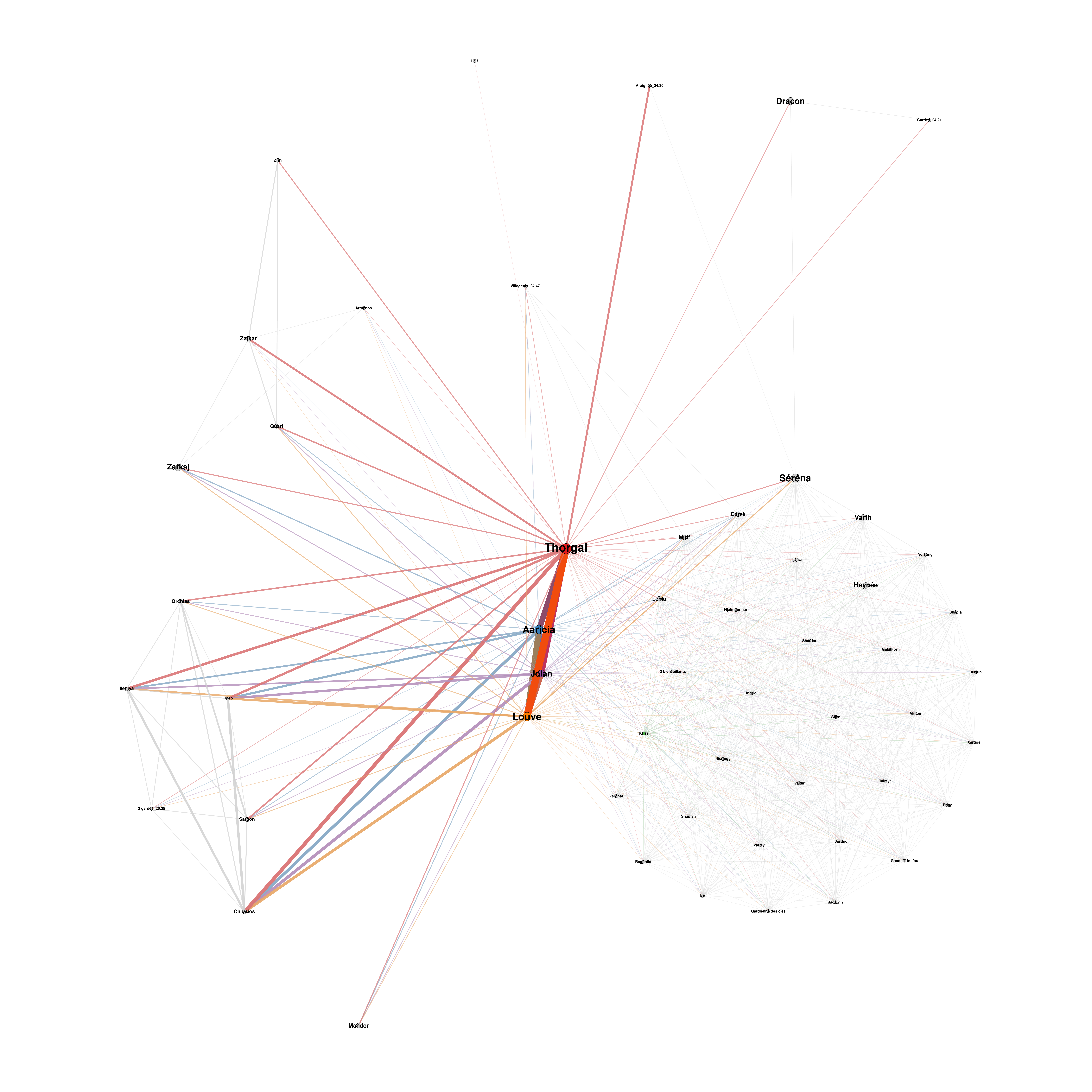}
                \phantomsubcaption\label{fig:NetTypeExArc9}
            \end{subfigure}~~
	        \begin{subfigure}[t]{0.49\textwidth}
                \includegraphics[height=5.6cm]{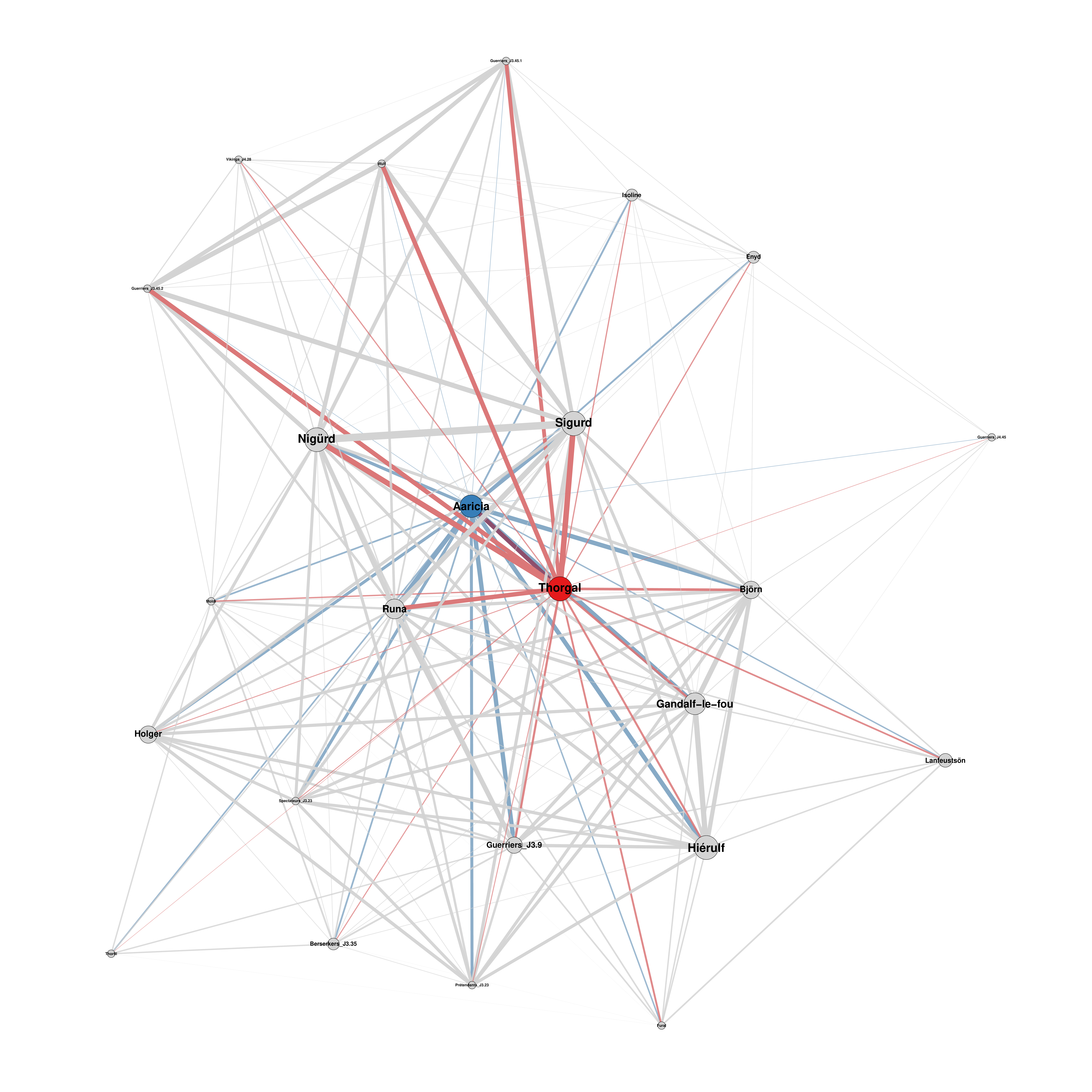}
                \phantomsubcaption\label{fig:NetTypeExArc18}
            \end{subfigure}
        };
        \node[anchor=west, circle, fill=white, inner sep=0.5pt] at (0.08,5.80) {\fontsize{8}{8}\selectfont{}\textbf{a)}};
        \node[anchor=west, circle, fill=white, inner sep=0.5pt] at (6.49,5.80) {\fontsize{8}{8}\selectfont{}\textbf{b)}};
        \node[anchor=west, circle, fill=white, inner sep=0.5pt] at (0.08,0.50) {\fontsize{8}{8}\selectfont{}\textbf{c)}};
        \node[anchor=west, circle, fill=white, inner sep=0.5pt] at (6.49,0.50) {\fontsize{8}{8}\selectfont{}\textbf{d)}};
    \end{tikzpicture}
    \vspace{-0.6cm}
    \caption{\color{black!60!blue} Selection of arc networks illustrating Rochat \& Triclot's typology. Like in Figure~\ref{fig:FilteredNet}, the five most frequent have a specific color; and vertex size and edge width are functions of betweenness and co-occurrences, respectively. The layout is different in each plot, to match the graph structure. (a)~Arc \#1 is \textit{unicentric}, as it introduces the main character, Thorgal. (b)~Arc \#4 is polycentric, because it is a collection of short stories. (c)~Arc \#9 exhibits a kernel constituted of Thorgal's close family. (d)~Arc \#18 is \textit{acentric} due to the uniform distribution of edges. Figure available at \href{https://doi.org/10.5281/zenodo.6573491}{10.5281/zenodo.6573491} under CC-BY license.}
    \label{fig:NetTypeEx}
\end{figure*}

\paragraph{Overall \textit{Thorgal} Network}
When defining their typology, Rochat \& Triclot experiment with various edge weight thresholds to remove minor characters~\cite{Rochat2017}, therefore when analyzing \textit{Thorgal}, we focus on our filtered network. As illustrated by Figure~\ref{fig:FilteredNet}, the filtered network extracted from the full series combines aspects of the \textit{kernel} and \textit{polycentric} types: the colored vertices corresponding to the five main characters constitute a kernel, but some of them are connected in an exclusive way to whole peripheral parts of the network. By comparison, Rochat \& Triclot describe the network of the \textit{Metabarons} graphic novel as acentric, and explain this by its serial nature. But \textit{Thorgal} is also a serial narrative, and furthermore it is much longer than \textit{Metabarons} ($63$ vs. $8$ volumes). The fact that it has a clear center nonetheless can be explained by the spatial (and therefore social) separation between the groups met by the main characters, who go all over the world during their many travels. 

\paragraph{Arc-Focused \textit{Thorgal} Networks}
In addition, \textit{Thorgal} can be broken down into $23$ narrative arcs, whose respective networks include instances of all four of Rochat \& Triclot's classes: kernel (3/23), unicentric (4), polycentric (13) and acentric (3). Table~\ref{tab:NarrArcs} lists the type of each arc, and Figure~\ref{fig:NetTypeEx} gives 4 examples. 

Arc \#1 (Figure~\ref{fig:NetTypeExArc1}) introduces the protagonist, Thorgal, and is completely built around him, constituting a typical case of unicentric network. Incidentally, the same observation holds for the first volumes of the spinoffs: Arc \#12 (Kriss of Valnor) and~\#14 (Louve). 

Arc \#4 (Figure~\ref{fig:NetTypeExArc4}) is a collection of short stories, and also has a \textit{story within a story} situation (a.k.a. \textit{hypodiegesis}), resulting in a polycentric structure. 

Arc \#5 takes place on a small island; it involves a handful of characters, and illustrates the behind-closed-doors acentric narrative. Arc \#18 (Figure~\ref{fig:NetTypeExArc18}) is also acentric, but it corresponds to the case where many characters are interconnected. 

Arc \#9 (Figure~\ref{fig:NetTypeExArc9}) exhibits a kernel structure, as it follows the story of Thorgal's family, whose members stay together (for once). 

Overall, the large prevalence of the polycentric structure reflects the fact that a majority of volumes tell several parallel stories at once, alternatively focusing on several main characters.

\section{Research Questions}
\label{sec:AddRq}
This Section treats an additional research question (Section~\ref{sec:AddRqSubstitute}), and provides extra figures regarding centrality distribution depending on character sex (Section~\ref{sec:AddRqSex}.

\subsection{Substitution of the Protagonist}
\label{sec:AddRqSubstitute}

\begin{figure*}[htb!]
    \centering
    \includegraphics[width=1\linewidth]{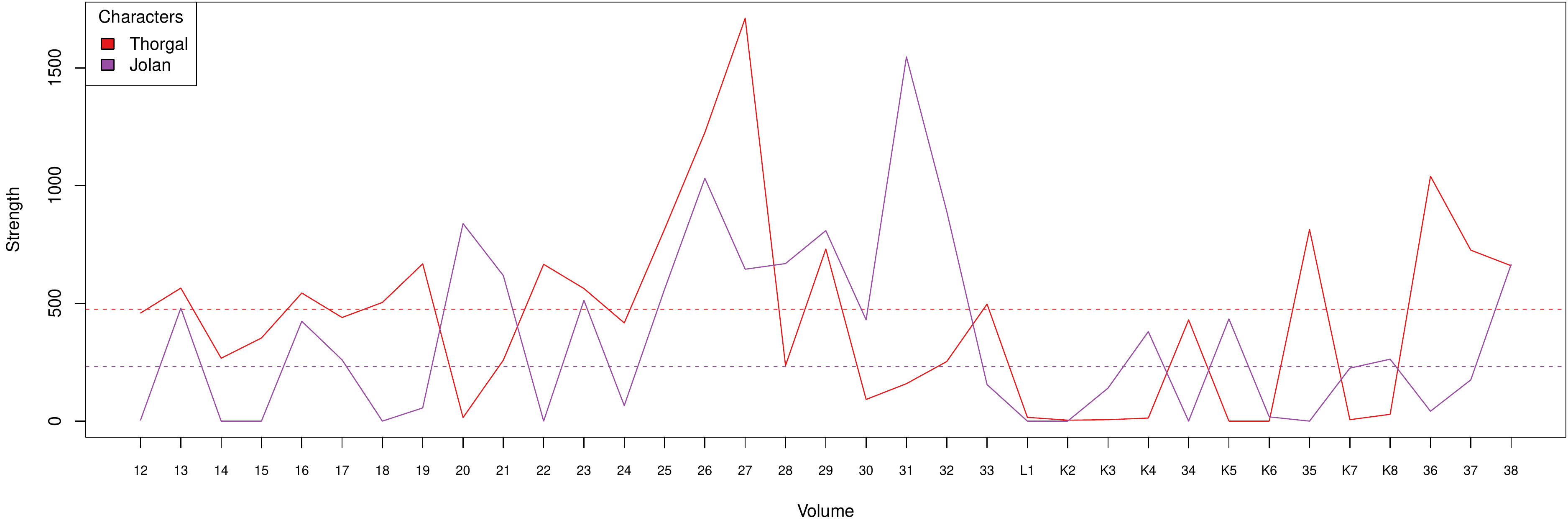}
    \caption{\color{black!60!blue} Evolution of the strength of Thorgal (red) and Jolan (purple) over the volumes. The dotted lines represent their respective average strength over the whole series, as a reference. Figure available at \href{https://doi.org/10.5281/zenodo.6573491}{10.5281/zenodo.6573491} under CC-BY license.}
    \label{fig:UnfilteredStrengthEvol}
\end{figure*}

As noted by Desfontaine \textit{et al}.~\cite{Desfontaine2018}, Thorgal and his elder son Jolan start sharing the stage in the volumes of the main series published right before their article. This corresponds to a time when Jolan reaches his late teens in the series. Figure~\ref{fig:UnfilteredStrengthEvol} shows the evolution of the strength of both characters, as a function of the volumes. Only the volumes in which at least one of them appears are represented. It appears indeed that Jolan overtakes Thorgal from volumes 28 to 32 in the main series. This event corresponds to an editorial choice from the authors, which became evident to the readers with volume 30, titled \textit{I, Jolan}. 

In the preceding volume (\textit{The Sacrifice}), Thorgal is critically ill. Jolan must accept the help of a mage, Manthor, to save him. In exchange, he swears to serve him for several years. This narrative trick is the occasion to give some independence to Jolan, which starts living his own adventures apart from his family, together with four other teenagers. This change matches the retirement of Jean Van Hamme, the historic writer of the series, and his replacement by new writer Yves Sente. It also corresponds to an effort by the publisher to rejuvenate the readership of the series. Indeed, as mentioned before, characters get older in \textit{Thorgal}, including the protagonist. Providing a teenage protagonist to the main series is a way to attract younger readers. This is also one of the motivations for the introduction of the \textit{Louve} spinoff, which tells the story of Thorgal's kid daughter. 

Interestingly, this situation only lasts a few volumes. After volume 33, Thorgal starts dominating again. This is due to a period of instability in the creative team, which saw several changes of writer in a short period of time. The idea of replacing Thorgal by a younger character as the new main protagonist was finally overturned, as shown in Figure~\ref{fig:UnfilteredStrengthEvol}.


\subsection{Centrality vs. Sex}
\label{sec:AddRqSex}
\begin{figure*}[htb!]
    \centering
    \begin{tikzpicture}
        \node[anchor=south west,inner sep=0,align=left] (image) at (0,0) {
	        \begin{subfigure}[t]{0.49\textwidth}
                 \includegraphics[width=1\textwidth]{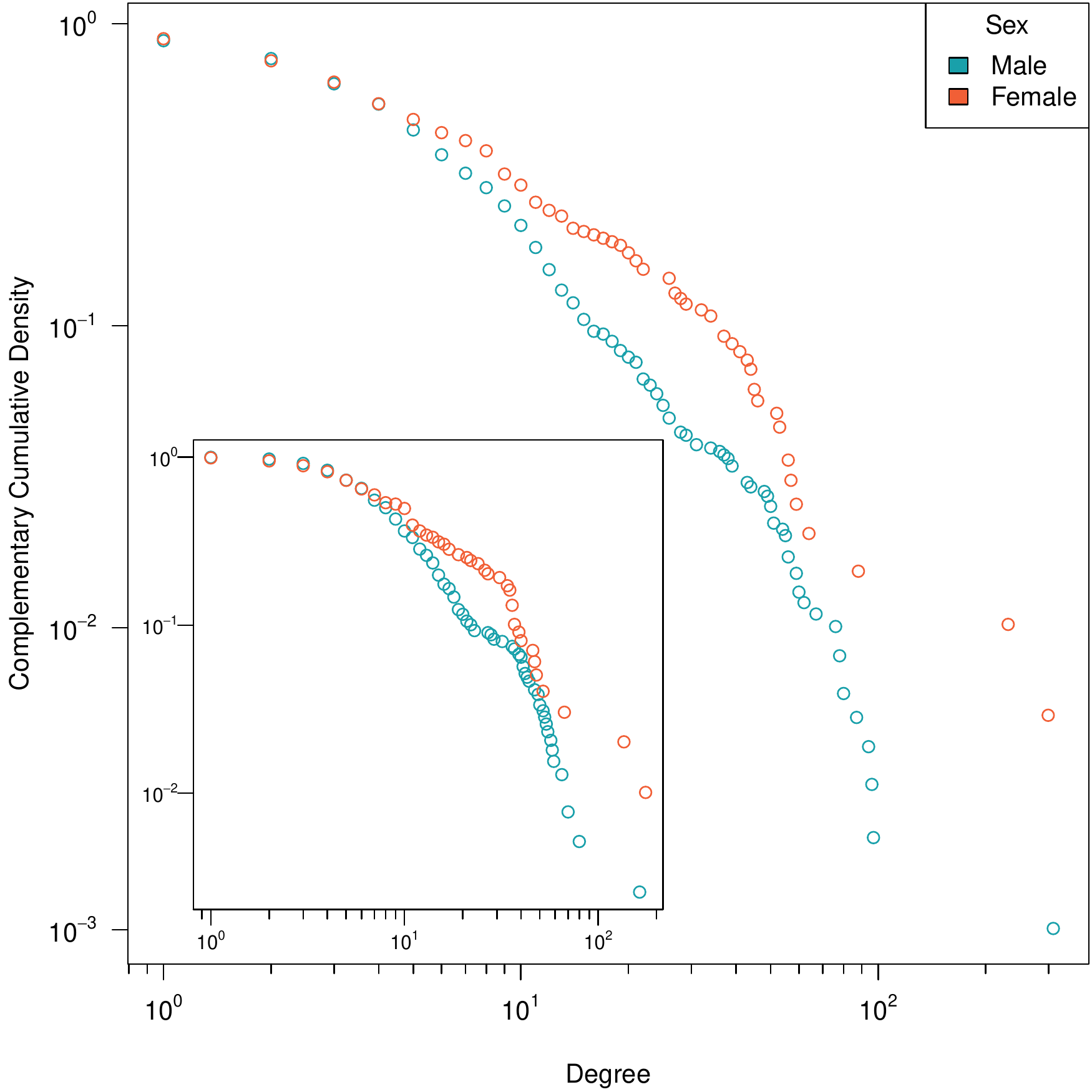}
                \phantomsubcaption\label{fig:DegreeDistribSex}
            \end{subfigure}~
	        \begin{subfigure}[t]{0.49\textwidth}
                \includegraphics[width=1\textwidth]{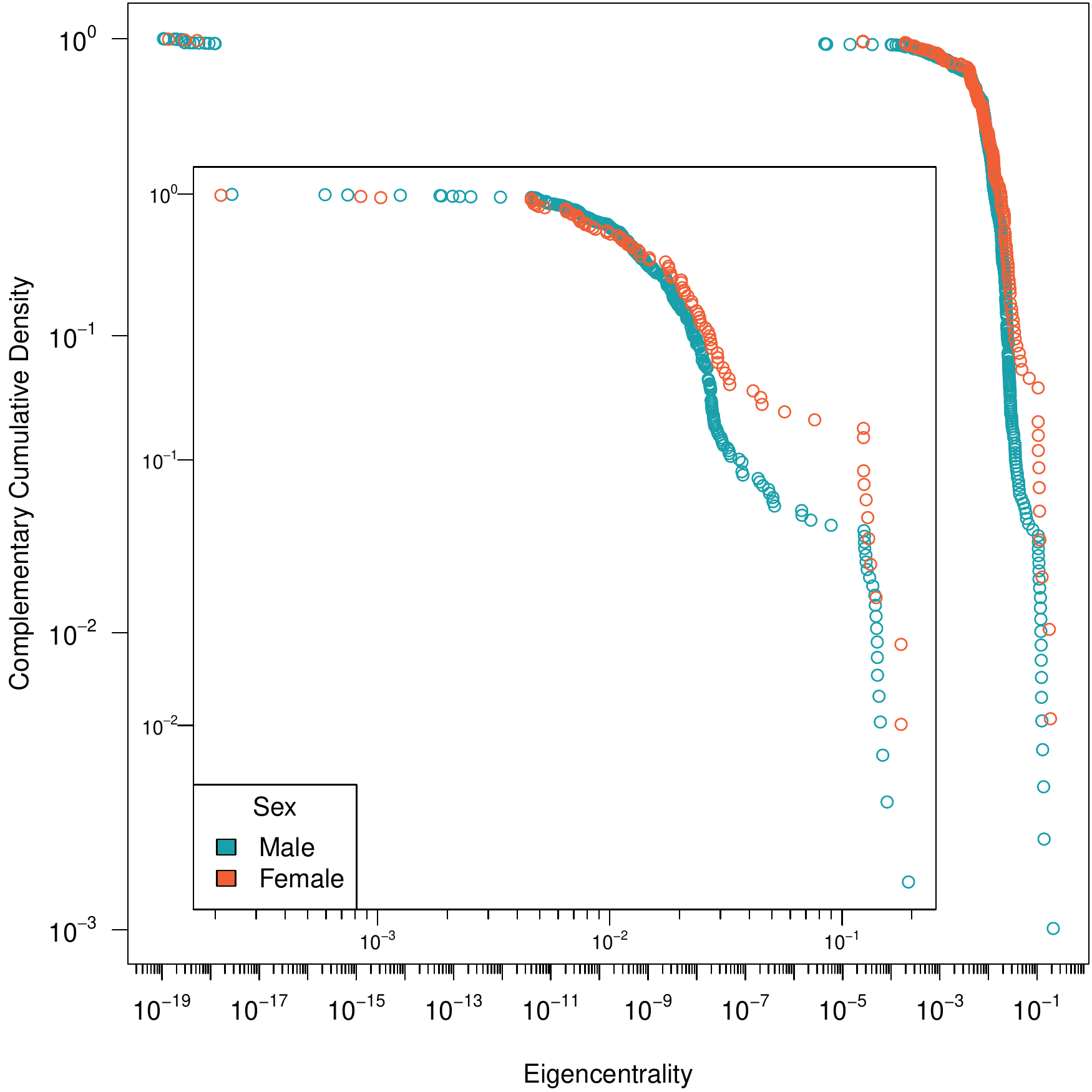}
                \phantomsubcaption\label{fig:EigenDistribSex}
            \end{subfigure}\\[-2mm]
	        \begin{subfigure}[t]{0.49\textwidth}
                 \includegraphics[width=1\textwidth]{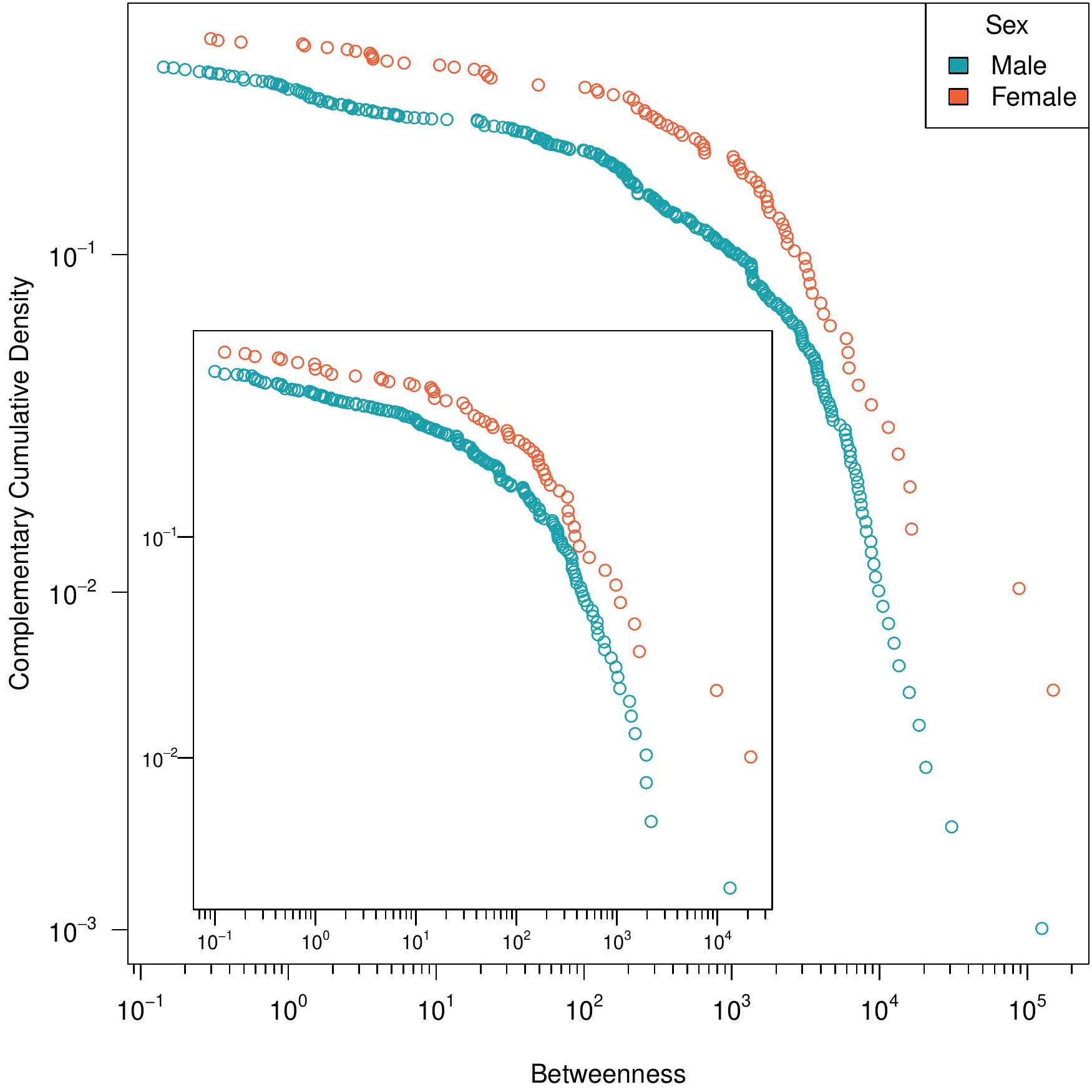}
                \phantomsubcaption\label{fig:BetwDistribSex}
            \end{subfigure}~
	        \begin{subfigure}[t]{0.49\textwidth}
                \includegraphics[width=1\textwidth]{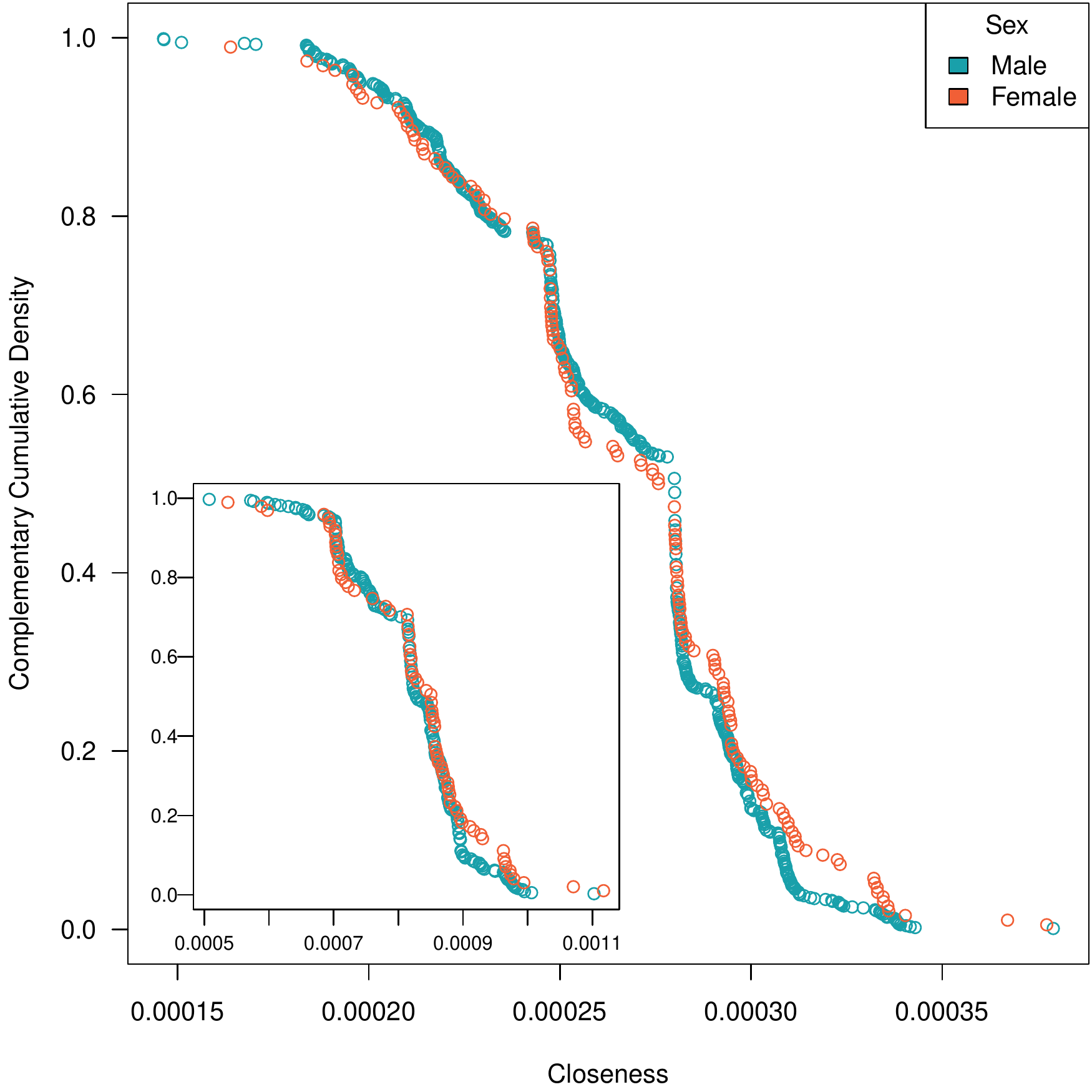}
                \phantomsubcaption\label{fig:CloseDistribSex}
            \end{subfigure}
        };
        \node[anchor=west] at (0.10,7.10) {\fontsize{8}{8}\selectfont{}\textbf{a)}};
        \node[anchor=west] at (6.60,7.10) {\fontsize{8}{8}\selectfont{}\textbf{b)}};
        \node[anchor=west] at (0.10,0.60) {\fontsize{8}{8}\selectfont{}\textbf{c)}};
        \node[anchor=west] at (6.60,0.60) {\fontsize{8}{8}\selectfont{}\textbf{d)}};
    \end{tikzpicture}
    \vspace{-0.8cm}
    \caption{\color{black!60!blue} Distribution of four centrality measures depending on character sex, for all characters (main plots) and after filtering them (inset). (a)~Degree. (b)~Eigenvector centrality. (c)~Betweenness. (d)~Closeness. All axes use a logarithmic scale. Figure available at \href{https://doi.org/10.5281/zenodo.6573491}{10.5281/zenodo.6573491} under CC-BY license.}
    \label{fig:CentrDistribSex}
\end{figure*}

Figure~\ref{fig:CentrDistribSex} shows the distribution of the four centrality measures, while distinguishing male and female characters (mixed and undetermined characters are not represented due to their small number). As before, the main plots show the distributions obtained for the unfiltered network, and the insets show the those obtained for the filtered networks. 

\section{Network Figures}
\label{sec:AddNets}
This section provides additional figures representing the whole network (Section~\ref{sec:AddNetsWhole}) as well as arc-specific networks (Section~\ref{sec:AddNetsArcs}).

\subsection{Whole Series}
\label{sec:AddNetsWhole}
\begin{figure*}[htb!]
    \centering
    \includegraphics[trim={5cm 6cm 4cm 5cm}, clip, width=1\linewidth]{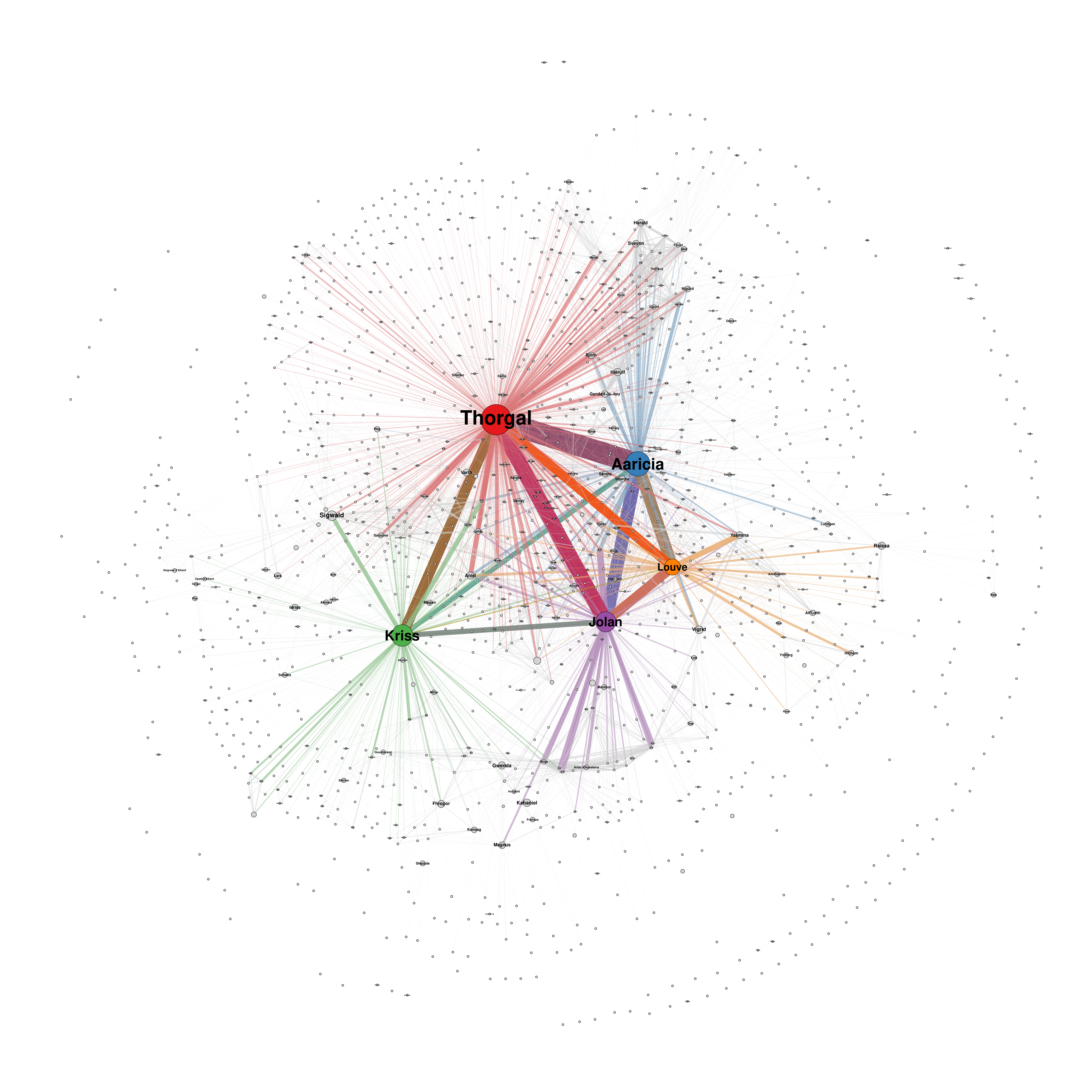}
    \caption{\color{black!60!blue} \textit{Unfiltered} network extracted from the whole \textit{Thorgal} series. Vertex size is a function of betweenness, and edge width is a function of the number of co-occurrences. The five most frequent characters (in terms of scenes) are shown in a specific color: Thorgal (red), his wife Aaricia (blue), their elder son Jolan (purple), their daughter Louve (Orange), and the antagonist Kriss of Valnor (green). See Figure~\ref{fig:FilteredNet} for the \textit{filtered} version of this plot. Figure available at \href{https://doi.org/10.5281/zenodo.6573491}{10.5281/zenodo.6573491} under CC-BY license.}
    \label{fig:UnfilteredNet}
\end{figure*}

\begin{figure*}[htb!]
    \centering
    \includegraphics[trim={5cm 6cm 4cm 5cm}, clip, width=1\linewidth]{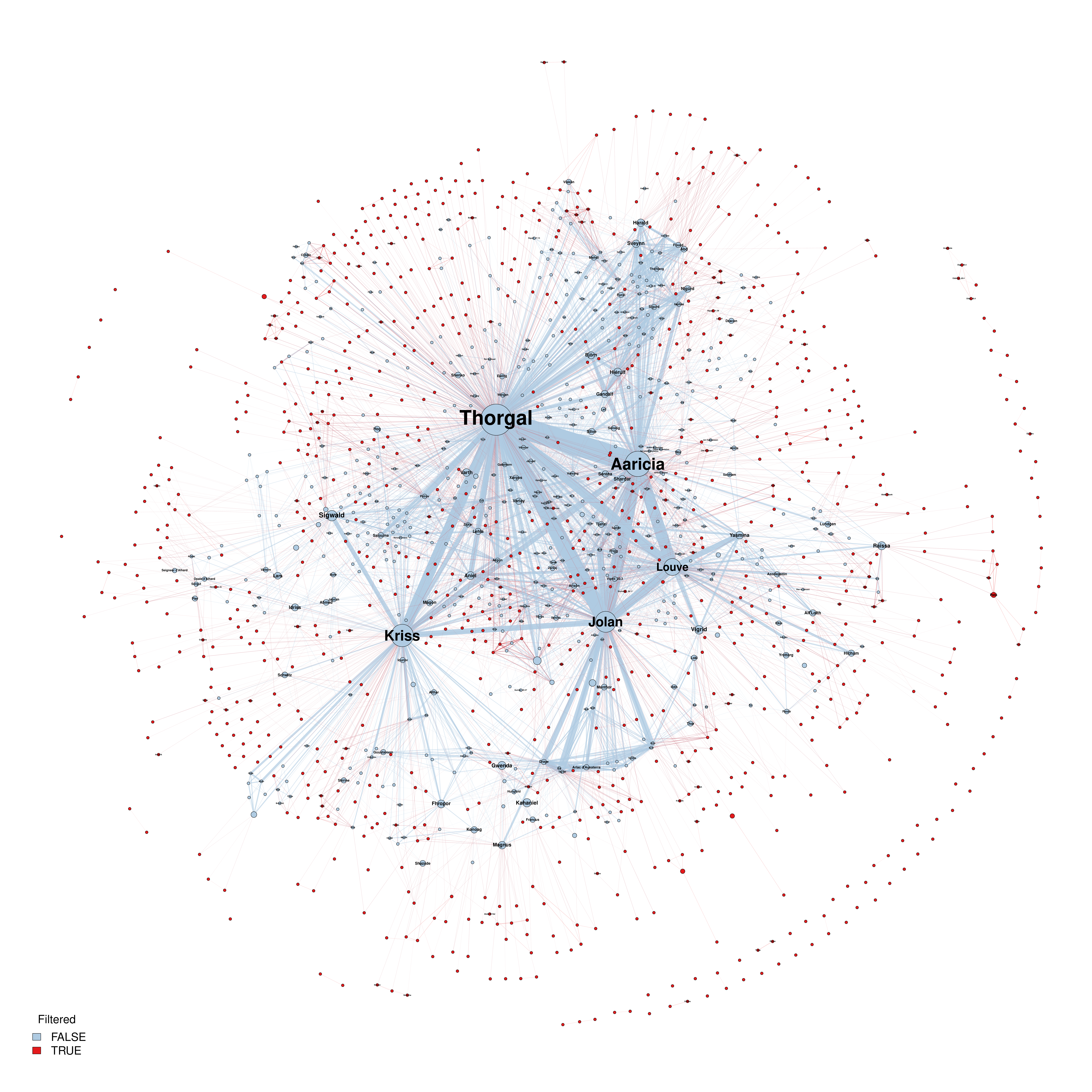}
    \caption{\color{black!60!blue} \textit{Unfiltered} network showing the filtered vertices (in red). Vertex size and edge width have the same meaning as in Figure~\ref{fig:UnfilteredNet}. Figure available at \href{https://doi.org/10.5281/zenodo.6573491}{10.5281/zenodo.6573491} under CC-BY license.}
    \label{fig:UnfilteredNetFilt}
\end{figure*}

Figure~\ref{fig:UnfilteredNet} shows the \textit{unfiltered} version of the Thorgal network, i.e. with all characters, even minor ones. By comparison, Figure~\ref{fig:FilteredNet} represents the \textit{filtered} network, focusing only on primary characters as defined in Section~\ref{sec:DataFiltering}. Figure~\ref{fig:UnfilteredNetFilt} highlights (in red) the vertices that are removed during the filtering step.

\subsection{Arc Networks}
\label{sec:AddNetsArcs}
\begin{figure*}[htb!]
    \centering
    \begin{tikzpicture}
        \useasboundingbox (0.0,0.0) rectangle (\textwidth,0.9\textheight);
        \node[anchor=south west, inner sep=0, text width=\textwidth] (image) at (0,0) {
        	\frame{\includegraphics[trim={15cm 20cm 18cm 16cm}, clip, width=0.32\textwidth]{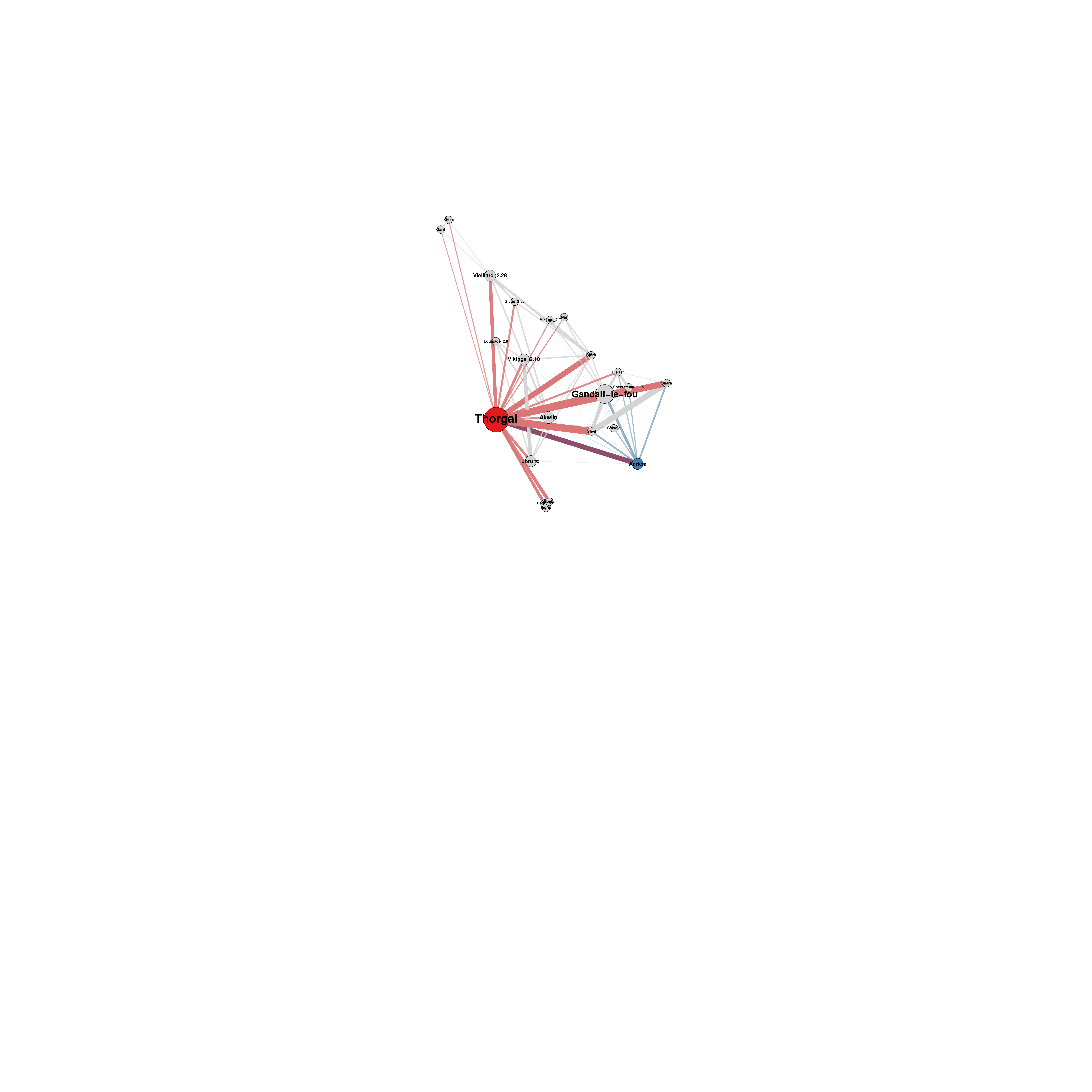}}~~
        	\frame{\includegraphics[trim={15cm 20cm 18cm 16cm}, clip, width=0.32\textwidth]{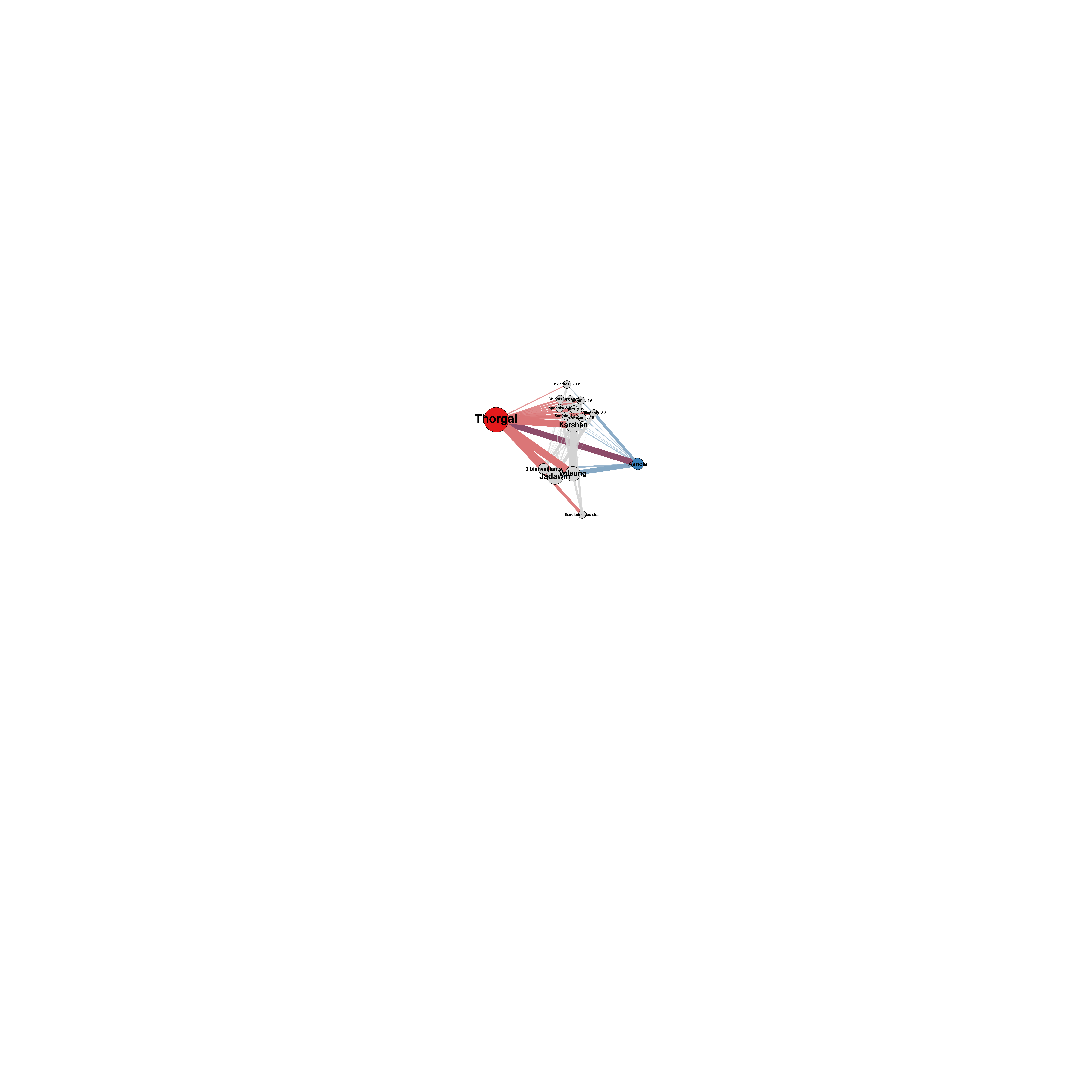}}~~
        	\frame{\includegraphics[trim={15cm 20cm 18cm 16cm}, clip, width=0.32\textwidth]{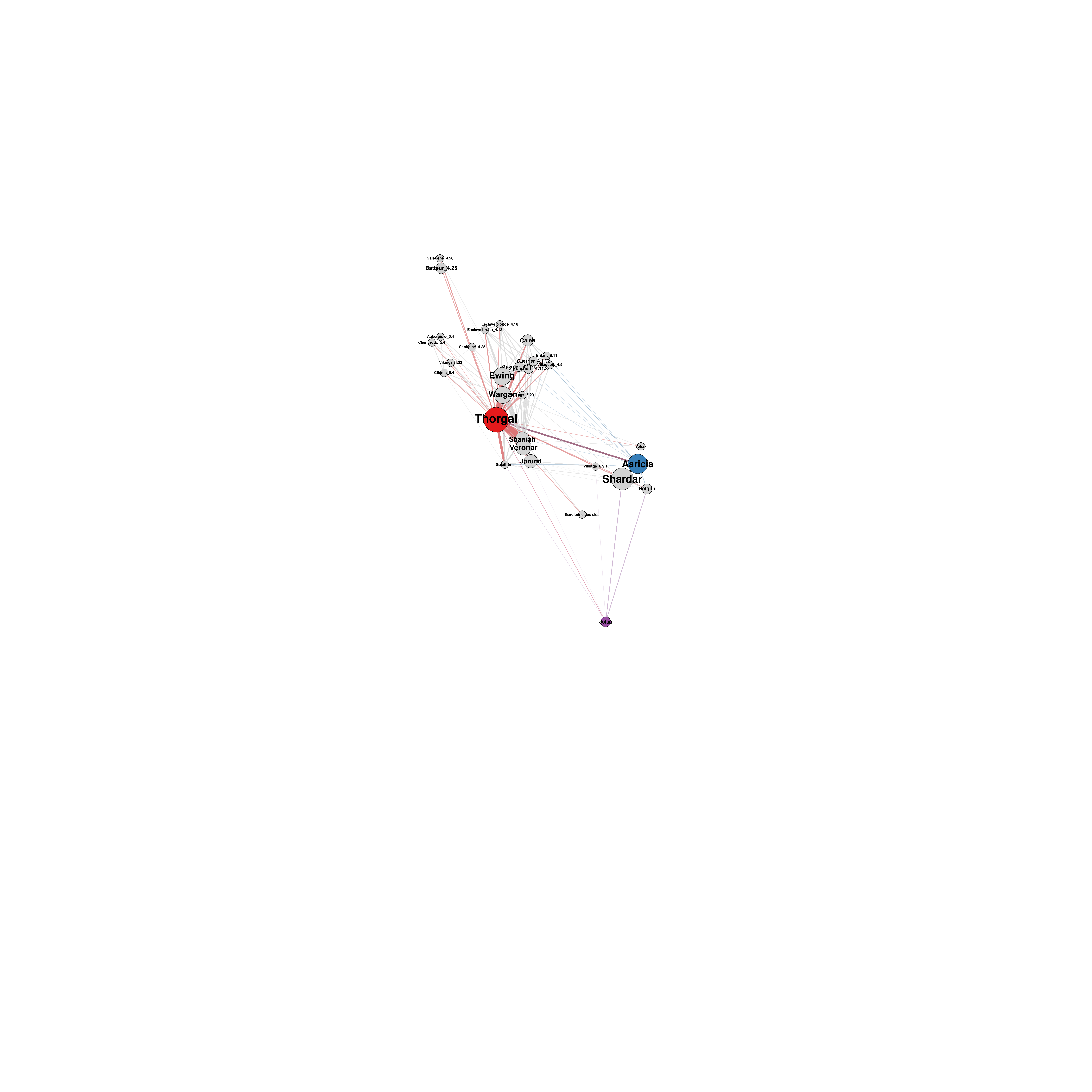}} \\[0.35cm]
        	\frame{\includegraphics[trim={15cm 20cm 18cm 16cm}, clip, width=0.32\textwidth]{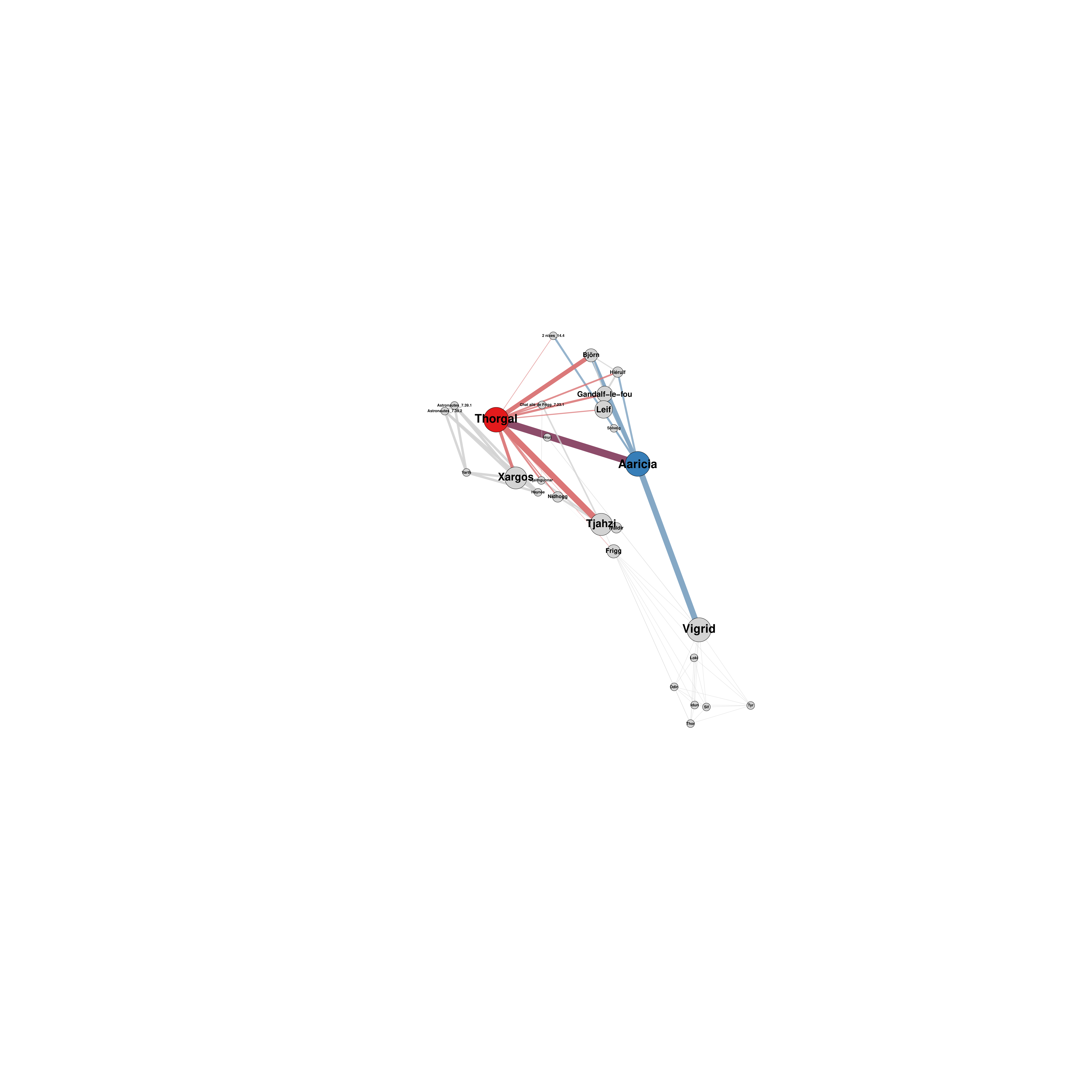}}~~
        	\frame{\includegraphics[trim={15cm 20cm 18cm 16cm}, clip, width=0.32\textwidth]{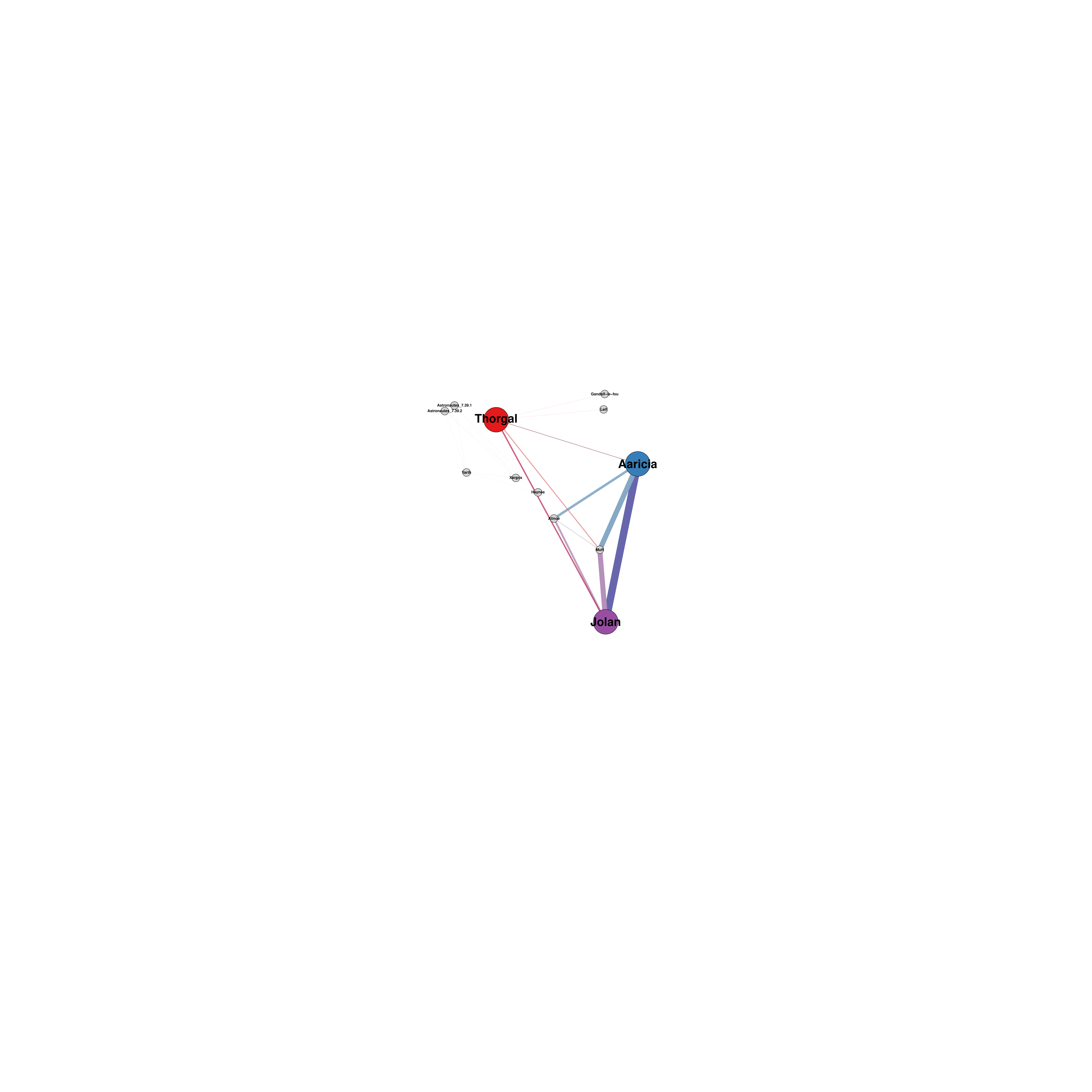}}~~
        	\frame{\includegraphics[trim={15cm 20cm 18cm 16cm}, clip, width=0.32\textwidth]{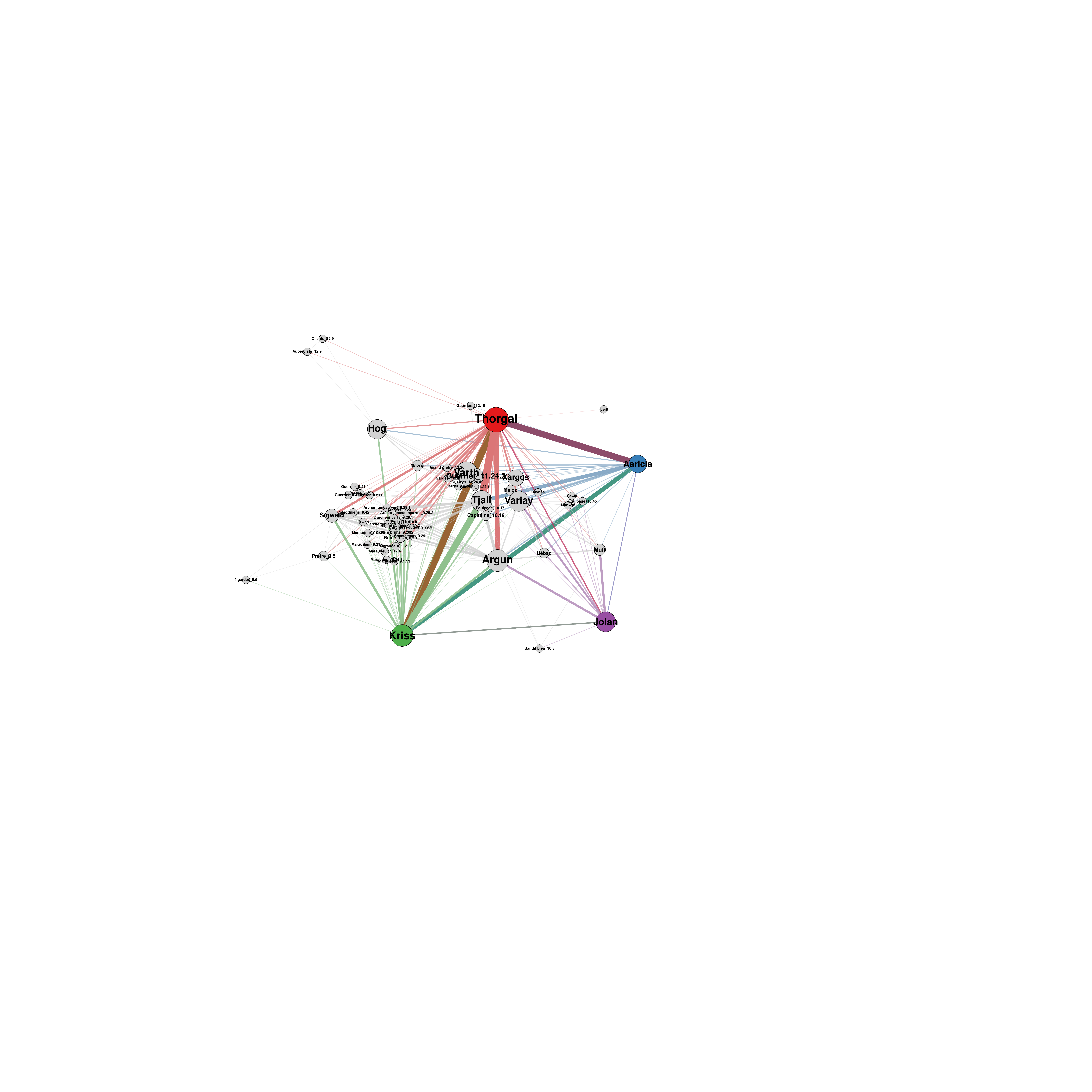}} \\[0.35cm]
        	\frame{\includegraphics[trim={15cm 20cm 18cm 16cm}, clip, width=0.32\textwidth]{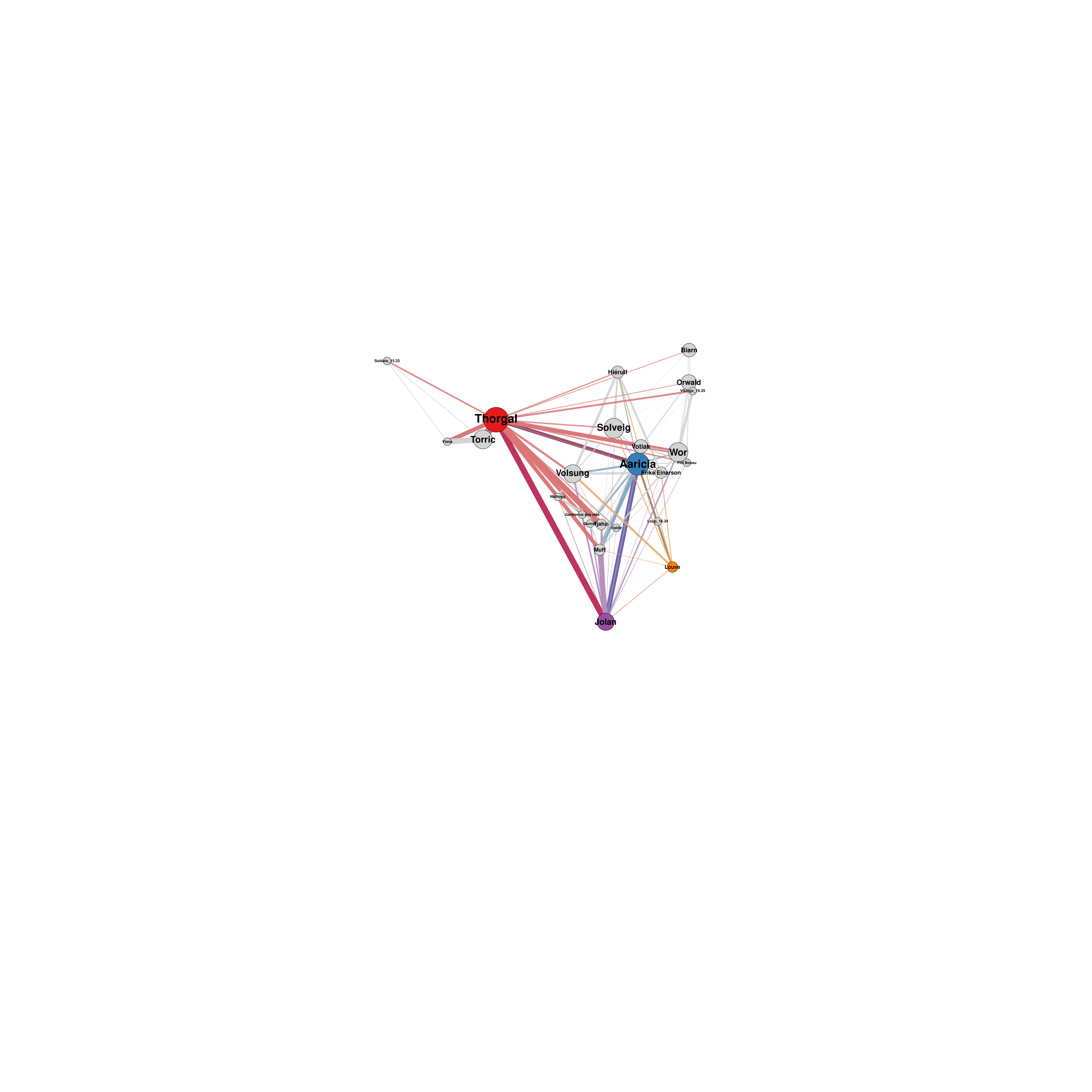}}~~
        	\frame{\includegraphics[trim={15cm 20cm 18cm 16cm}, clip, width=0.32\textwidth]{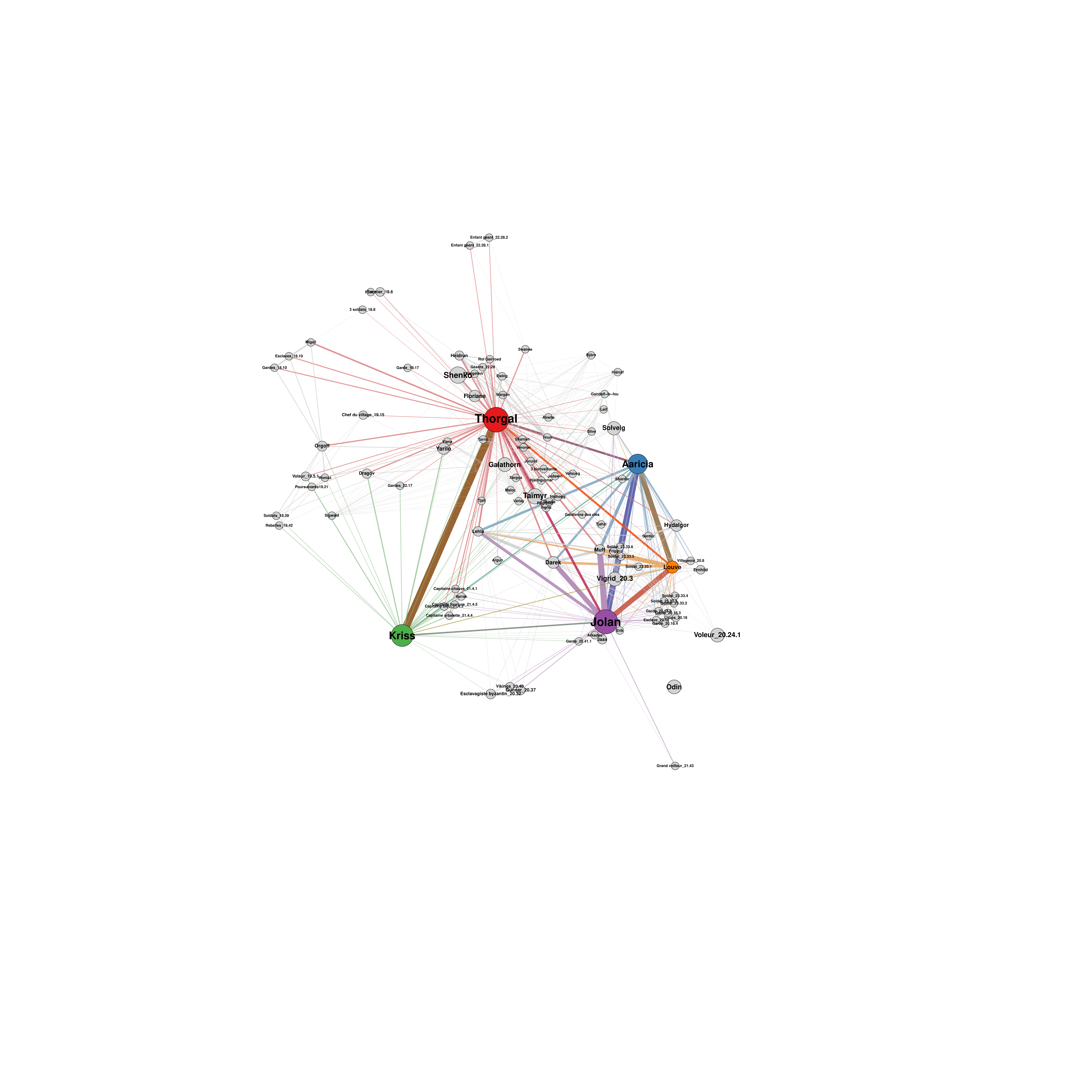}}~~
        	\frame{\includegraphics[trim={15cm 20cm 18cm 16cm}, clip, width=0.32\textwidth]{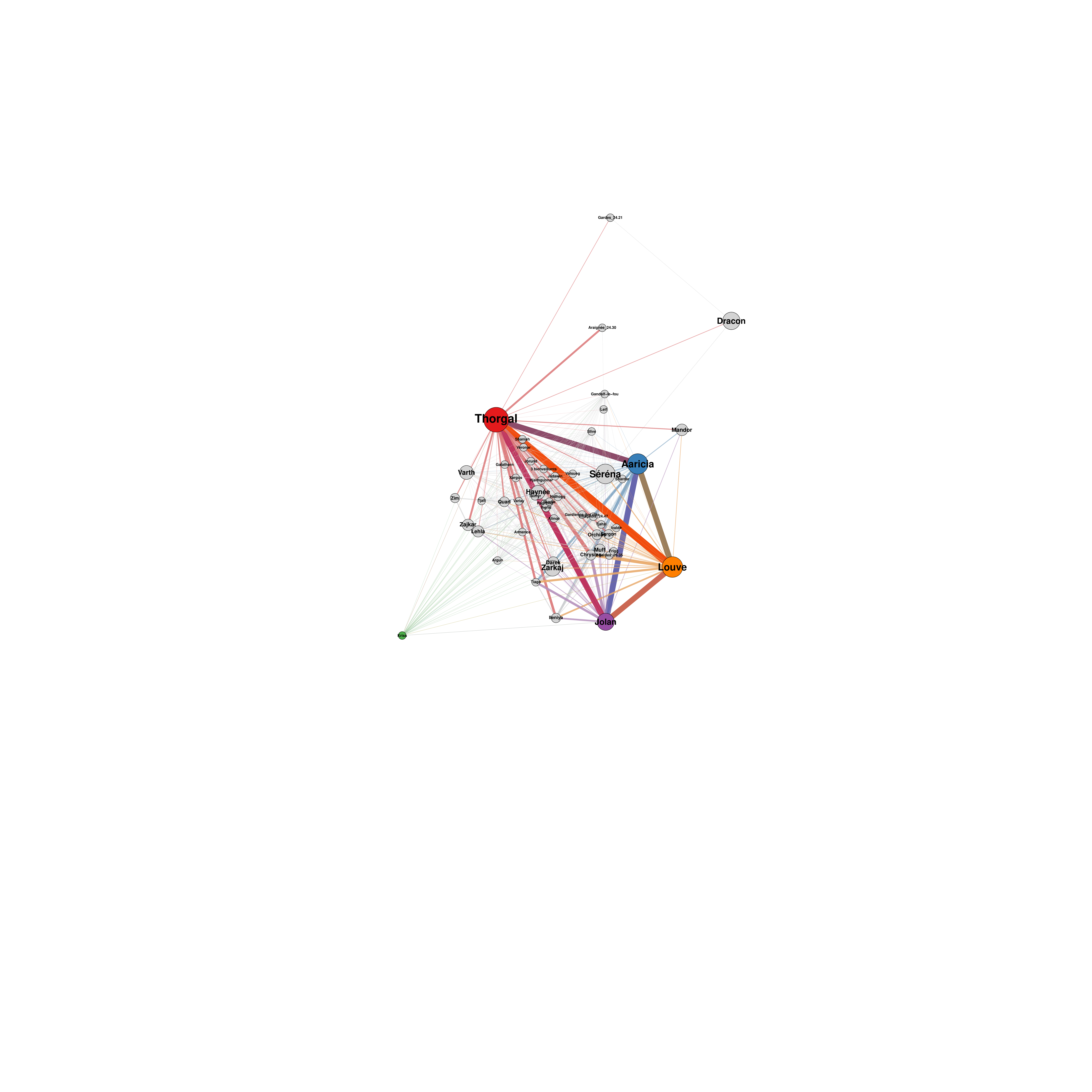}} \\[0.35cm]
        	\frame{\includegraphics[trim={15cm 20cm 18cm 16cm}, clip, width=0.32\textwidth]{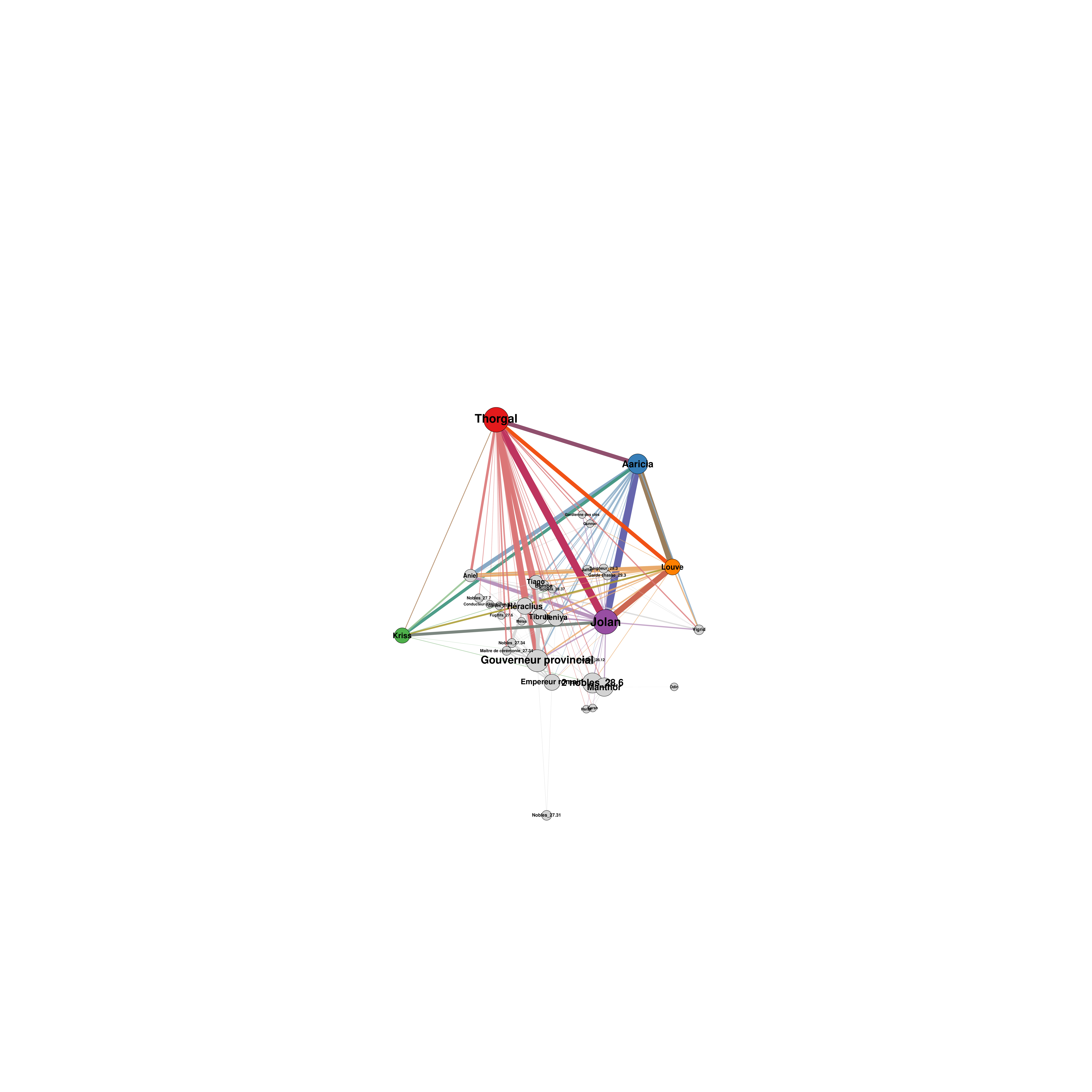}}~~
        	\frame{\includegraphics[trim={15cm 20cm 18cm 16cm}, clip, width=0.32\textwidth]{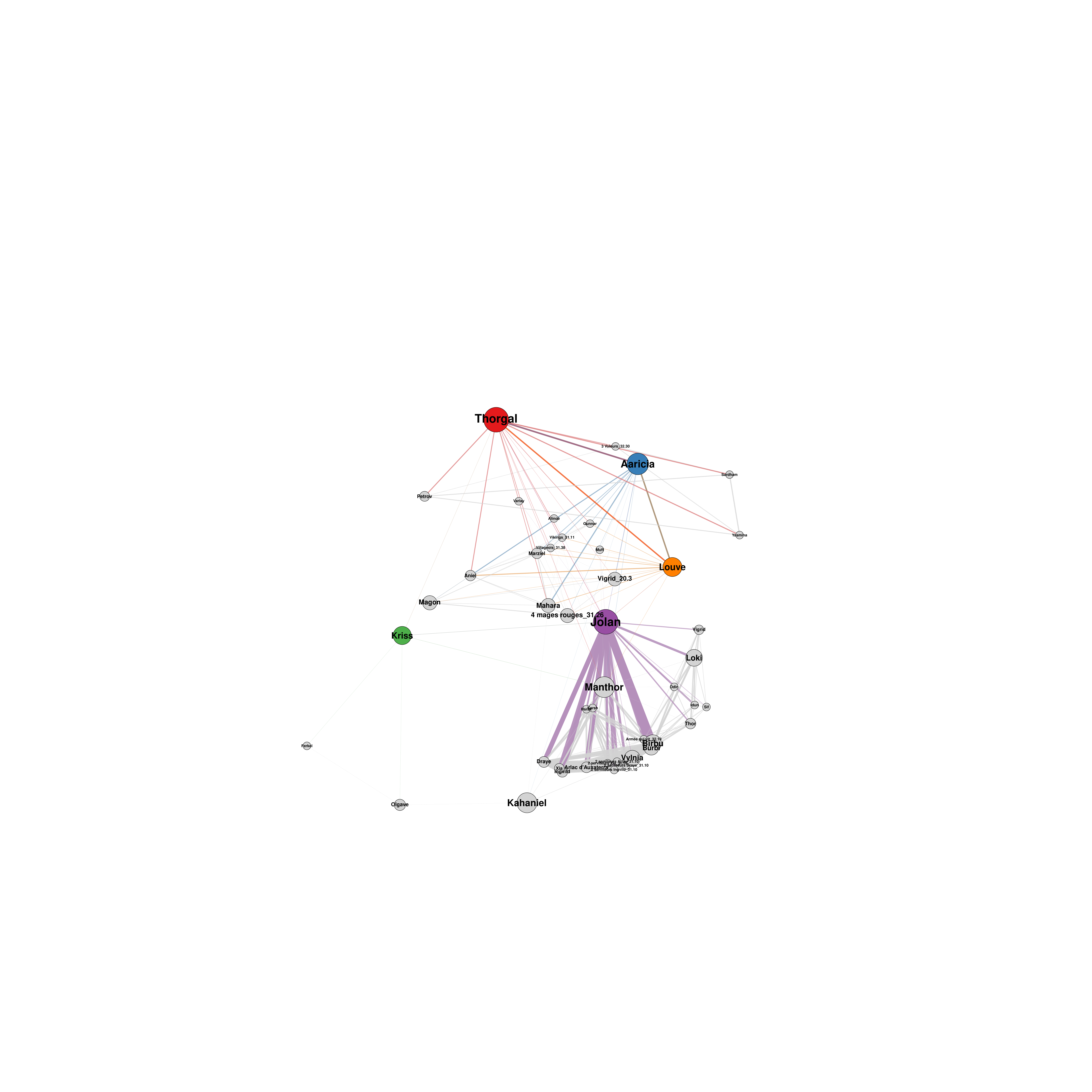}}~~
        	\frame{\includegraphics[trim={15cm 20cm 18cm 16cm}, clip, width=0.32\textwidth]{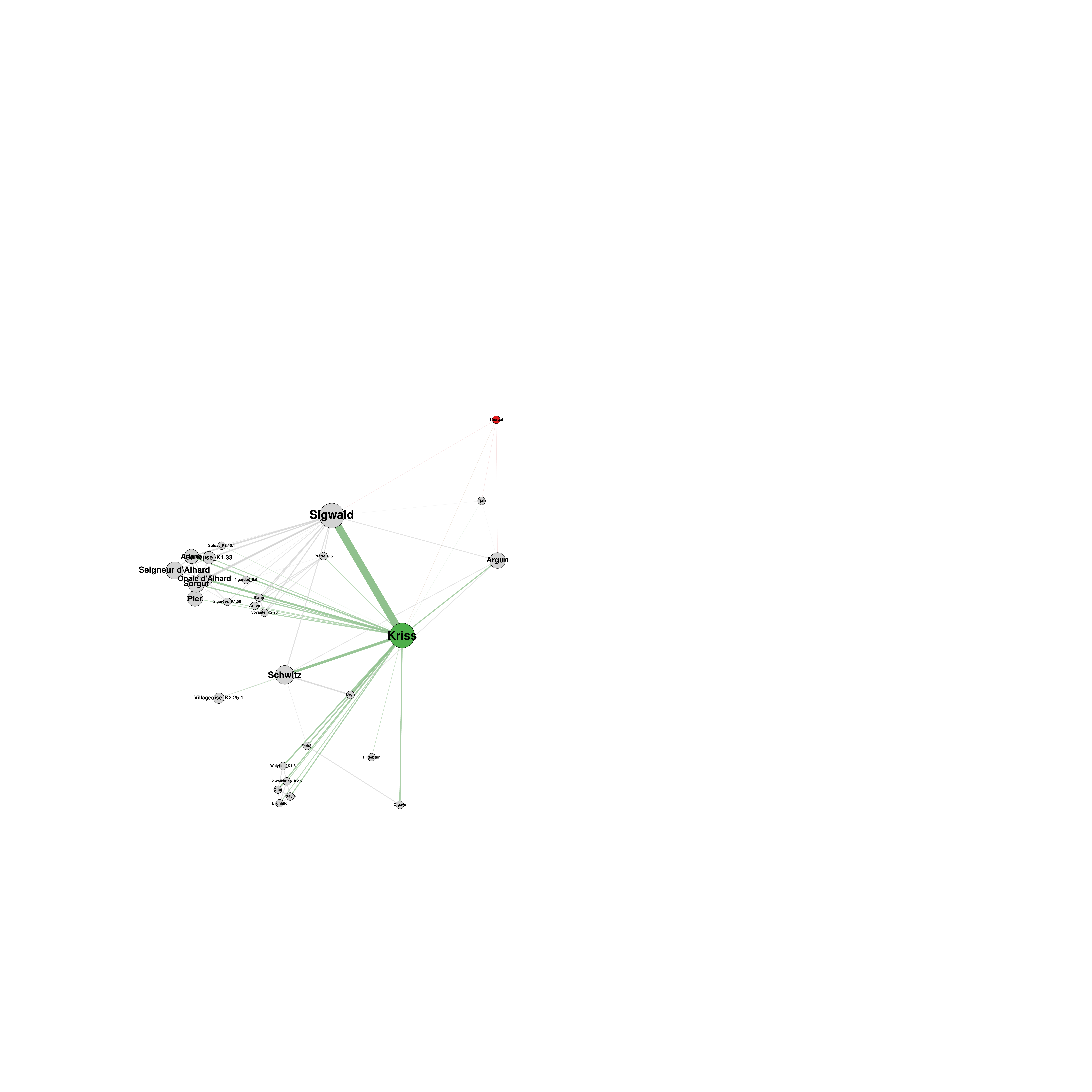}}
        };
        \node[anchor=west] at ( 0.00,16.40) {\fontsize{8}{8}\selectfont{}\textbf{Arc \#1}};
        \node[anchor=west] at ( 4.40,16.40) {\fontsize{8}{8}\selectfont{}\textbf{Arc \#2}};
        \node[anchor=west] at ( 8.85,16.40) {\fontsize{8}{8}\selectfont{}\textbf{Arc \#3}};
        \node[anchor=west] at ( 0.00,12.15) {\fontsize{8}{8}\selectfont{}\textbf{Arc \#4}};
        \node[anchor=west] at ( 4.40,12.15) {\fontsize{8}{8}\selectfont{}\textbf{Arc \#5}};
        \node[anchor=west] at ( 8.85,12.15) {\fontsize{8}{8}\selectfont{}\textbf{Arc \#6}};
        \node[anchor=west] at ( 0.00, 7.90) {\fontsize{8}{8}\selectfont{}\textbf{Arc \#7}};
        \node[anchor=west] at ( 4.40, 7.90) {\fontsize{8}{8}\selectfont{}\textbf{Arc \#8}};
        \node[anchor=west] at ( 8.85, 7.90) {\fontsize{8}{8}\selectfont{}\textbf{Arc \#9}};
        \node[anchor=west] at ( 0.00, 3.65) {\fontsize{8}{8}\selectfont{}\textbf{Arc \#10}};
        \node[anchor=west] at ( 4.40, 3.65) {\fontsize{8}{8}\selectfont{}\textbf{Arc \#11}};
        \node[anchor=west] at ( 8.85, 3.65) {\fontsize{8}{8}\selectfont{}\textbf{Arc \#12}};
    \end{tikzpicture}
    \caption{\color{black!60!blue} Filtered characters involved in narrative arcs \#1--12 of the \textit{Thorgal} series. As in Figures~\ref{fig:FilteredNet} \&~\ref{fig:UnfilteredNet}, vertex size is a function of betweenness, edge width is a function of the number of co-occurrences, and the five most frequent characters are shown in a specific color: Thorgal (red), his wife Aaricia (blue), their elder son Jolan (purple), their daughter Louve (Orange), and the antagonist Kriss of Valnor (green). Vertex positions are fixed, to ease visual comparison. See Figure~\ref{fig:Arcs2} for narrative arcs \#13--23, and Table~\ref{tab:NarrArcs} for the arc titles. Figure available at \href{https://doi.org/10.5281/zenodo.6573491}{10.5281/zenodo.6573491} under CC-BY license.}
    \label{fig:Arcs1}
\end{figure*}

\begin{figure*}[htb!]
    \centering
    \begin{tikzpicture}
        \useasboundingbox (0.0,0.0) rectangle (\textwidth,0.9\textheight);
        \node[anchor=south west, inner sep=0, text width=\textwidth] (image) at (0,0) {
        	\frame{\includegraphics[trim={15cm 20cm 18cm 16cm}, clip, width=0.32\textwidth]{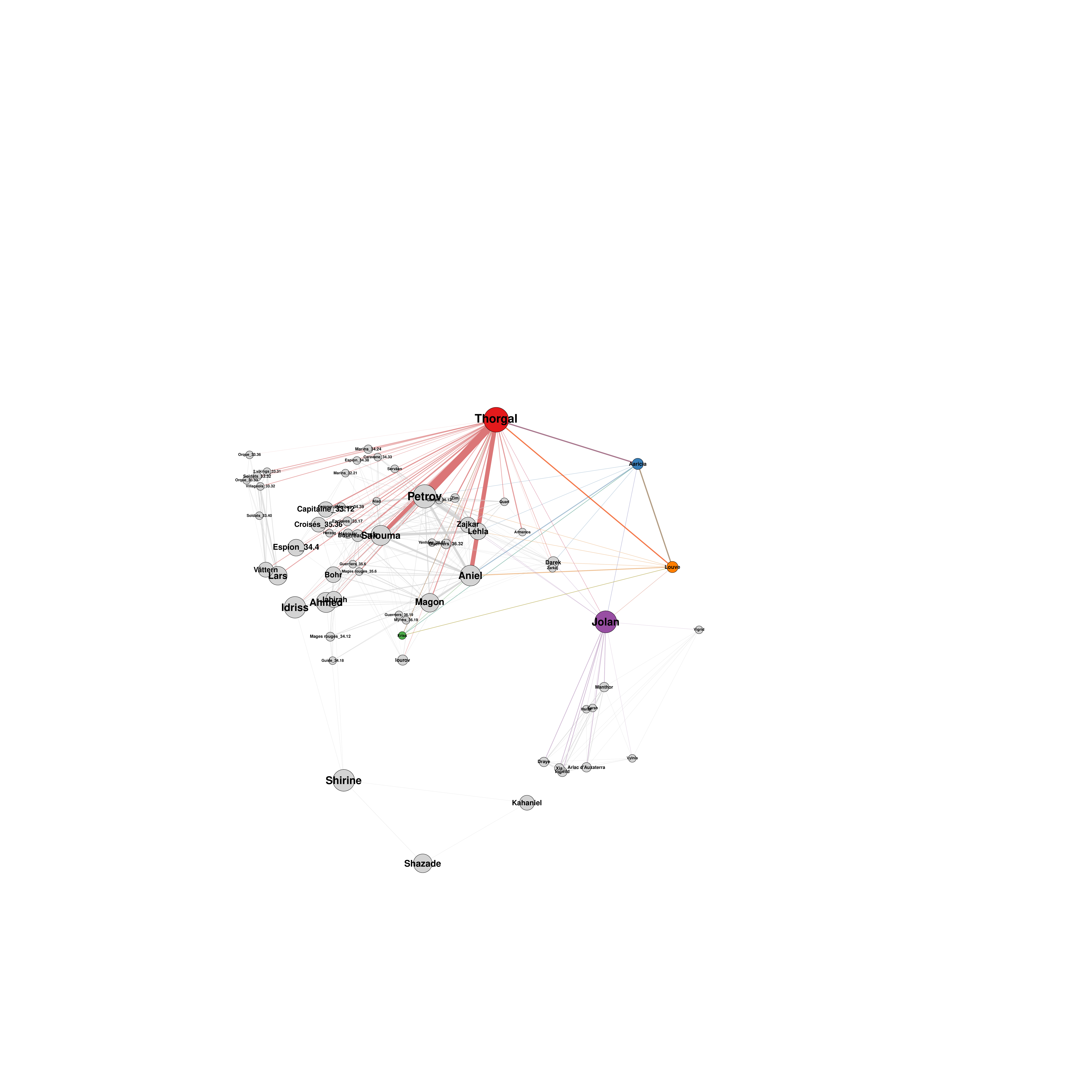}}~~
        	\frame{\includegraphics[trim={15cm 20cm 18cm 16cm}, clip, width=0.32\textwidth]{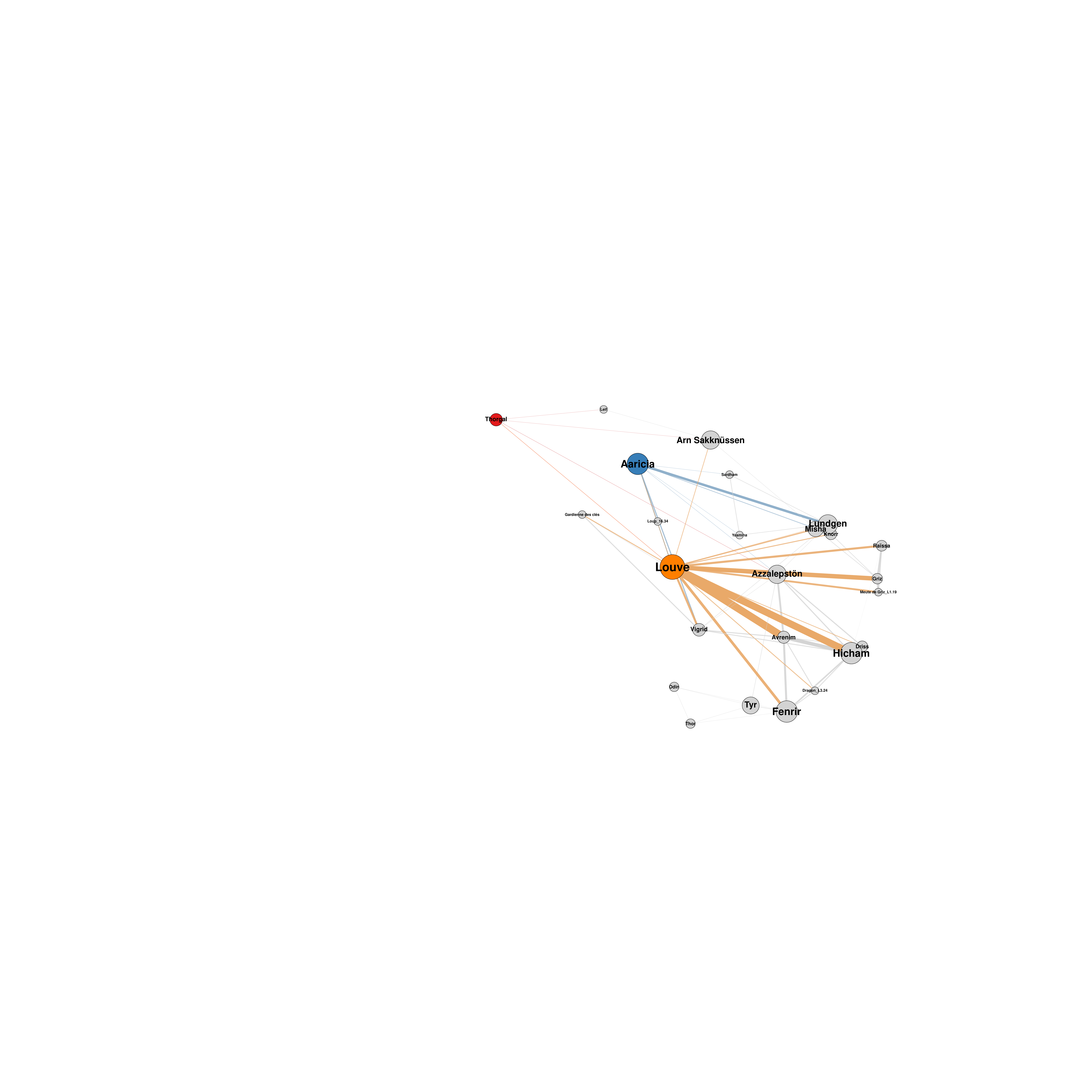}}~~
        	\frame{\includegraphics[trim={15cm 20cm 18cm 16cm}, clip, width=0.32\textwidth]{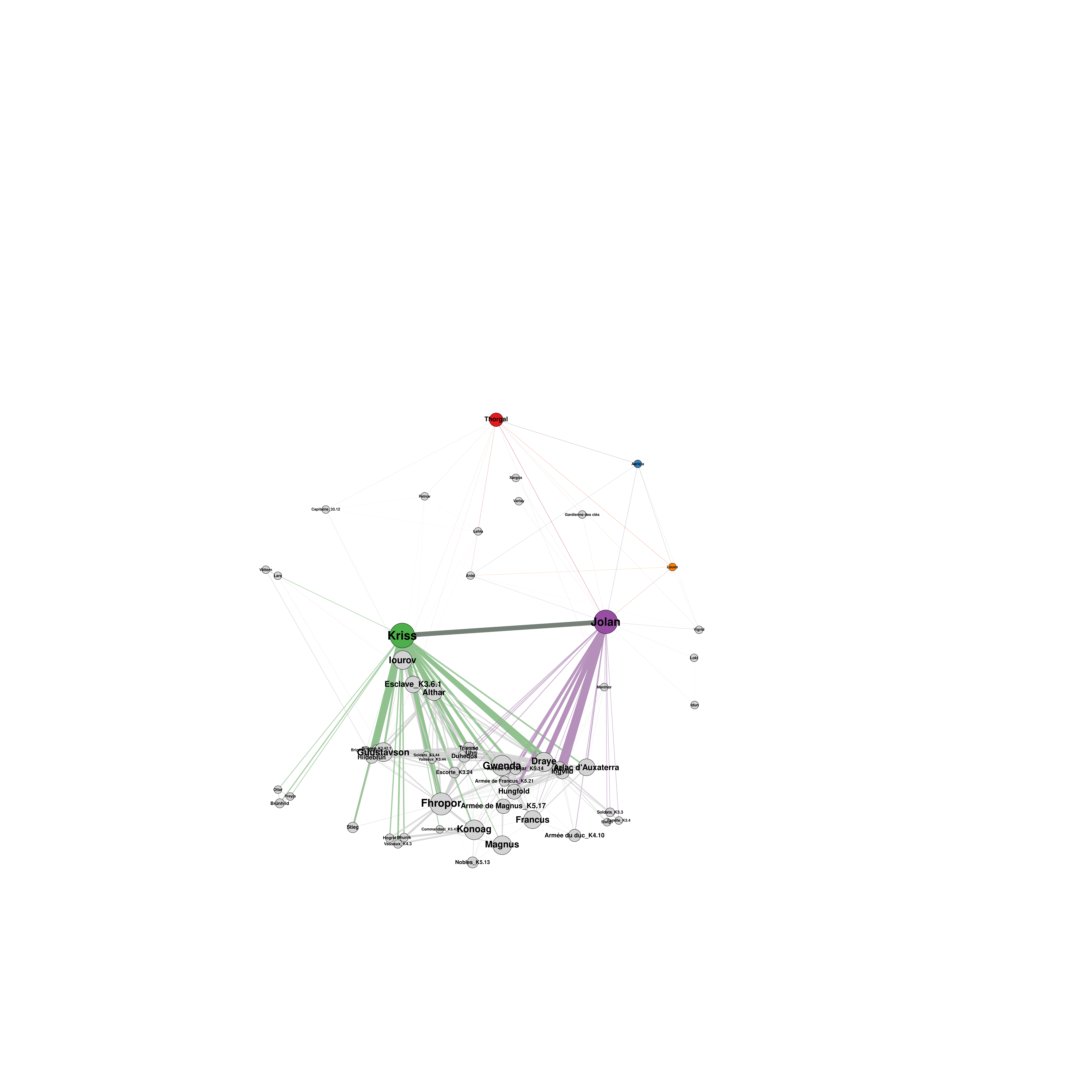}} \\[0.35cm]
        	\frame{\includegraphics[trim={15cm 20cm 18cm 16cm}, clip, width=0.32\textwidth]{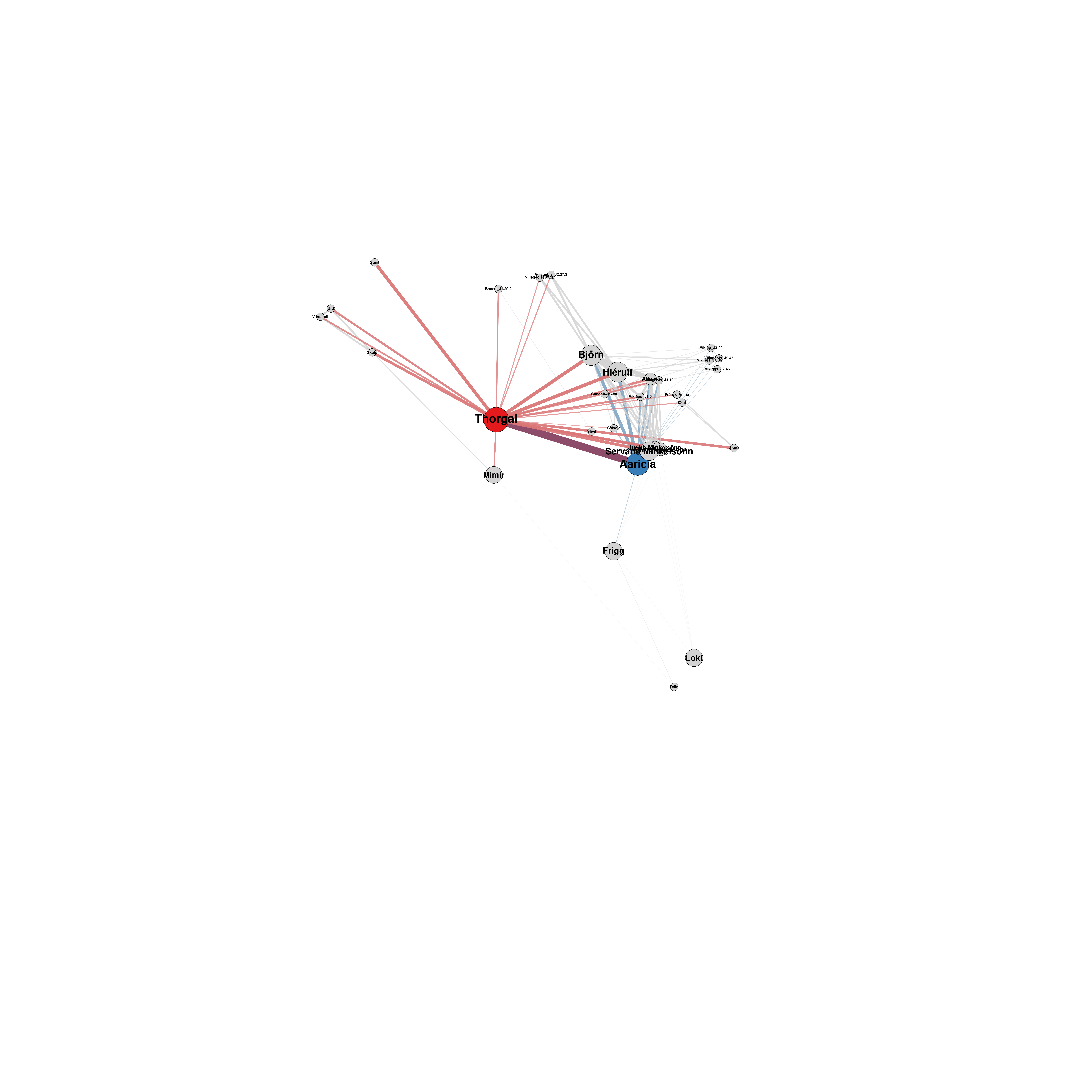}}~~
        	\frame{\includegraphics[trim={15cm 20cm 18cm 16cm}, clip, width=0.32\textwidth]{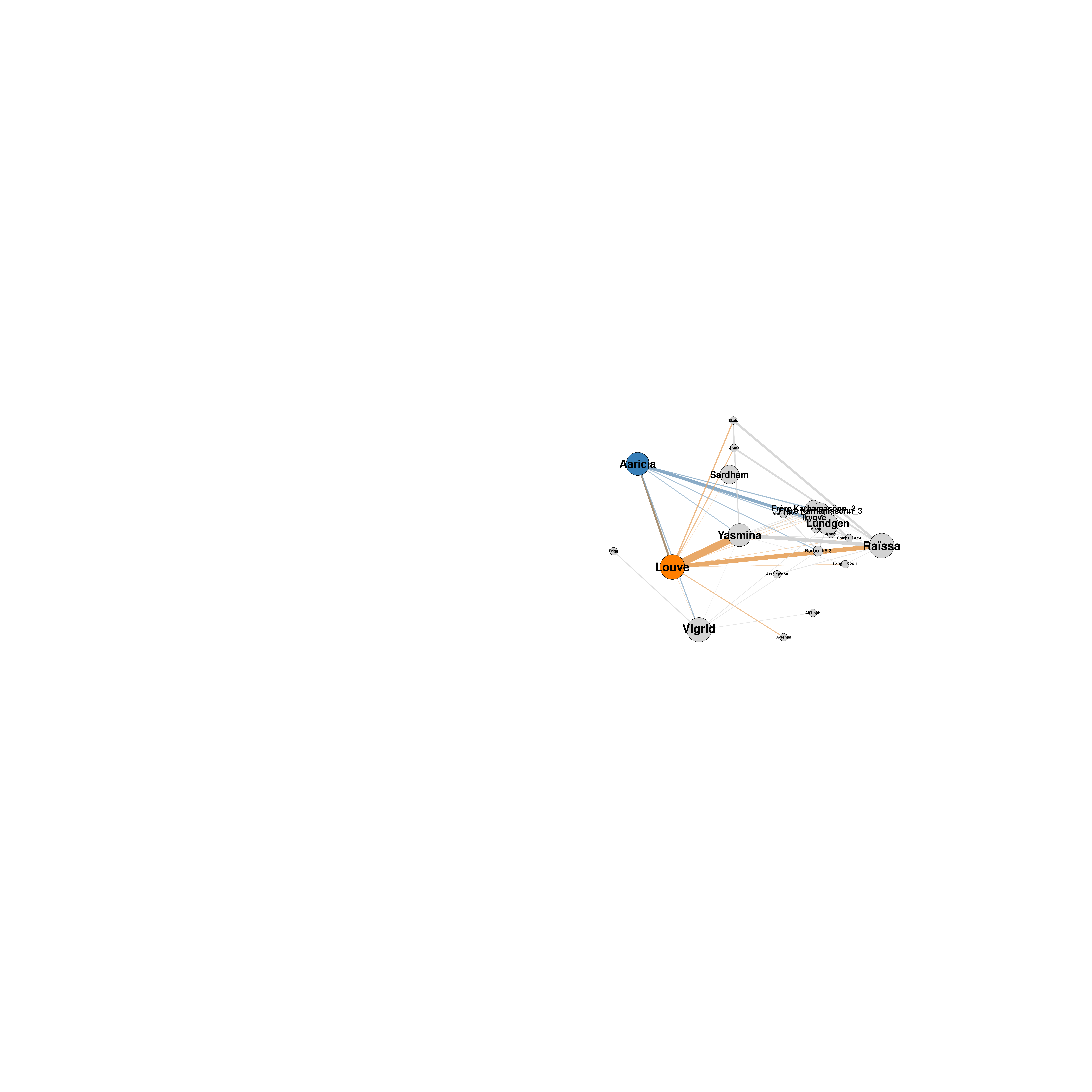}}~~
        	\frame{\includegraphics[trim={15cm 20cm 18cm 16cm}, clip, width=0.32\textwidth]{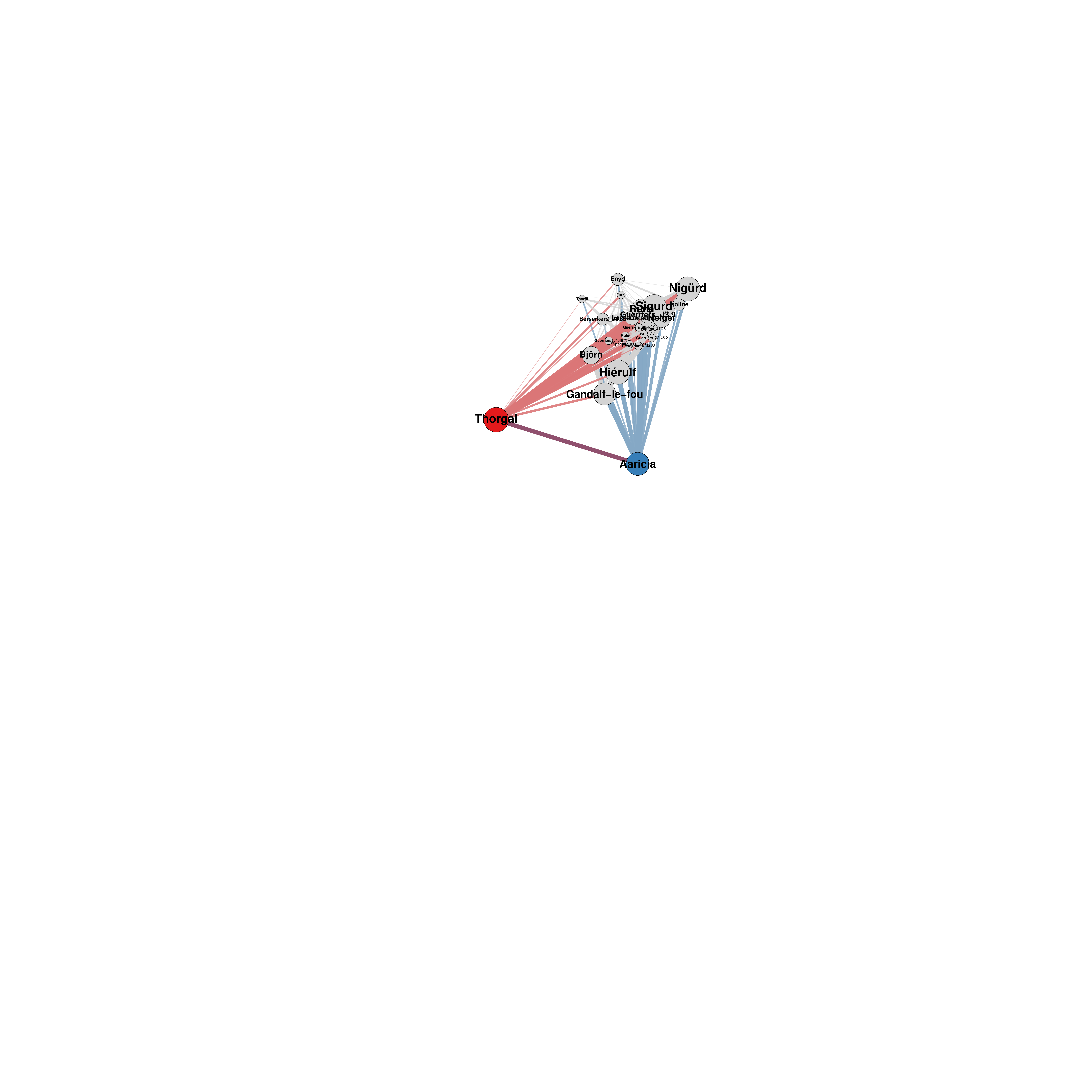}} \\[0.35cm]
        	\frame{\includegraphics[trim={15cm 20cm 18cm 16cm}, clip, width=0.32\textwidth]{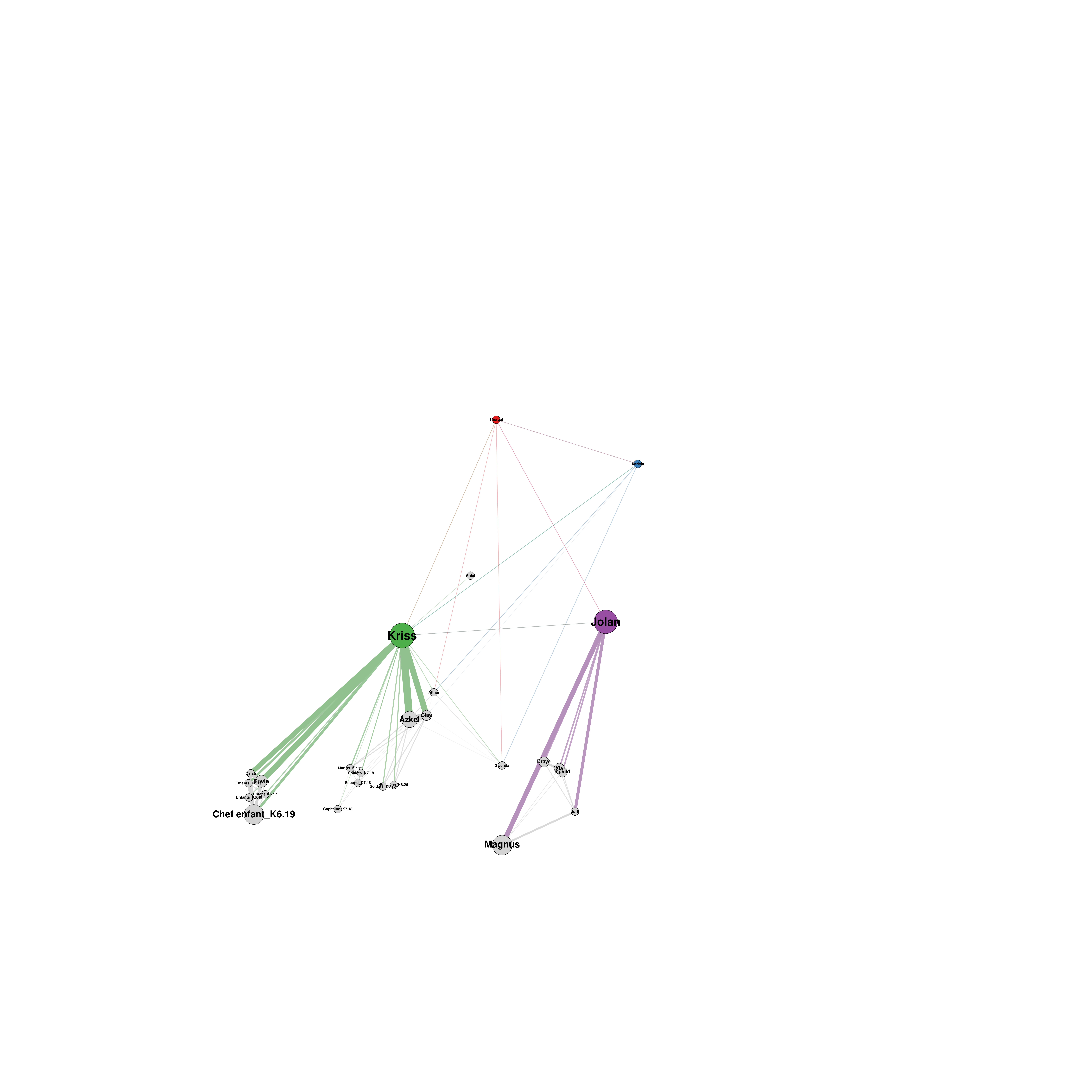}}~~
        	\frame{\includegraphics[trim={15cm 20cm 18cm 16cm}, clip, width=0.32\textwidth]{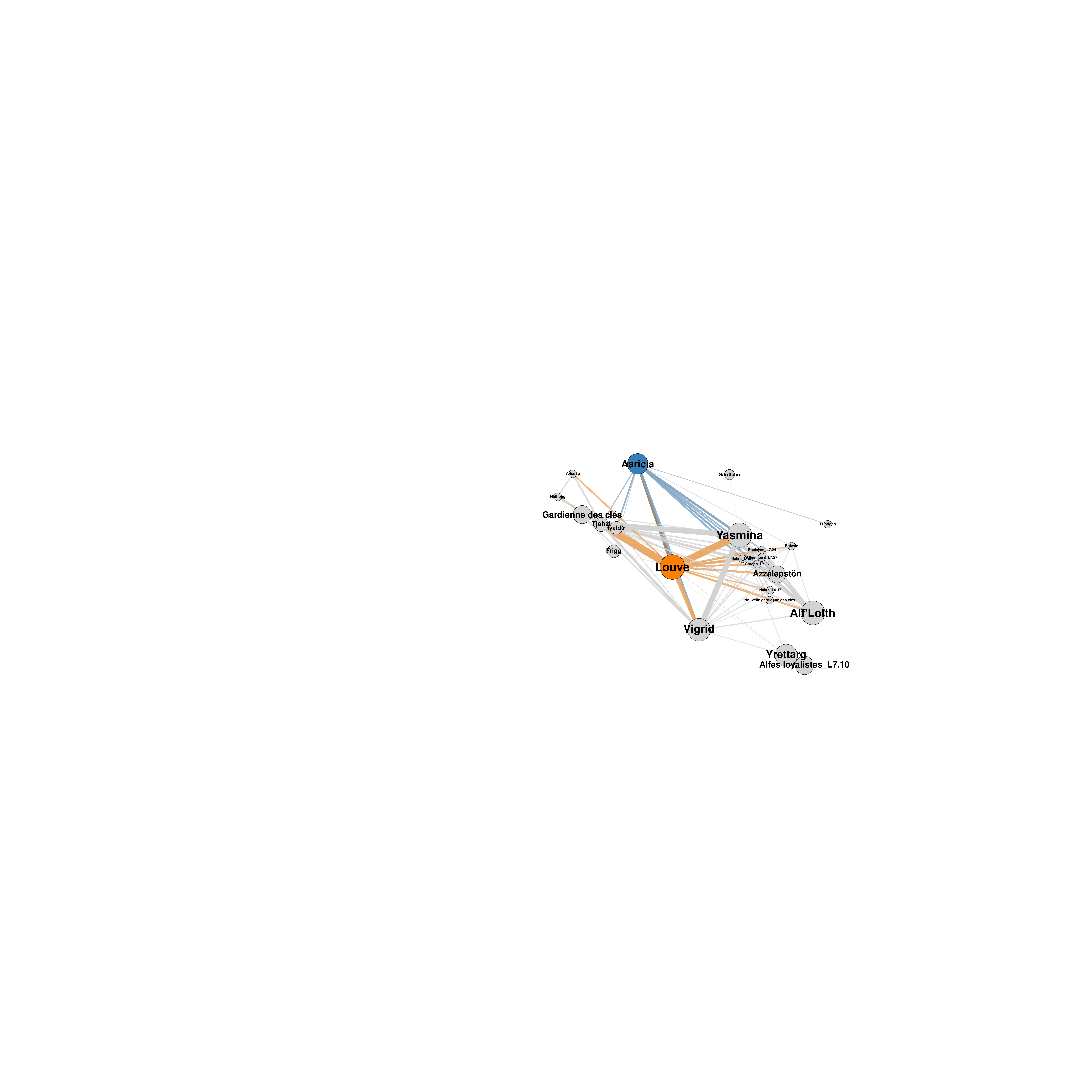}}~~
        	\frame{\includegraphics[trim={15cm 20cm 18cm 16cm}, clip, width=0.32\textwidth]{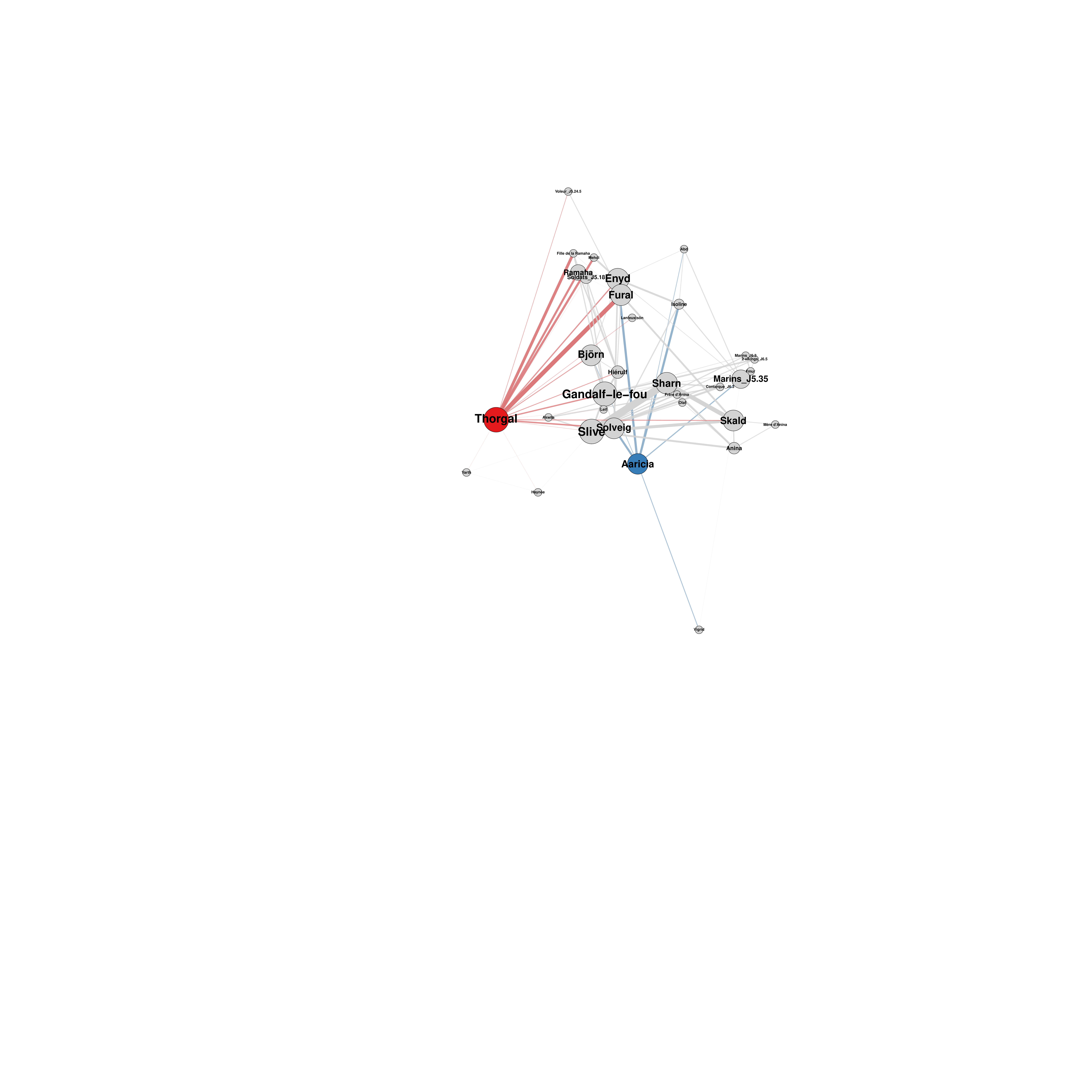}} \\[0.35cm]
        	\frame{\includegraphics[trim={15cm 20cm 18cm 16cm}, clip, width=0.32\textwidth]{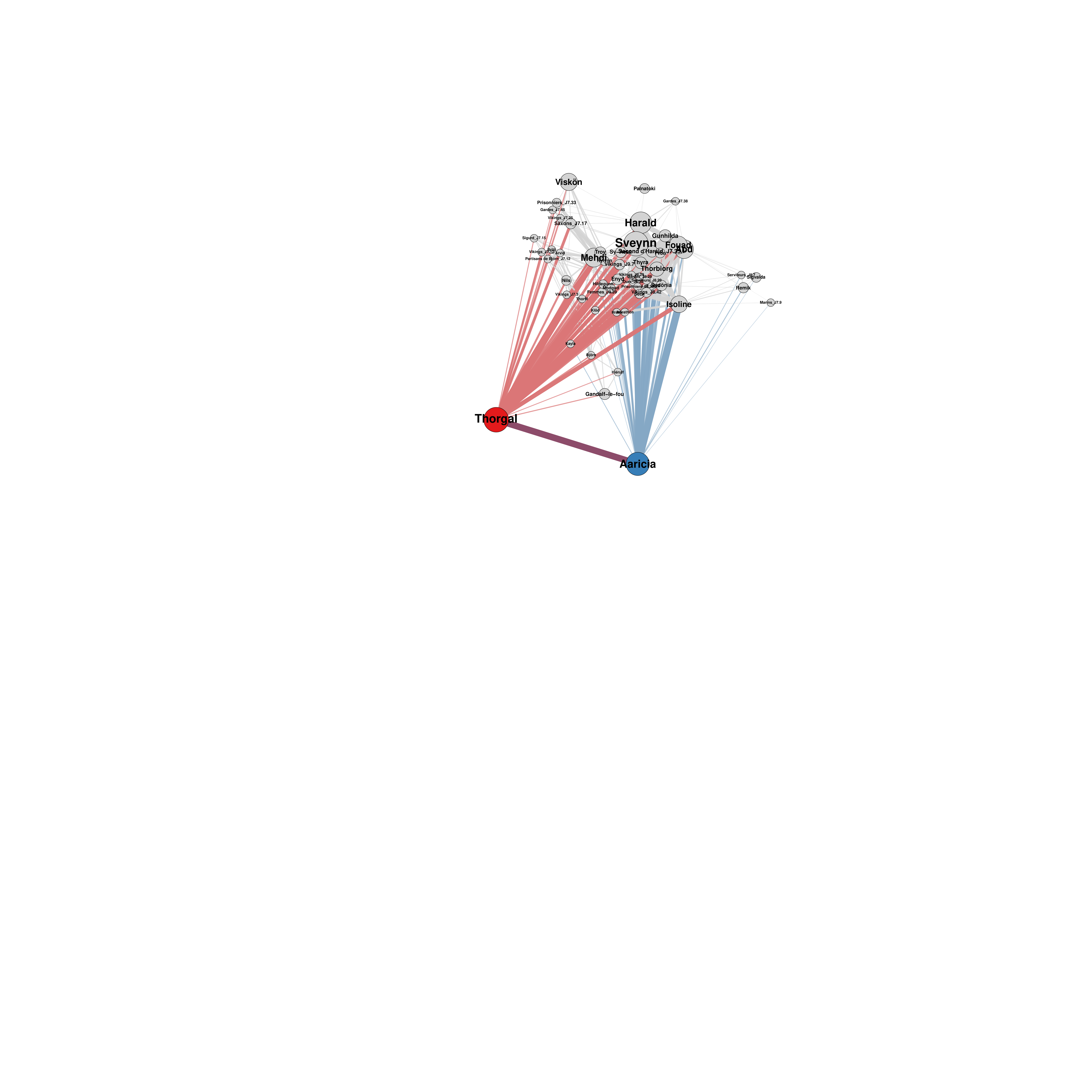}}~~
        	\frame{\includegraphics[trim={15cm 20cm 18cm 16cm}, clip, width=0.32\textwidth]{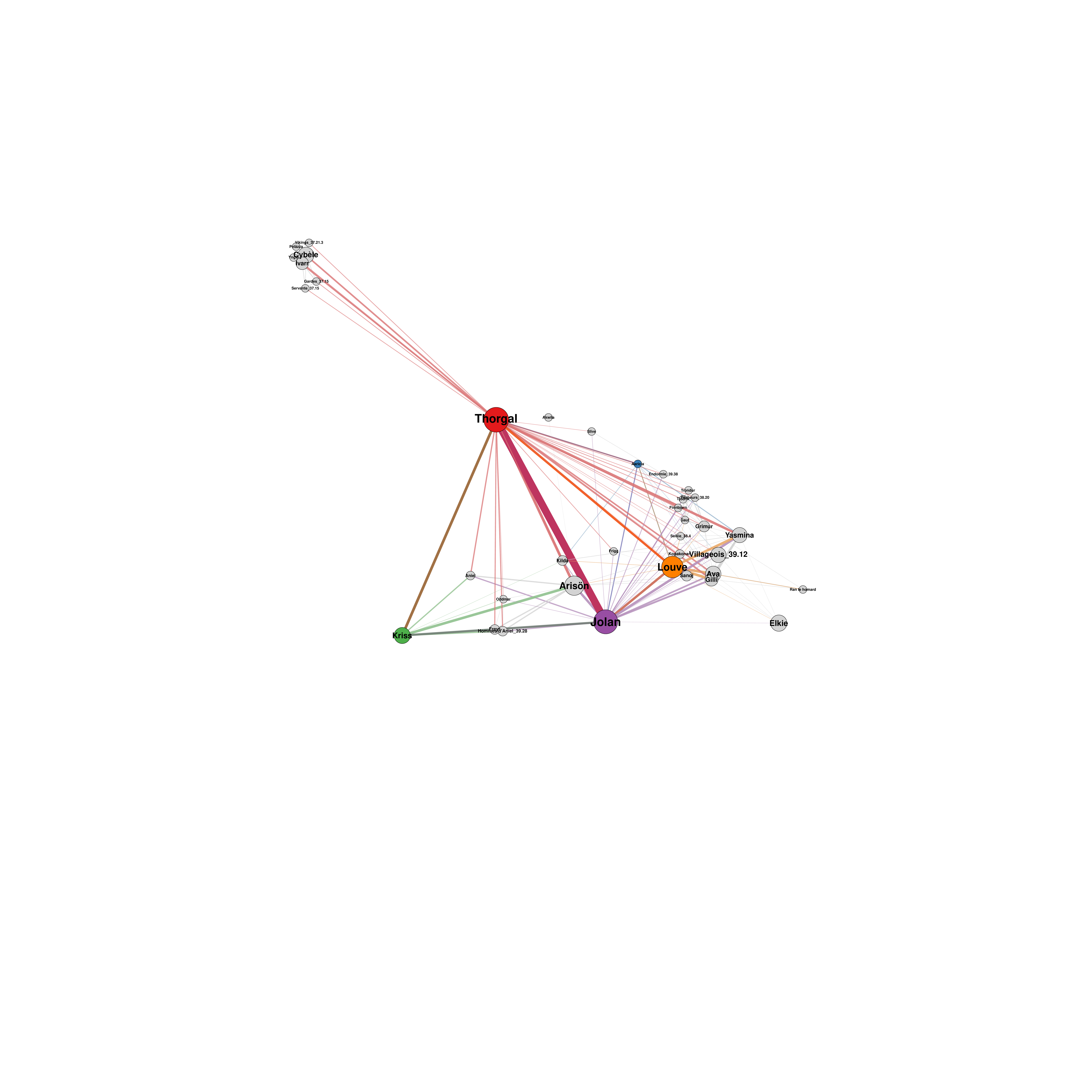}}~~
        };
        \node[anchor=west] at ( 0.00,16.40) {\fontsize{8}{8}\selectfont{}\textbf{Arc \#13}};
        \node[anchor=west] at ( 4.40,16.40) {\fontsize{8}{8}\selectfont{}\textbf{Arc \#14}};
        \node[anchor=west] at ( 8.85,16.40) {\fontsize{8}{8}\selectfont{}\textbf{Arc \#15}};
        \node[anchor=west] at ( 0.00,12.15) {\fontsize{8}{8}\selectfont{}\textbf{Arc \#16}};
        \node[anchor=west] at ( 4.40,12.15) {\fontsize{8}{8}\selectfont{}\textbf{Arc \#17}};
        \node[anchor=west] at ( 8.85,12.15) {\fontsize{8}{8}\selectfont{}\textbf{Arc \#18}};
        \node[anchor=west] at ( 0.00, 7.90) {\fontsize{8}{8}\selectfont{}\textbf{Arc \#19}};
        \node[anchor=west] at ( 4.40, 7.90) {\fontsize{8}{8}\selectfont{}\textbf{Arc \#20}};
        \node[anchor=west] at ( 8.85, 7.90) {\fontsize{8}{8}\selectfont{}\textbf{Arc \#21}};
        \node[anchor=west] at ( 0.00, 3.65) {\fontsize{8}{8}\selectfont{}\textbf{Arc \#22}};
        \node[anchor=west] at ( 4.40, 3.65) {\fontsize{8}{8}\selectfont{}\textbf{Arc \#23}};
    \end{tikzpicture}
    \caption{\color{black!60!blue} Filtered characters involved in narrative arcs \#13--23 of the \textit{Thorgal} series. As in Figures~\ref{fig:FilteredNet} \&~\ref{fig:UnfilteredNet}, vertex size is a function of betweenness, edge width is a function of the number of co-occurrences, and the five most frequent characters are shown in a specific color: Thorgal (red), his wife Aaricia (blue), their elder son Jolan (purple), their daughter Louve (Orange), and the antagonist Kriss of Valnor (green). Vertex positions are fixed, to ease visual comparison. See Figure~\ref{fig:Arcs1} for narrative arcs \#1--12, and Table~\ref{tab:NarrArcs} for the arc titles. Figure available at \href{https://doi.org/10.5281/zenodo.6573491}{10.5281/zenodo.6573491} under CC-BY license.}
    \label{fig:Arcs2}
\end{figure*}

Figures~\ref{fig:Arcs1} and~\ref{fig:Arcs2} show the character networks obtained for each narrative arc of \textit{Thorgal}, as listed in Table~\ref{tab:NarrArcs}. To ease visual comparison, all plots use the same layout, so that the same vertex is always located at the same place.

\end{document}